\documentclass[aps,prb,footinbib,showpacs,twocolumn,showkeys,amssymb,amsmath,superscriptaddress,groupedaddress,reprint]{revtex4-1}
\pdfoutput=1

\usepackage{hyperref}

\usepackage{amsmath}
\usepackage{braket}
\usepackage{graphicx}
\usepackage{graphics}
\usepackage{color}
\usepackage{float}

\usepackage{verbatim}

\begin{document}

\title{Study of counterintuitive transport properties in the Aubry-Andr\'e-Harper model via entanglement entropy and persistent current}

\author{Nilanjan Roy}
\author{Auditya Sharma}
\affiliation{Department of Physics, Indian Institute of Science Education and Research, Bhopal, Madhya Pradesh 462066, India}

\date{\today}

\begin{abstract}
The single particle eigenstates of the Aubry-Andr\'e-Harper model are
known to show a delocalization-localization transition at a finite
strength of the quasi-periodic disorder. In this work, we point out
that an intimate relationship exists between the sub-band structure of the spectrum
and transport properties of the model.  To capture the transport
properties we have not only used a variety of single-particle measures
like inverse participation ratio, and von Neumann entropy, but also
many-particle measures such as persistent current and its variance,
and many body entanglement entropy. The many-particle measures are
very sensitive to the sub-band structure of the spectrum. Even in the
delocalized phase, surprisingly the entanglement entropy is
substantially suppressed when the Fermi level is in the band gaps
whereas the persistent current is vanishingly small for the same
locations of the Fermi level.  The entanglement entropy seems to
follow area-law exclusively for these special locations of Fermi level
or filling fractions of free fermions.  A study of the standard deviation of persistent current
offers further distinguishing features for the special fillings. In the delocalized phase, the
standard deviation vs. mean persistent current curves are
discontinuous for the non-special values of filling fractions and
continuous (closed) for the special values of filling fractions
whereas in the localized phase, these curves become straight lines for
both types of filling fractions. We have also discussed how the results depend on the system size. Our results, specially on the
persistent current, can potentially be tested experimentally using the
present day set-ups based on ultra-cold atoms.
\end{abstract}

\maketitle

\section{Introduction}
Over the last few decades, an immense effort has been expended to understand
properties of quantum systems in the presence of quasi-periodic
potentials~\cite{rmp,kohmoto_rev,maci}. In one dimension, in contrast to the phenomenon of Anderson
localization which is seen even in the presence of an infinitesimally tiny random
potential~\cite{anderson}, a substantial strength of the quasi-periodic potential is required before localization
sets in. Therefore, the one-dimensional (1D) quasi-periodic potential model~\cite{aubry,harper}, also known as the
Aubry-Andr\'e-Harper (AAH) model, admits a ``delocalization-localization transition" at finite strength of the potential. 
The spectrum of the AAH model is known to show self-similar Cantor set
structure~\cite{kohmoto2,hofstadter}. The energy spectrum~\cite{pichard,kohmoto2,kohmoto3}shows band gaps (to be discussed later)
at certain locations, which is related to the quasi-periodicity in the
system.

The phase transition in the model leads to several interesting
transport properties which have been addressed in many theoretical and
experimental
works~\cite{modugno2009exponential,jstat,divakaran,lahini,Archak_anomalous}. A buzz of
activity in recent times~\cite{eisert,laflorencie,alet} has
established the profitability of the study of quantum entanglement
whenever striking transport properties~\cite{sharma2015landauer,nehra2018manybody,bhakuni2018characteristic} lie
underneath. Despite the extensive literature on the AAH model, the
correlation between the transport properties of the model and the band
gap structure of its spectrum has not been discussed anywhere, to the
best of our knowledge. There have been only very few mentions of such studies 
in the literature~\cite{ASinha2018,pnas2017}. In this work, we have made an attempt to
explore the relationship between these band gaps and the single
particle and many fermionic equilibrium transport properties. There are special eigenstates in the spectra that show a drastically different localization properties as compared to the others as it is captured by the inverse participation ratio and von Neumann entropy for a
single particle when the system size is a non-Fibonacci number (explained later). Although the localization properties of all the single particle eigenstates become identical if one chooses the system size to be a Fibonacci number. This effect was not explicitly revealed earlier. There are large energy gaps at the top of these special eigenstates, the effect of which on many-particle equilibrium transport properties in ground state is also explored. We have numerically calculated the entanglement entropy and  
persistent current for spinless non-interacting fermions in the AAH potential.
All the quantities seem to capture the effect of the band gaps and agree with each other. We have obtained vanishingly
small current and substantially suppressed entanglement entropy when
the Fermi level is set near the location of the band gap, even in the
delocalized phase where typically one obtains high current
and entanglement. The filling fraction-dependent many particle results are qualitatively independent of the choice of the system size (Fibonacci or non-Fibonacci) unlike the single particle results. 

Also, we have studied relations between the persistent current and its
fluctuations as a function of the strength of the AAH potential and
filling fractions. This kind of relationship has been investigated in
Ref.~\onlinecite{pc_metcalf} but for a translationally invariant one
 dimensional model. In the delocalized phase of the AAH model, the
 standard deviation vs mean persistent current curves are
 discontinuous for the regular filling fractions and continuous
 (closed) for the special fillings whereas in the localized phase,
 these curves become straight lines for both types of filling
 fractions. One of the important findings of our work is that none of
 the many-particle quantities we have studied changes drastically
 across the delocalization-localization transition point in case of
 special filling fractions in sharp contrast to the case of
 non-special filling fractions. Such non-trivial filling-fraction
 dependent transport properties appear not to have been explored in
 any other work. 

The paper is organized as follows. In section II, we have described
the delocalization-localization transition in the AAH model and briefly
discussed the interesting self-similar structure of the energy spectrum
and locations of the band gaps. Thereafter in section III, we have
numerically studied the single particle properties, where we have
calculated the inverse participation ratio (IPR) and the von Neumann
entropy for each single particle eigenstate. In section IV, we have
calculated the entanglement entropy in subsection A, the persistent
current in subsection B and the variance of current in subsection C,
for noninteracting spinless fermions to capture the effect of the
regular and special filling fractions (the band gaps) on the transport
properties of the fermionic system. The relations between the mean
persistent current and its standard deviation are studied in the
subsection D. At the end, we have rendered our conclusions in section
V.

\section{The Aubry-Andr\'e-Harper model}
\begin{figure*}[htpb]
\centering
\hspace{0.2cm}
\includegraphics[width=0.6\columnwidth]{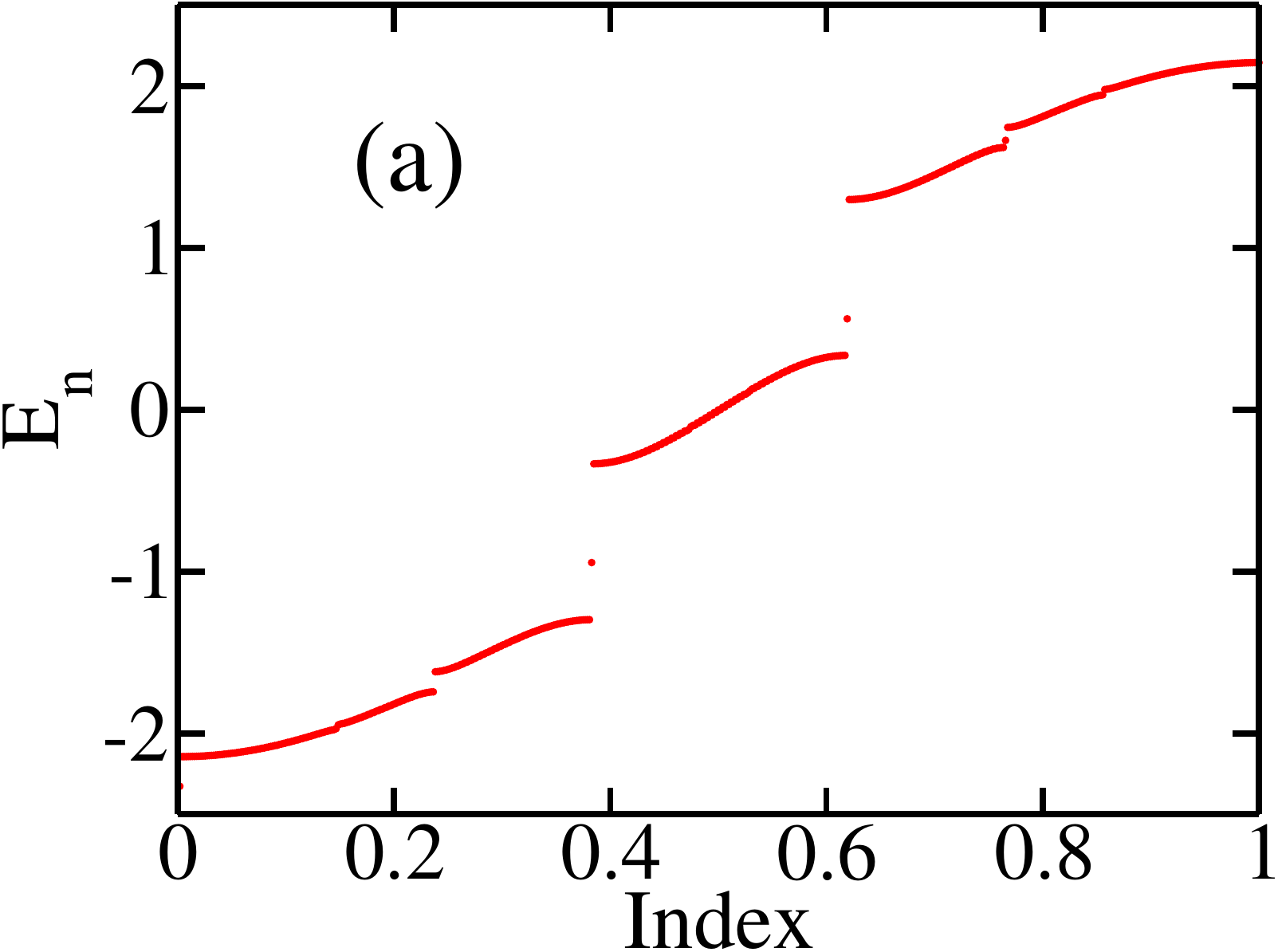}
\hspace{0.2cm}
\includegraphics[width=0.6\columnwidth]{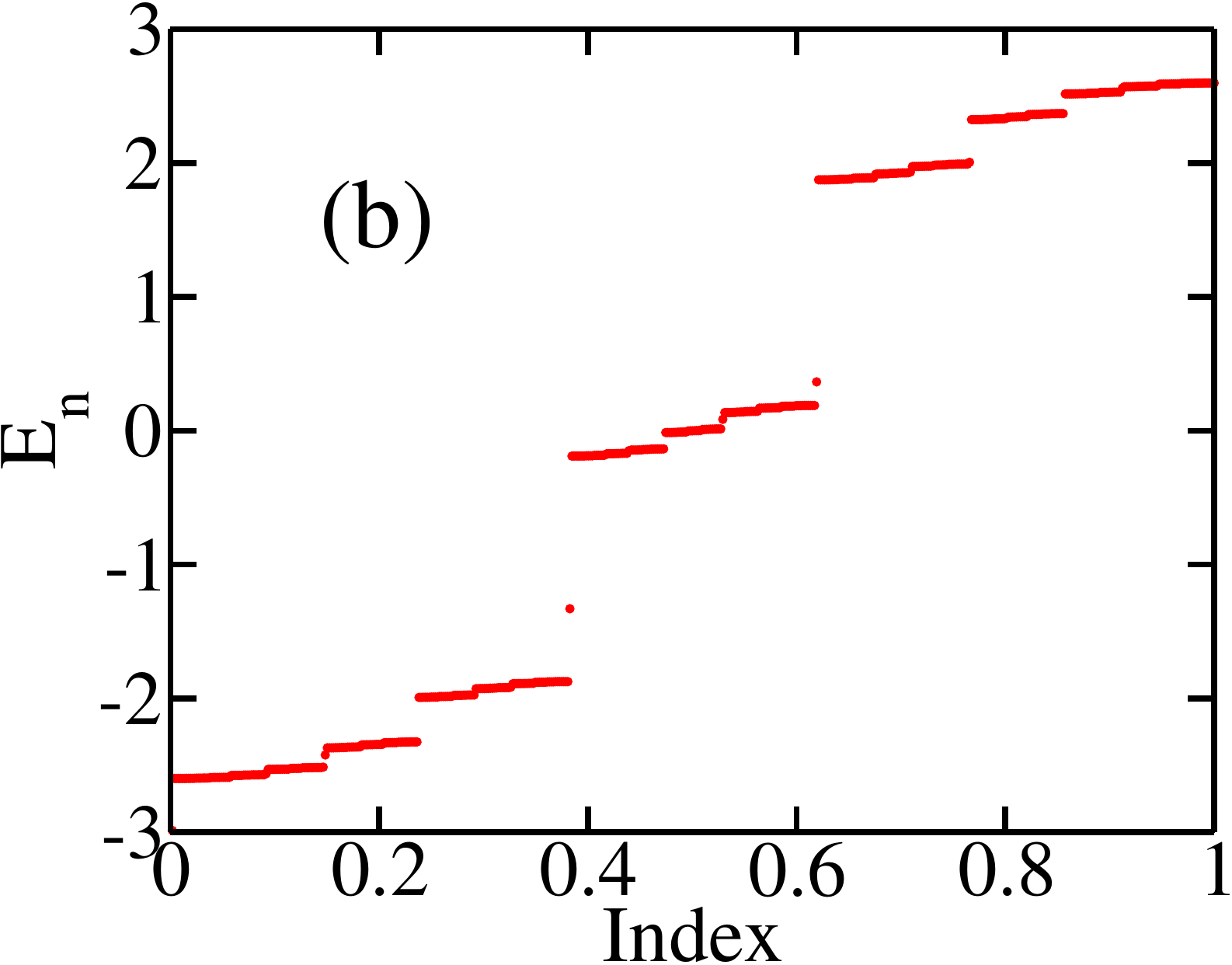}
\hspace{0.15cm}
\includegraphics[width=0.6\columnwidth]{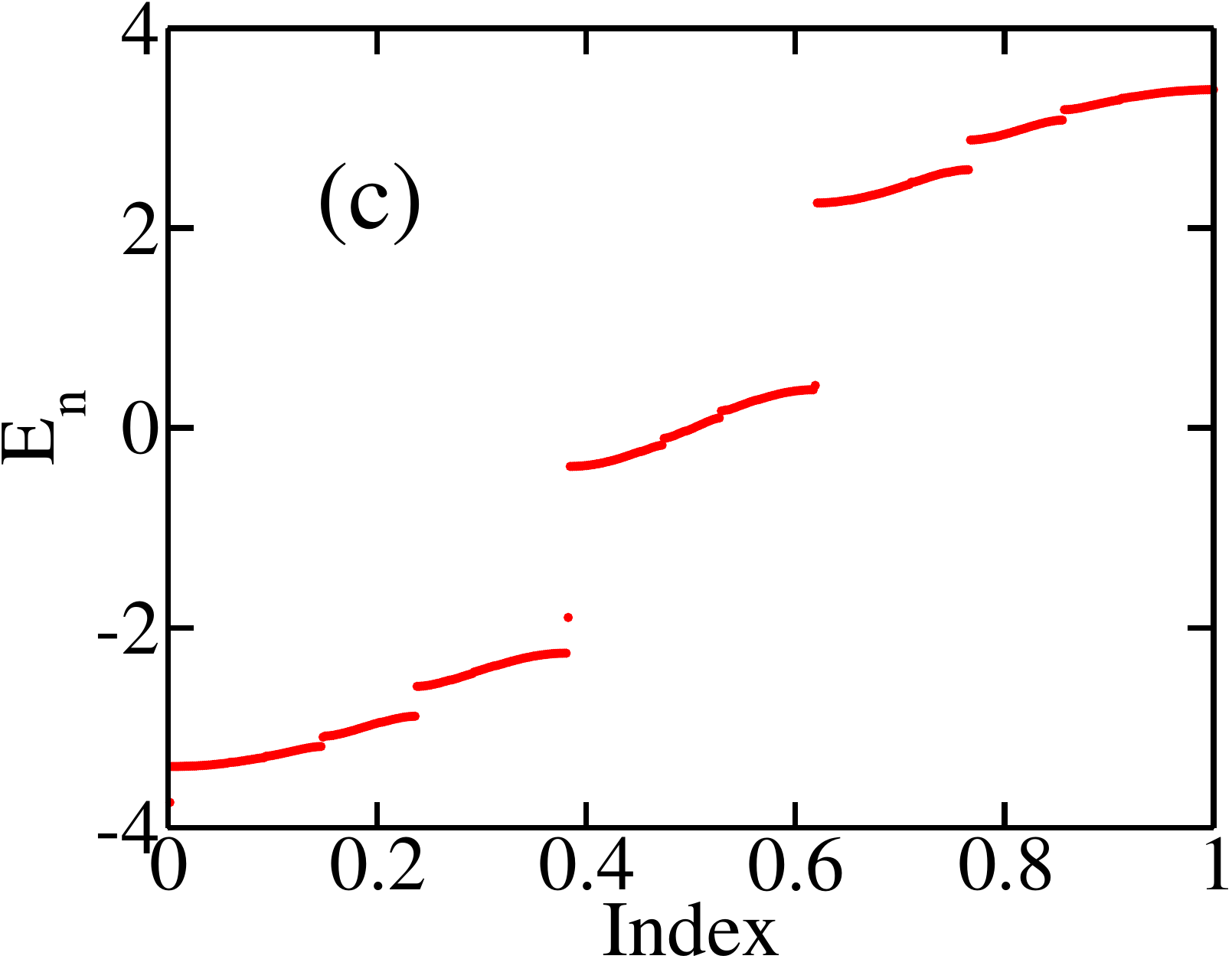}
\includegraphics[width=0.63\columnwidth]{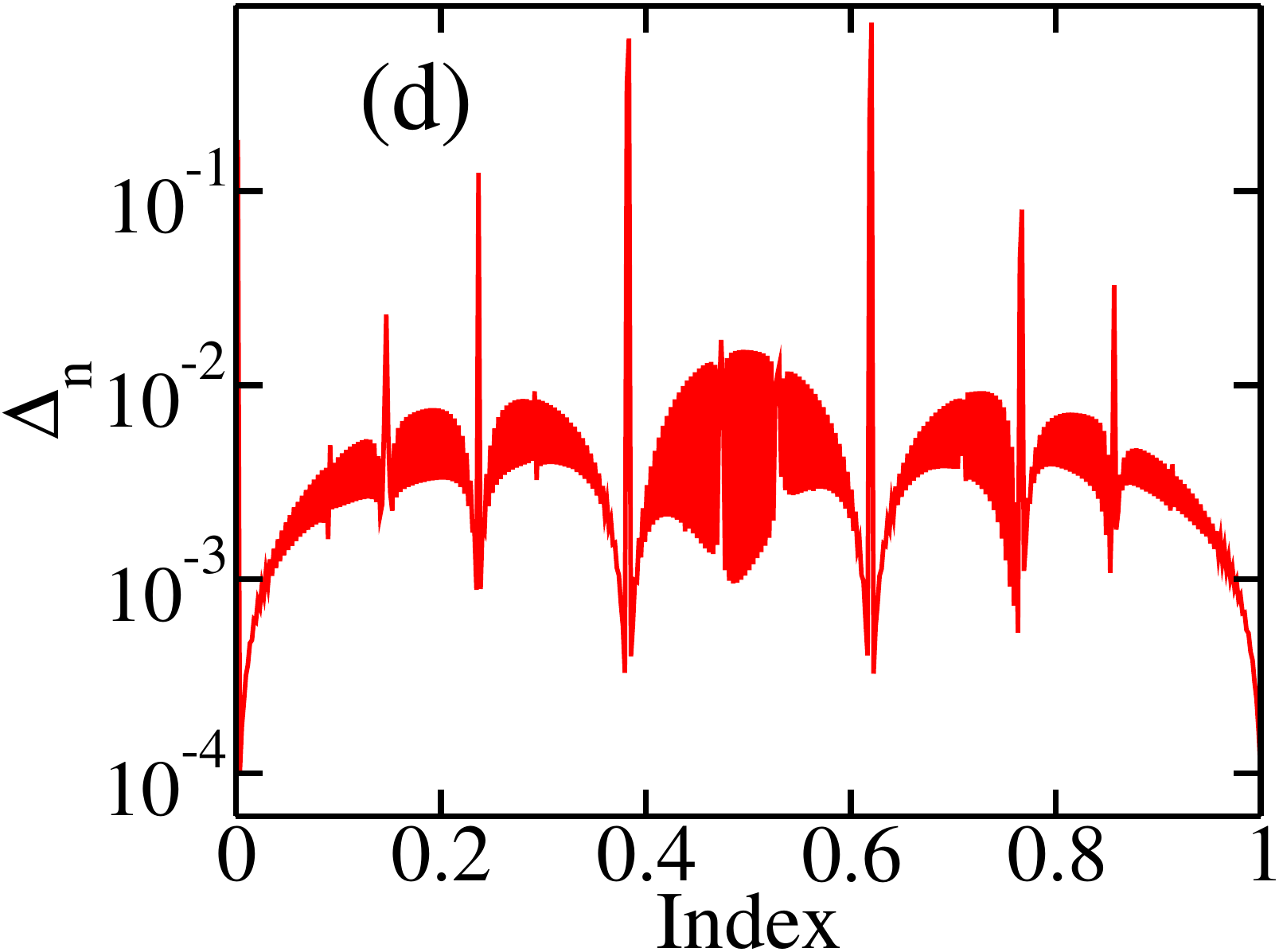}
\includegraphics[width=0.63\columnwidth]{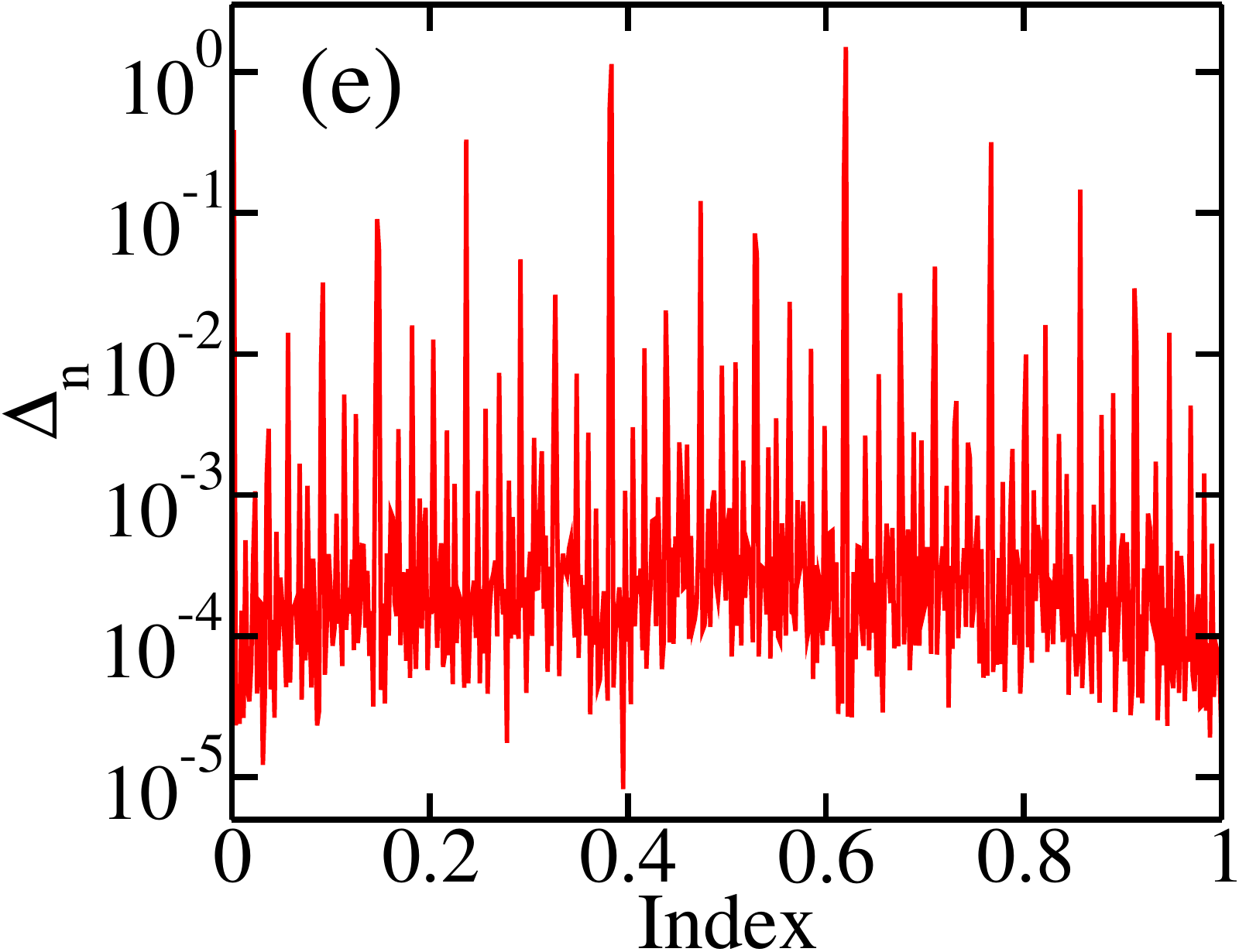}
\includegraphics[width=0.63\columnwidth]{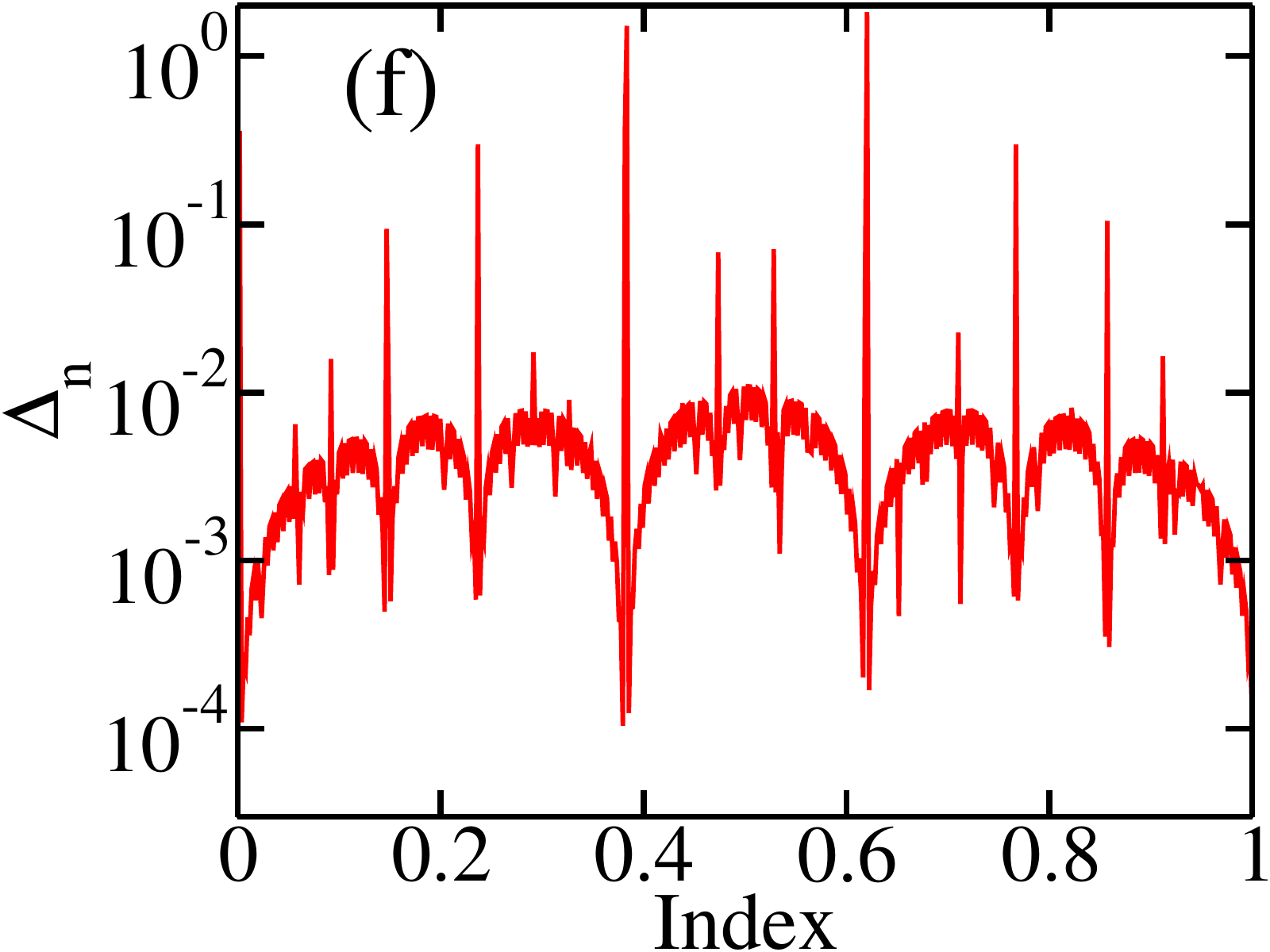}
\caption{(a-c) Single particle energy spectra $E_n$ for $\lambda=1.0,2.0$, and $3.0$ respectively. (d-f) The corresponding level-spacing spectra $\Delta_n$ of the AAH model in logscale for $\lambda=1.0,2.0$ and $3.0$ respectively. For all plots $N=512$. Here index is the serial number of energy (gap) levels  divided by the total number of energy (gap).}
\label{spectra}
\end{figure*}
The Aubry-Andr\'e-Harper (AAH) model in one dimension is given by the Hamiltonian~\cite{aubry,harper}:
\begin{eqnarray}
H = &&-J\sum\limits_{i}^{N} ( c_i^\dagger c_{i+1} + H.c.)\nonumber\\ &&+ \sum\limits_{i}^{N}\lambda \cos(2\pi\alpha i+\theta_p)c_i^\dagger c_i,
\end{eqnarray}
where $c_i^\dagger$ ($c_i$) represents the single particle creation (annihilation) operator at site
$i$. We consider a lattice on a circular ring of total number of sites
$N$. Here $\lambda$ is the strength of the quasi-periodic potential
with quasi-periodicity $\alpha$, an irrational number and an
arbitrary phase $\theta_p$. The strength of the nearest-neighbor
hopping is $J$. When the irrational number $\alpha$ is chosen to be a
Diophantine number, all the single particle eigenstates of the AAH model
become delocalized for $\lambda<2J$ and localized for $\lambda>2J$,
$\lambda=2J$ being the quantum critical point, where all the
eigenstates are multifractal~\cite{modugno2009exponential}. Also at
$\lambda=2J$, the AAH model in position space maps to itself in the
momentum space, thus making the model self-dual at
$\lambda=2J$~\cite{thouless_AAH}. In this work, we will assume $J=1$
and $\alpha=(\sqrt{5}-1)/2$, which is inverse of the golden mean,
unless otherwise stated.
\begin{figure*}[!ht]
\includegraphics[height=3.95cm,width=5.2cm]{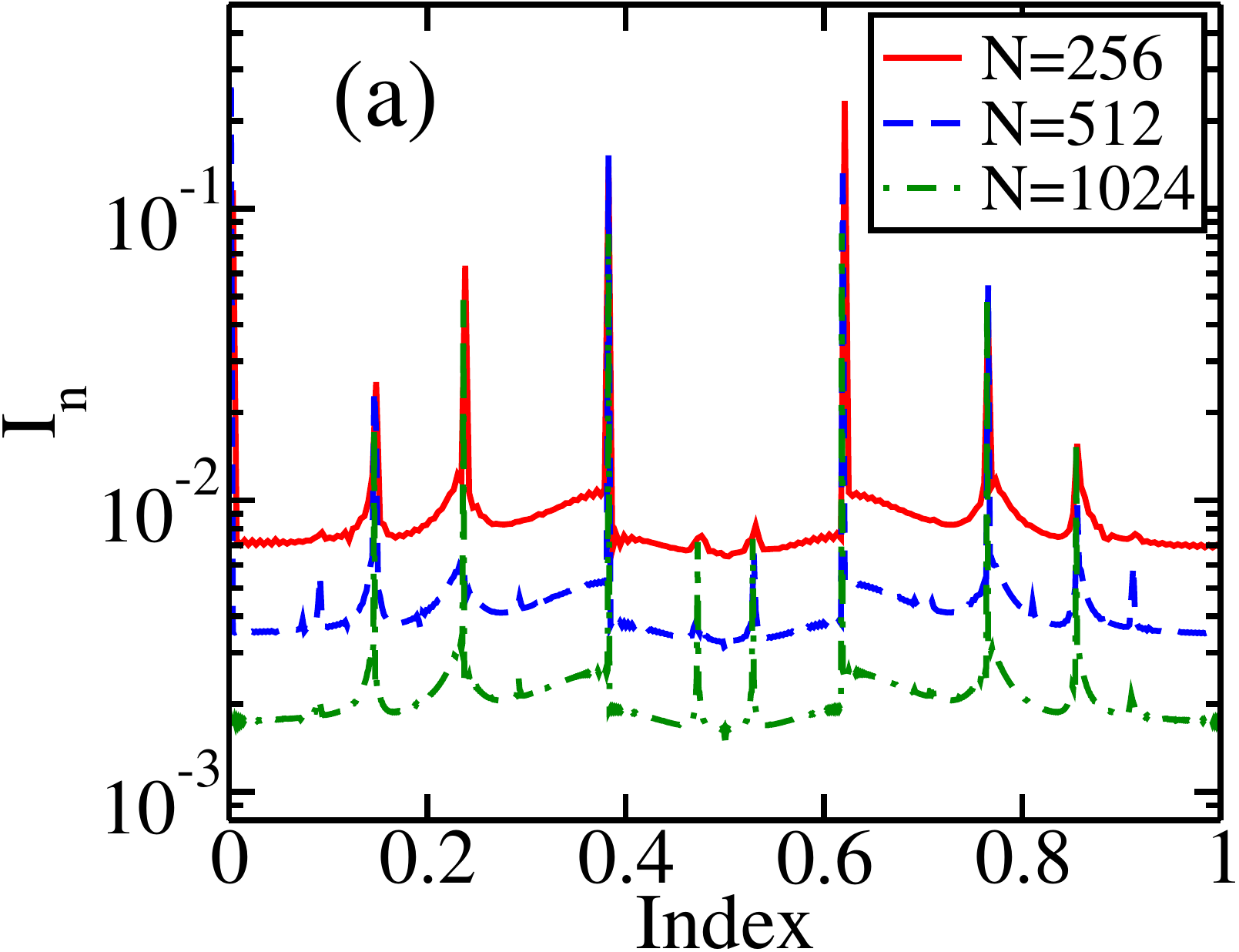}
\includegraphics[width=0.6\columnwidth]{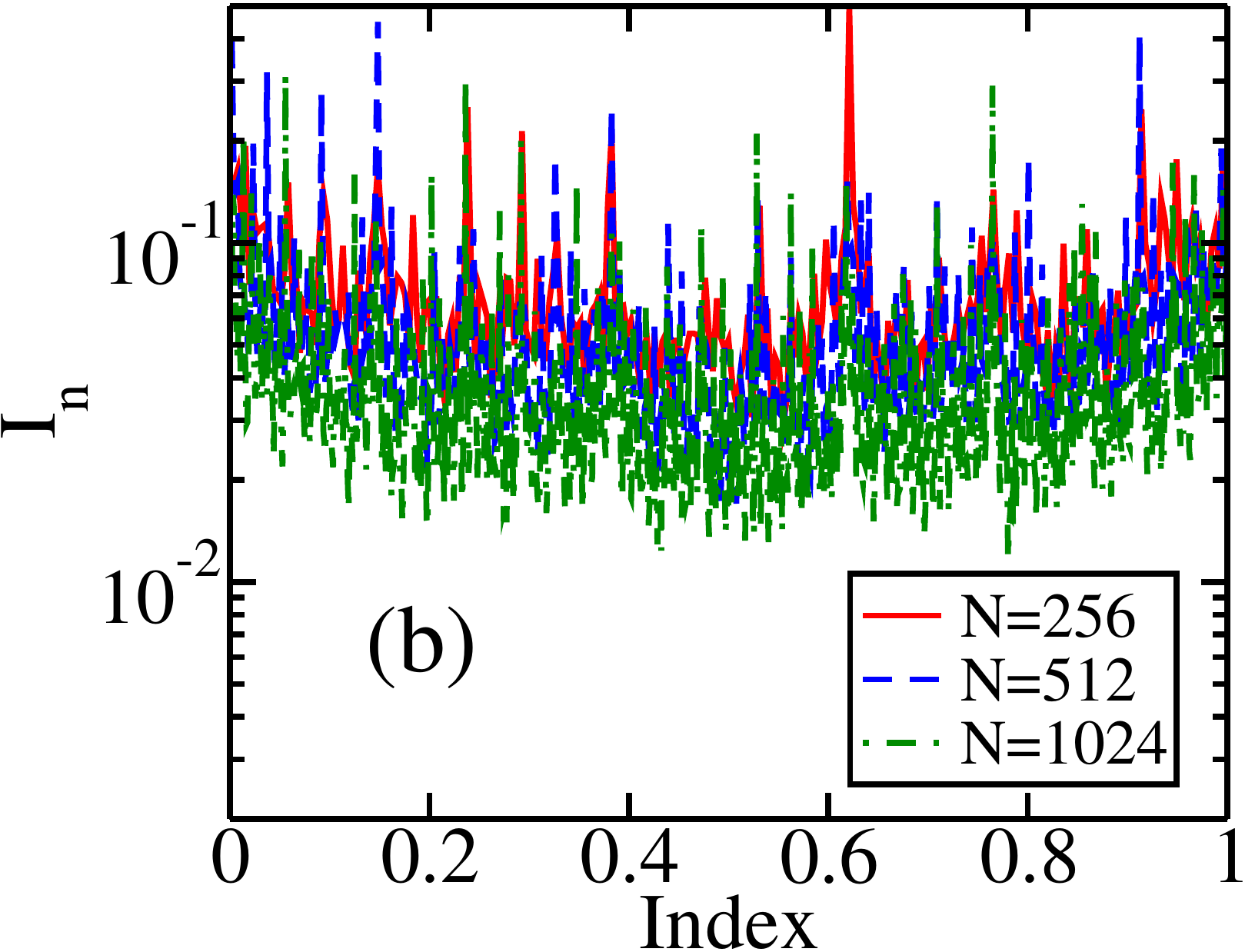}
\includegraphics[width=0.6\columnwidth]{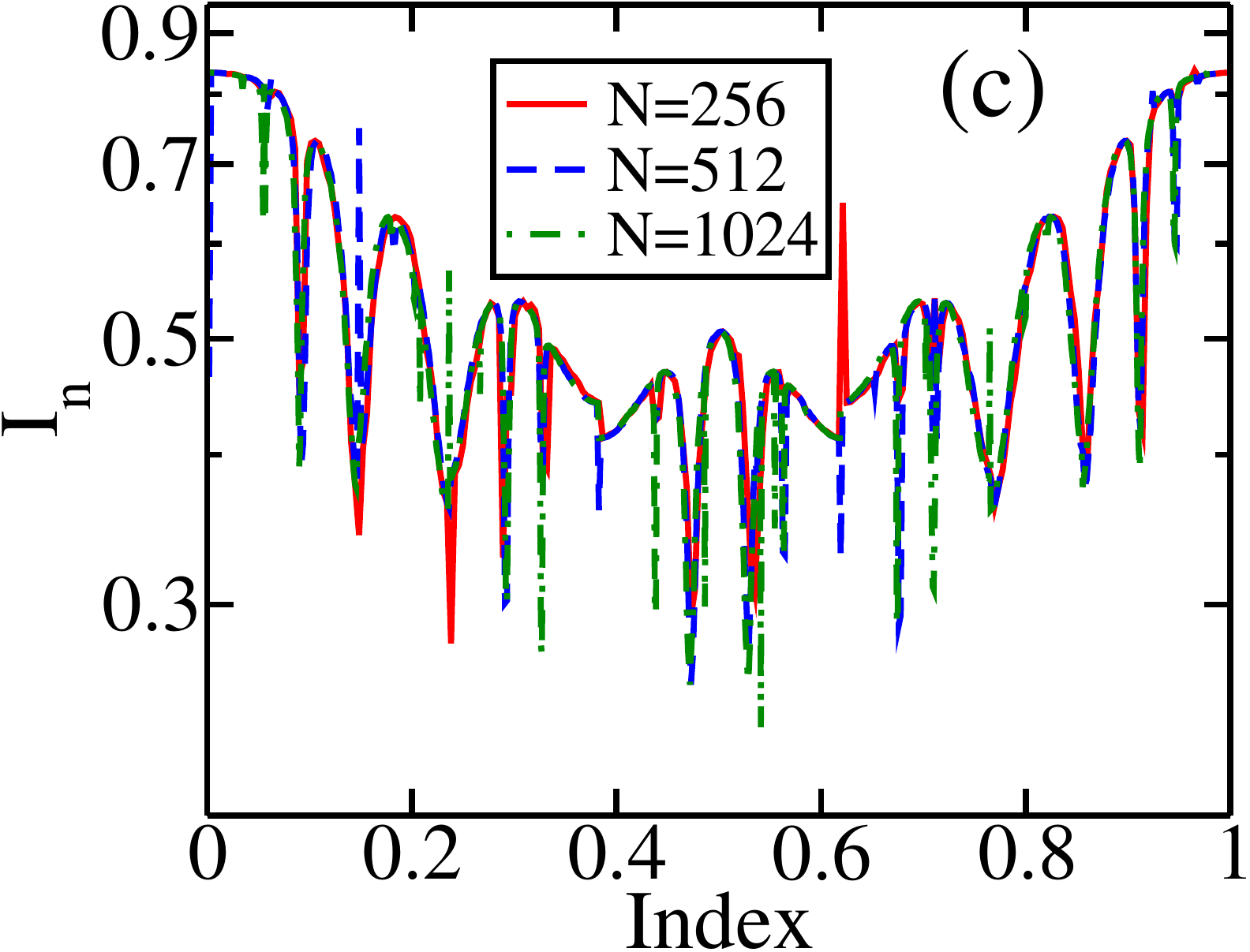}\\
\hspace{0.0cm}
\includegraphics[height=3.95cm,width=5cm]{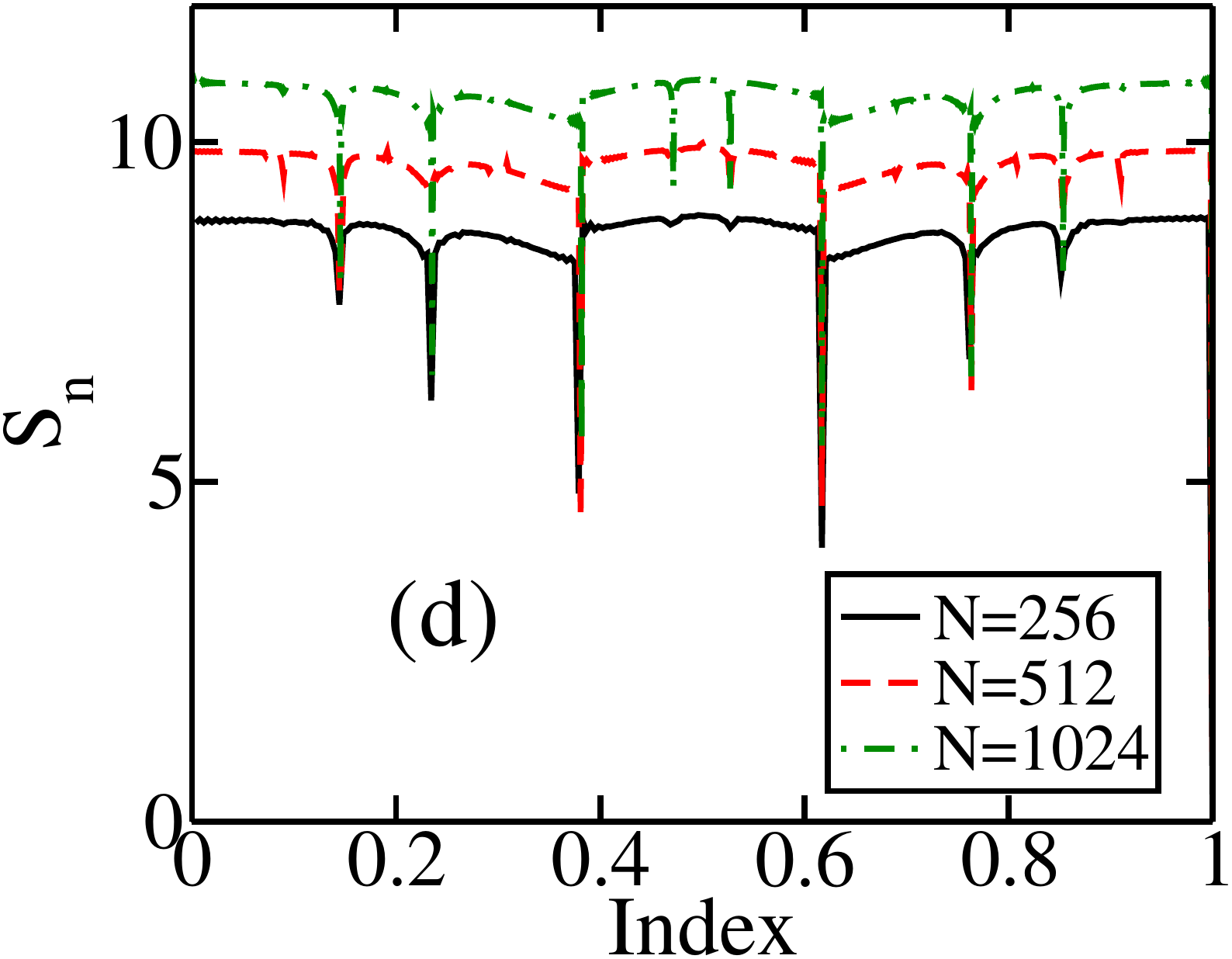}
\hspace{0.25cm}
\includegraphics[height=3.95cm,width=4.9cm]{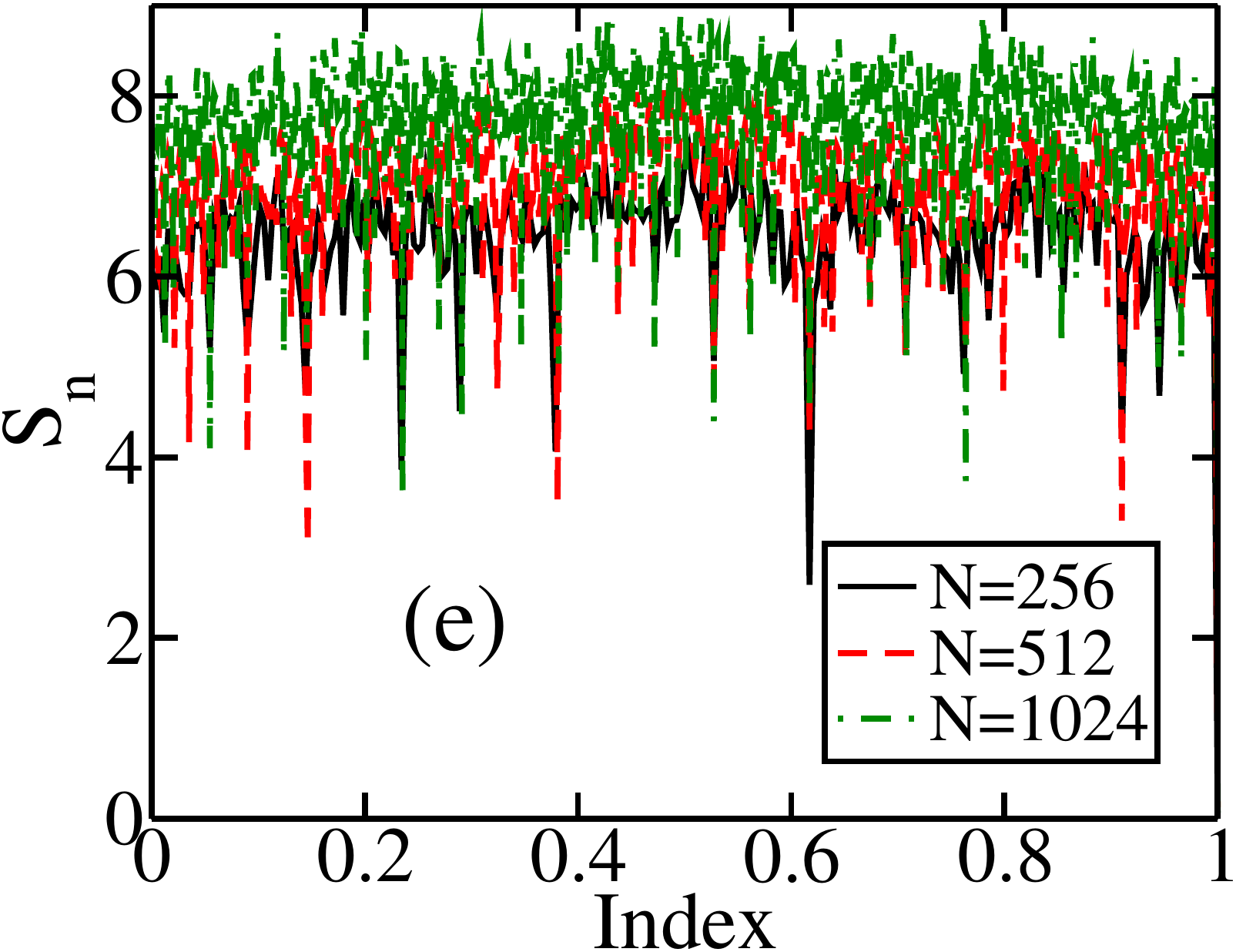}
\hspace{0.15cm}
\includegraphics[height=3.95cm,width=4.9cm]{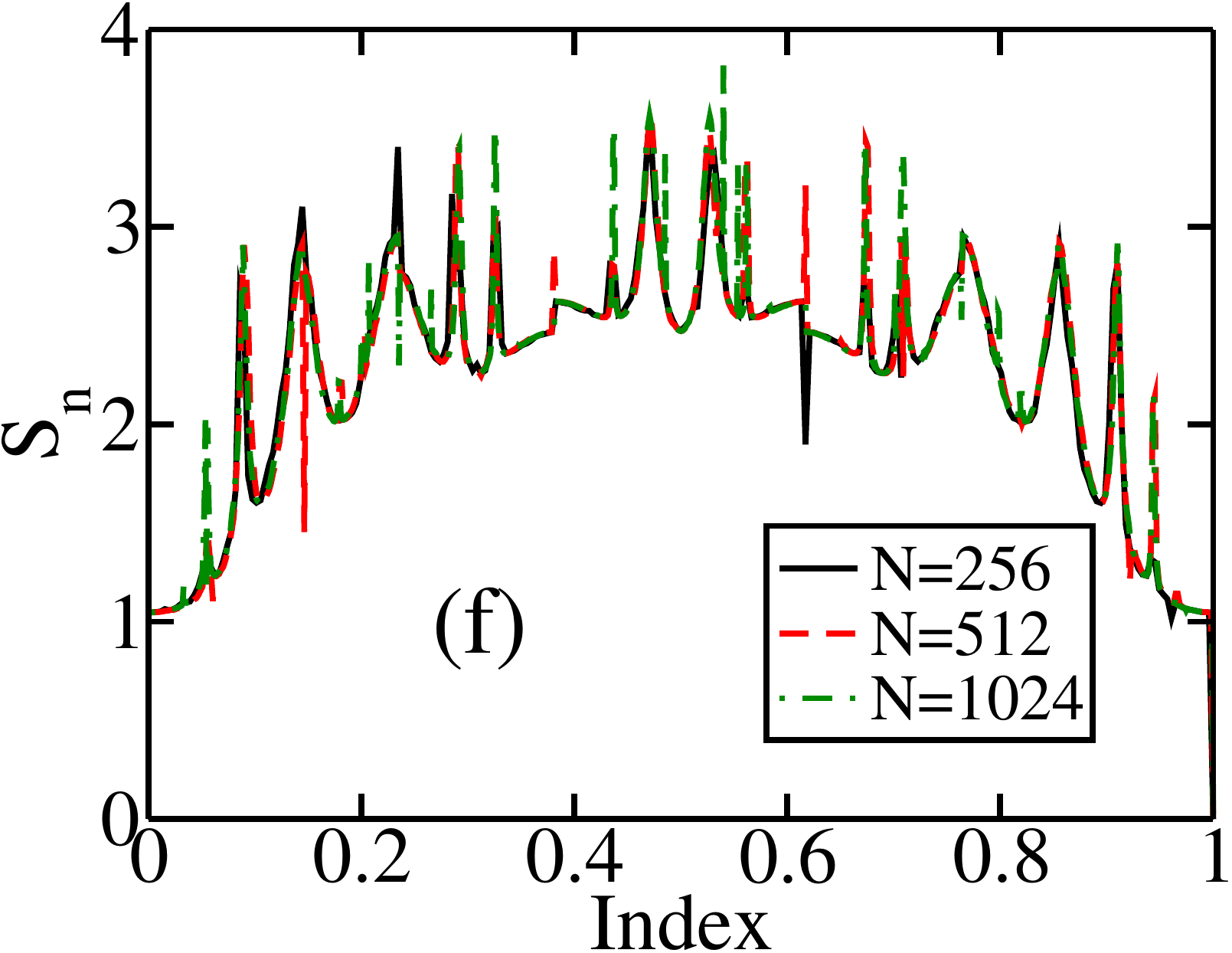}
\caption{(a-c) IPR of all the single particle eigenstates in logscale
  for $\lambda=1.0,2.0$ and $3.0$ respectively. (d-f) von Neumann
  entropy $S_n$ of all the single particle eigenstates for
  $\lambda=1.0,2.0$ and $3.0$ respectively. In all the plots the
  dependence on system size $N$ is studied. Here index is the serial number of eigenstate divided by the total number $N$.}
\label{spee}
\end{figure*}

\subsection{Generics of spectrum}
The single particle energy spectrum $E_n$ and the nearest level spacing $\Delta_n=E_{n+1}-E_n$ of the AAH model have already been
investigated in seminal works~\cite{kohmoto3,pichard,takada,machida}. It can be mathematically shown that the
energy spectrum forms a Cantor set-like self-similar structure~\cite{svetlana}. The energy level-spacing distribution of the AAH model is not Wigner-like in the delocalized phase~\cite{pichard} whereas it is Poissonian in the localized phase~\cite{machida,takada}. At the critical point, the level-spacing distribution satisfies an inverse power law~\cite{takada,pichard}. The energy spectra
and the corresponding level-spacing spectra for the delocalized,
multifractal and localized phases are shown in Fig.~\ref{spectra}. Due to the Cantor set
structure of the spectra, there are many gaps between the subbands Fig.~\ref{spectra}(a-c),
whose specific locations, are related to the irrational number $\alpha$, to be discussed next. There are isolated states in the spectrum in the large gaps, represented by the isolated dots in~\ref{spectra}(a). However, these isolated states vanish if one chooses system size $N$ to be a Fibonacci number, defined later in Equation~\ref{fibo}. 
\subsection{Special locations of band gaps}
In Figs.~\ref{spectra}(a) and \ref{spectra}(d) the large gaps are
apparent when the fractional index is $\approx \alpha, \alpha^2, \alpha^3,
\alpha^4(\approx0.618, 0.382, 0.236, 0.145)$ etc. The fluctuations in
the gap become maximum at the quantum critical or the multifractal
point $\lambda=2$ (Fig.~\ref{spectra}(e))~\cite{pichard} and the
magnitude of the gaps increases as $\lambda$
increases(Fig.~\ref{spectra}(f)). In this work, we explore the effect
of such gaps on the transport properties at the single and many
particle levels, especially in the delocalized phase. Also the
localization properties of the single-particle eigenstates near the
locations of these band gaps may be very different from the other
eigenstates, which we will explore in the next section.

\section{Single-particle transport properties}
\begin{figure}
\includegraphics[width=0.85\columnwidth]{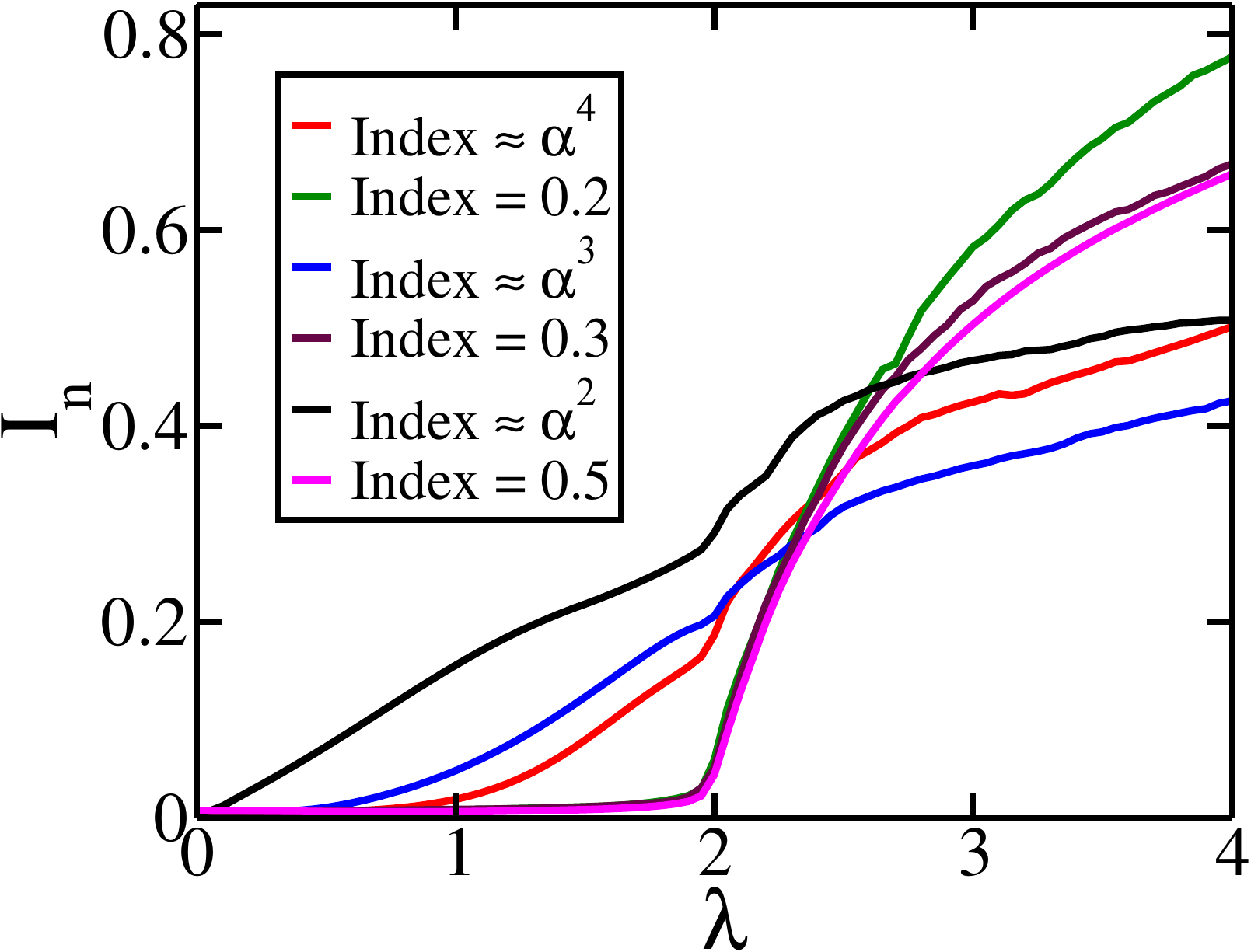}
\caption{Variation of IPR with $\lambda$ of the single particle
  eigenstates near the locations of the large band gaps (Index
  $\approx \alpha^4,\alpha^3,\alpha^2$) compared to the other
  eigenstates (with Index $=0.2,0.3,0.5$ )respectively. For all the
  plots $N=256$. To smooth out the curve, an average over $100$
  values of $\theta_p$ (running from $0$ to $2\pi$ in uniform steps) is
  performed.}
\label{ipr_l}
\end{figure}

In order to study the transport properties of a single particle in AAH
potential, we have analyzed the inverse
participation ratio (IPR) and the von Neumann entropy. For simplicity all
the results presented in this section are calculated assuming
$\theta_p=0$ unless mentioned. We briefly review these quantities and present our
results in the following.
\subsection{Inverse participation ratio}
The inverse participation ratio (IPR) is a key quantity for studying delocalization-localization transitions, which is defined as:
\begin{eqnarray}
I_n = \sum_{i=1}^{N} |\psi_n(i)|^4,
\end{eqnarray}
where the $n\textsuperscript{th}$ normalized single-particle
eigenstate $\ket{\psi_n}=\sum_{i=1}^{N}\psi_n(i) \ket{i}$ is written
in terms of the Wannier basis $\ket{i}$, representing the state of a
single particle localized at the site $i$ of the lattice. For a
perfectly delocalized eigenstate $I_n=1/N$ whereas $I_n=1$ for a
single-site localized eigenstate. In the critical phases $I_n$ is expected to show an intermediate behavior. In Fig.~\ref{spee}(a-c) the IPR for
all the eigenstates can be seen for $\lambda=1$(delocalized),
$\lambda=2.0$(multifractal) and $\lambda=3.0$(localized)
respectively. The IPR shows sudden jumps exactly where the sub-band
gaps can be found as can be seen from Fig.~\ref{spee}(a). In the
delocalized phase typically IPR$\sim1/N$ except at these special points where IPR behaves anomalously with $N$. At the multifractal point,
the fluctuations become maximum and the IPR shows
anomalous scaling with $N$ almost all over the
spectrum~\ref{spee}(b). In the localized phase, the IPR overlap
with each other for different values of $N$~\ref{spee}(c). However, in this phase one obtains dips, instead of peaks, at the positions of large band gaps. 

\subsection{von Neumann entropy}
It is now well established that entanglement entropy is a good
measure to explore localization phenomena in quantum
systems~\cite{}. In this work we aim to calculate von Neumann entropy
connected to a single-site. As a single particle has two local states
$\ket{0}_i$ and $\ket{1}_i$ at the site $i$, the local density matrix
$\rho_{n,i}$ for the $n\textsuperscript{th}$ eigenstate can be written
as~\cite{SChakravarty}:
\begin{eqnarray}
\rho_{n,i}= |\psi_n(i)|^2 \ket{1}_i\bra{1}_i + (1 - |\psi_n(i)|^2) \ket{0}_i\bra{0}_i.
\end{eqnarray}
The von Neumann entropy associated with site $i$ is then given by~\cite{gong}:
\begin{eqnarray}
S_{n,i} = &&-|\psi_n(i)|^2 \log_2(|\psi_n(i)|^2) \nonumber \\
&&- (1 - |\psi_n(i)|^2) \log_2 (1 - |\psi_n(i)|^2).
\end{eqnarray}
In a delocalized eigenstate $|\psi_n(i)|^2=1/N$ and hence $S_{n,i}\approx \frac{1}{N}\log_2 N + \frac{1}{N}$ for large value of $N$ whereas for a single-site localized state $S_{n,i}=0$.
The contributions from all sites for a single-particle eigenstate are given by:
\begin{eqnarray}
S_n=\sum_{i=1}^{N} S_{n,i}.
\end{eqnarray}
For large values of $N$, $S_n\approx(\log_2 N + 1)$ in the delocalized
phase whereas $S_n\approx0$ in the extremely (single-site) localized
phase. For the critcal phases $S_n$ can take any intermediate values. As
we can see from Fig.~\ref{spee}(a) the single particle von
Neumann entropy $S_n$ has higher value in the delocalized phase and
varies as $\log N$ but it shows sudden fall at the special points
where IPR shows jumps and anomalous dependence on $N$. In the
localized phase, as expected $S_n$ takes smaller values and shows no
dependence on $N$ but instead of a sudden fall one obtains peaks at the special points [see Fig.~\ref{spee}(c)]. At the critical point,
$S_n$ shows wide fluctuations and picks up intermediate values showing anomalous scaling with
$N$, which is shown in Fig.~\ref{spee}(b).

To contrast the localization properties of the special eigenstates
near the large band gaps with the non-special eigenstates, the $IPR$
is plotted as function of $\lambda$ in Fig.~\ref{ipr_l}.  As we can
see from the figure, the non-special eigenstates undergo
delocalization-localization transition at $\lambda=2$ as IPR changes
abruptly, whereas the special eigenstates, with isolated energy levels, display different behaviour
as IPR vs. $\lambda$ curves do not reflect criticality at
$\lambda=2$. The contrasting localization properties of the special and non-special single particle eigenstates are consistent with the breakdown of the single-parameter scaling at the localization transition in quasiperiodic quantum systems, recently found in the literature~\cite{IISc_breakdown}. This kind of special feature of not reflecting the
criticality of the model is present even in the behavior of the
many-particle quantities when Fermi level is set in those band
gaps. We discuss this in the following section.
However, if one chooses the system size from the Fibonacci sequence such as $N=34,144,610$ (commensurate) as described in Eq. (8) of the paper, the isolated energy levels, as shown in Fig.~\ref{spectra}(a), disappear and all the single particle eigenstates become delocalized for $\lambda<2$ and localized for $\lambda>2$. So the peaks in $I_n$ (and corresponding falls in $S_n$) as shown in Fig.~\ref{spee}(a) for $\lambda<2$ disappear. Also the distinction between the special states and non-special states in  Fig.~\ref{ipr_l} vanishes in this case.
\section{Many-particle transport properties}
In this section, we consider non-interacting spinless fermions in the
system. Since we have seen the surprising effect of energy gaps in the
single-particle picture, we expect to see such effects even in the
system of many fermions. We calculate many-fermionic entanglement
entropy and persistent current and look at the behavior of these
quantities as a function of the filling fraction $\nu=N_p/N$, where
$N_p$ is the number of fermions in a periodic ring with $N$
sites. Entanglement entropy and persistent current have been
studied very recently for the AAH model at half
filling~\cite{prb,jstat}. In what follows, we briefly describe the
quantities and discuss our results.
\subsection{Entanglement entropy}
\begin{figure*}[!ht]
\centering
\includegraphics[width=0.6\columnwidth]{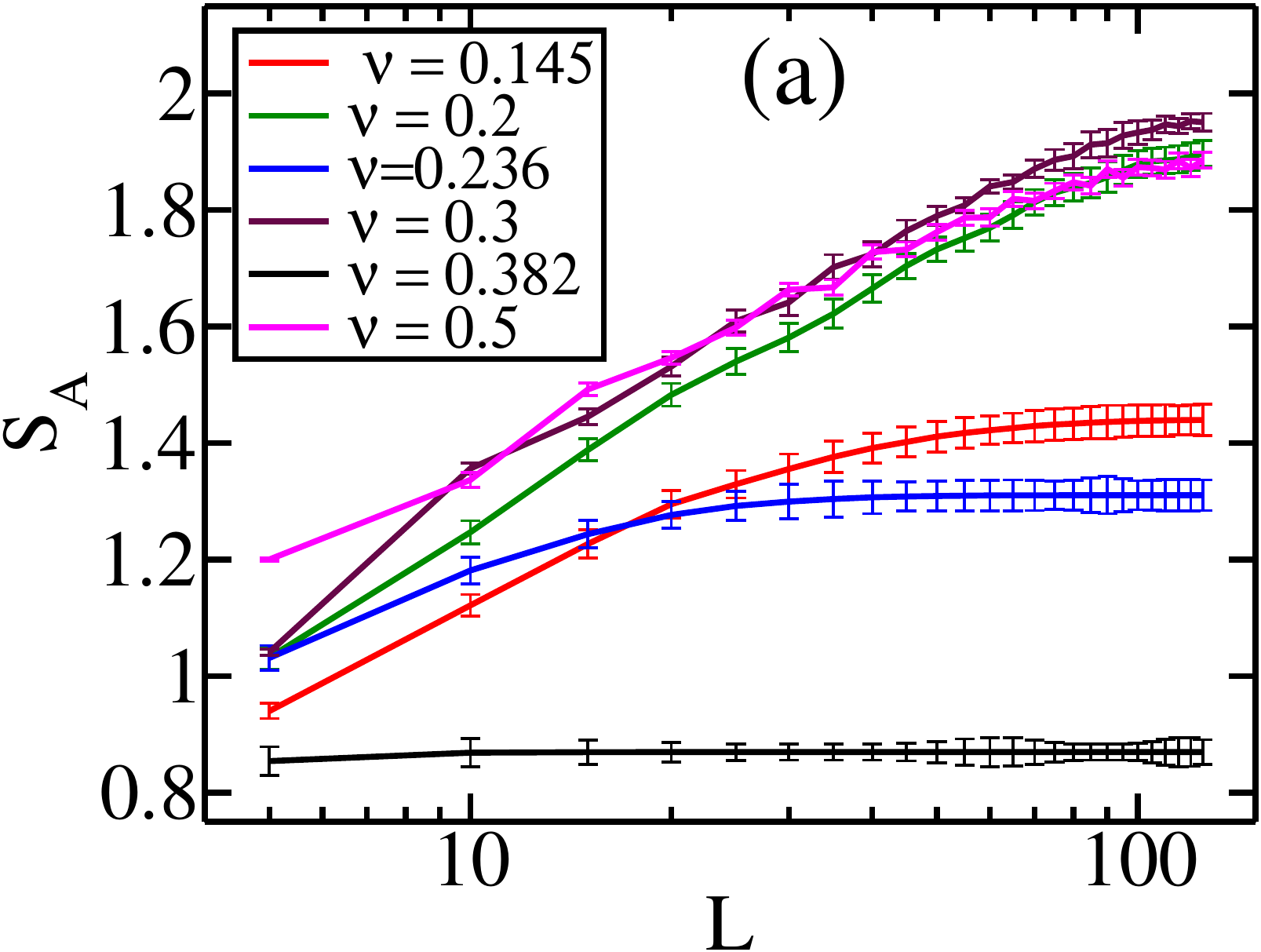}
\includegraphics[width=0.6\columnwidth]{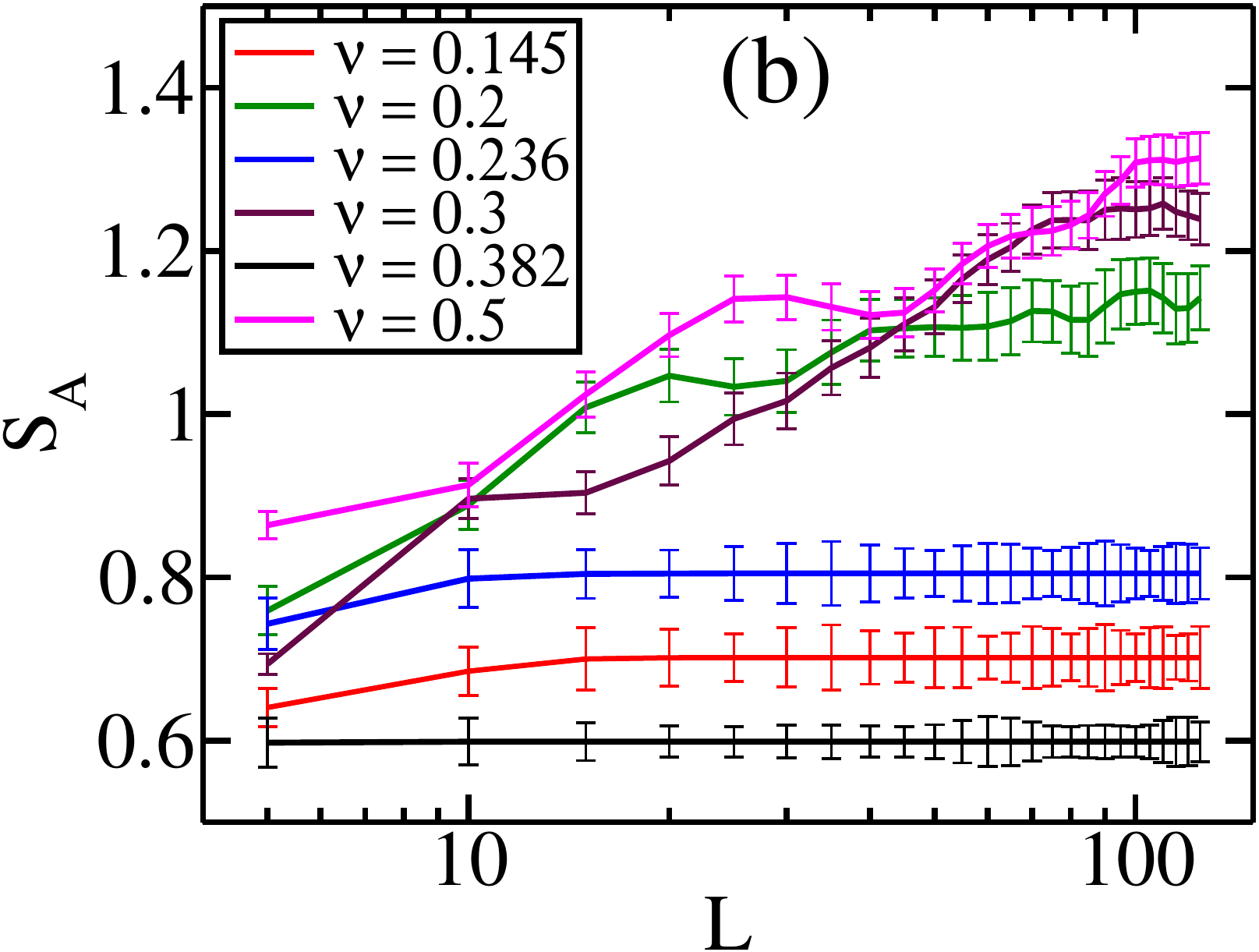}
\includegraphics[width=0.6\columnwidth]{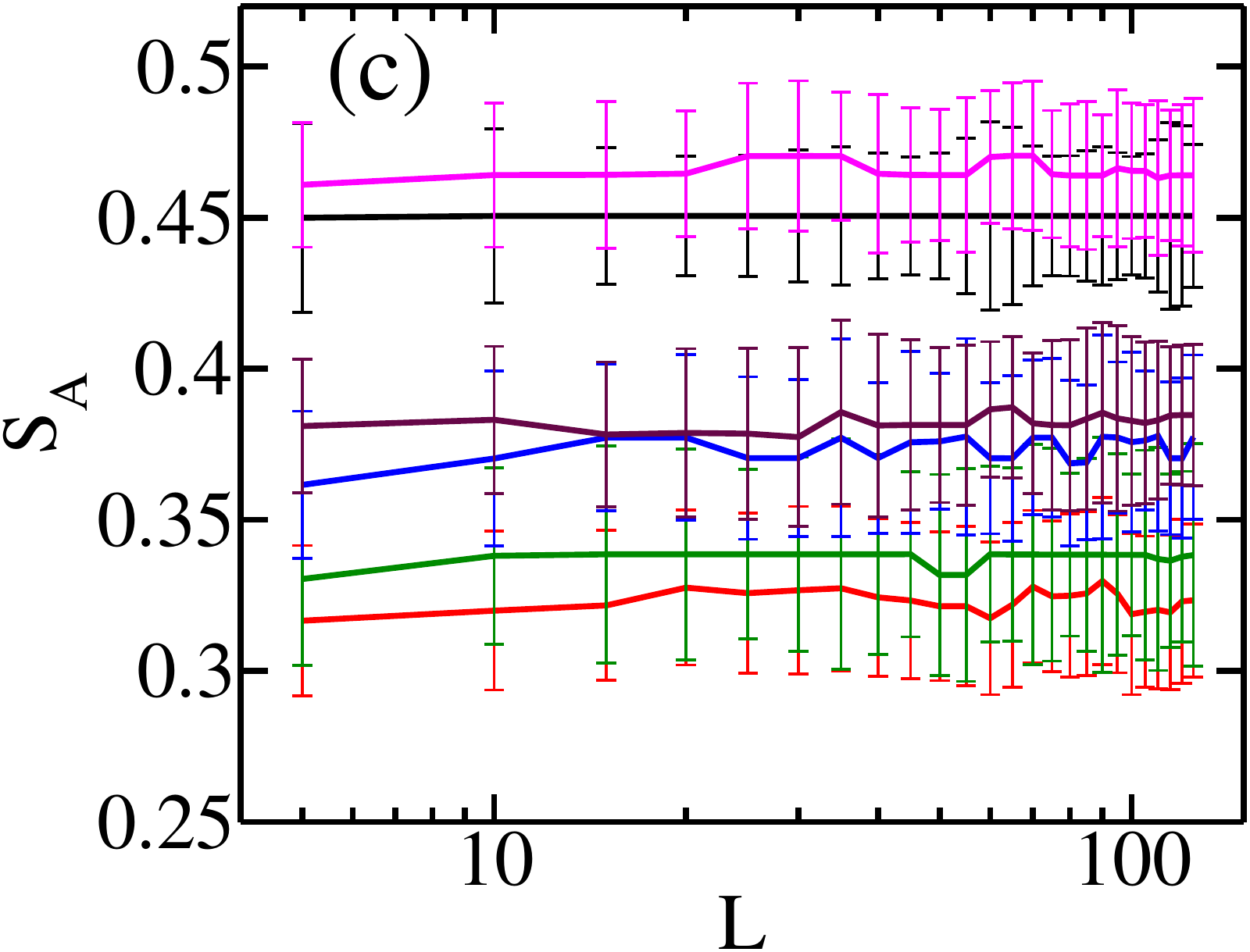}
\includegraphics[width=0.6\columnwidth]{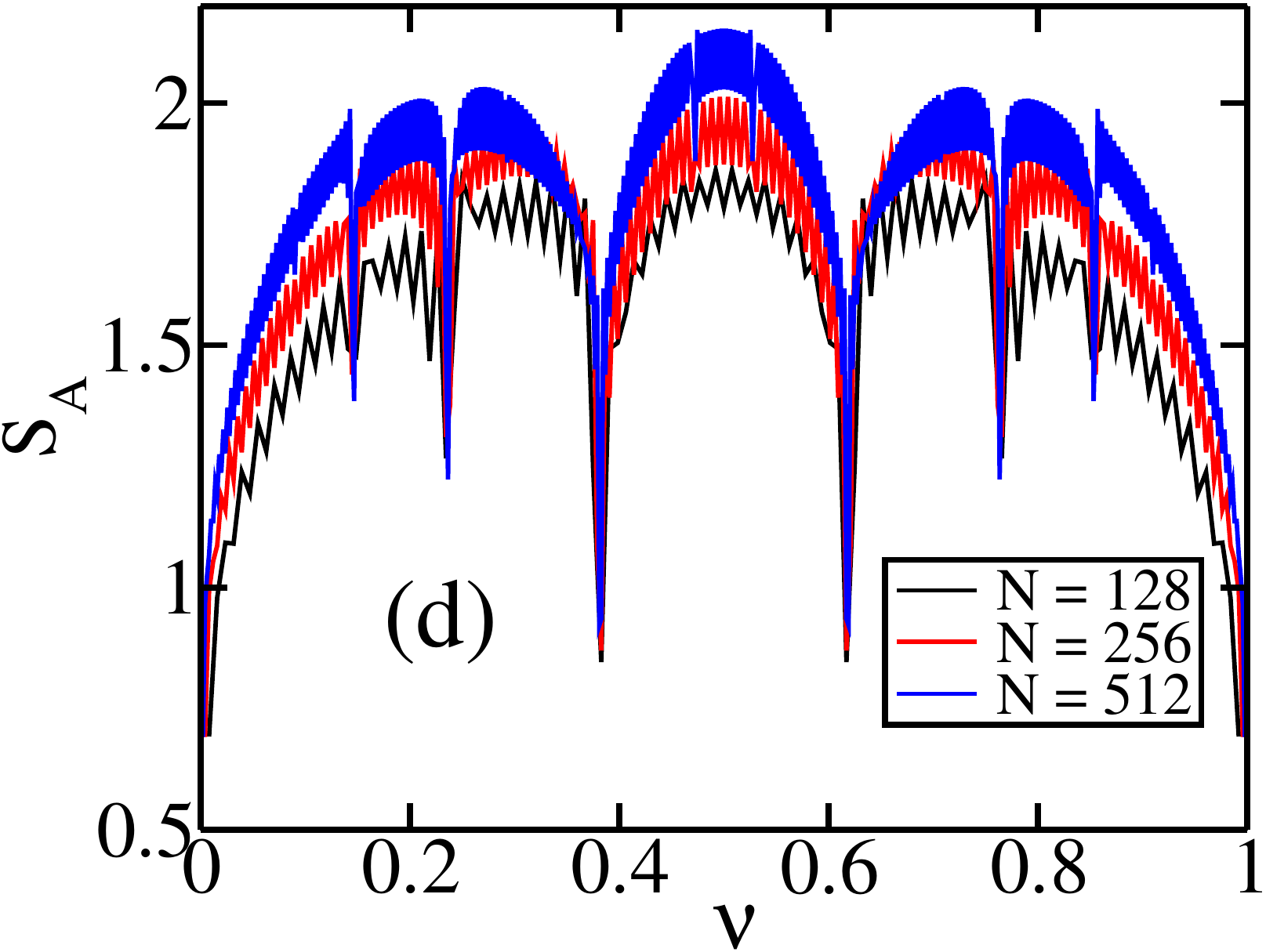}
\includegraphics[width=0.6\columnwidth]{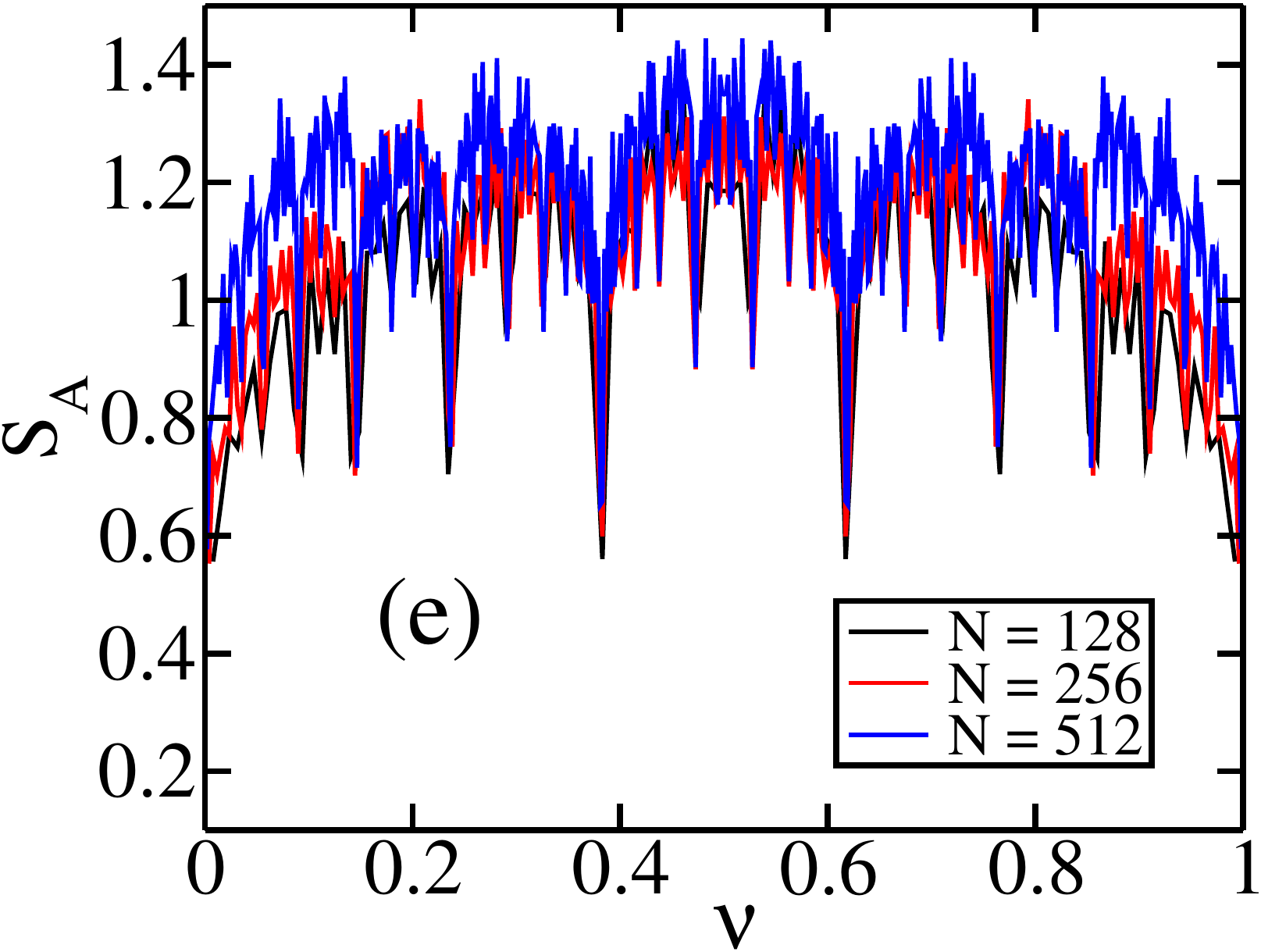}
\includegraphics[width=0.6\columnwidth]{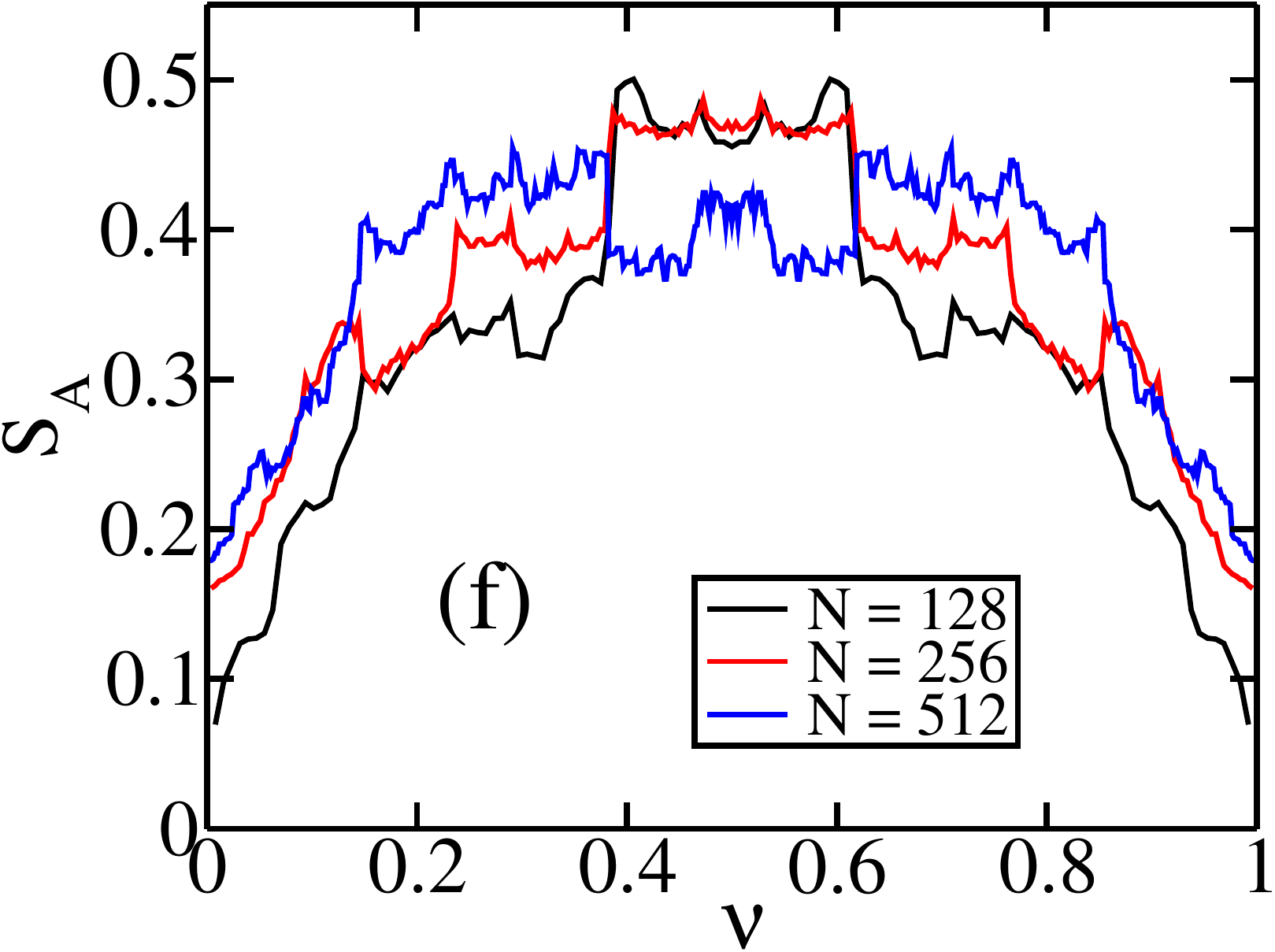}
\caption{(a-c) Dependence of entanglement entropy $S_A$ of the
  many-body ground state on subsystem size $L$ (in logscale) for
  increasing values of the filling of fermions $\nu$ for
  $\lambda=1.0$(a), $\lambda=2.0$(b) and $\lambda=3.0$(c)
  respectively.  For figures (a-c) $N=256$. (d-f) Variation of $S_A$ with $\nu$ for
  increasing system size $N$ for $\lambda=1.0$(d), $\lambda=2.0$(e) and
  $\lambda=3.0$(f) respectively. For the last three plots $L=N/2$. For
  all the plots, an average over $100$ values of $\theta_p$ (running from $0$ to $2\pi$ in uniform steps) has been carried out.}
\label{ee}
\end{figure*}
\begin{figure}
\includegraphics[width=0.85\columnwidth]{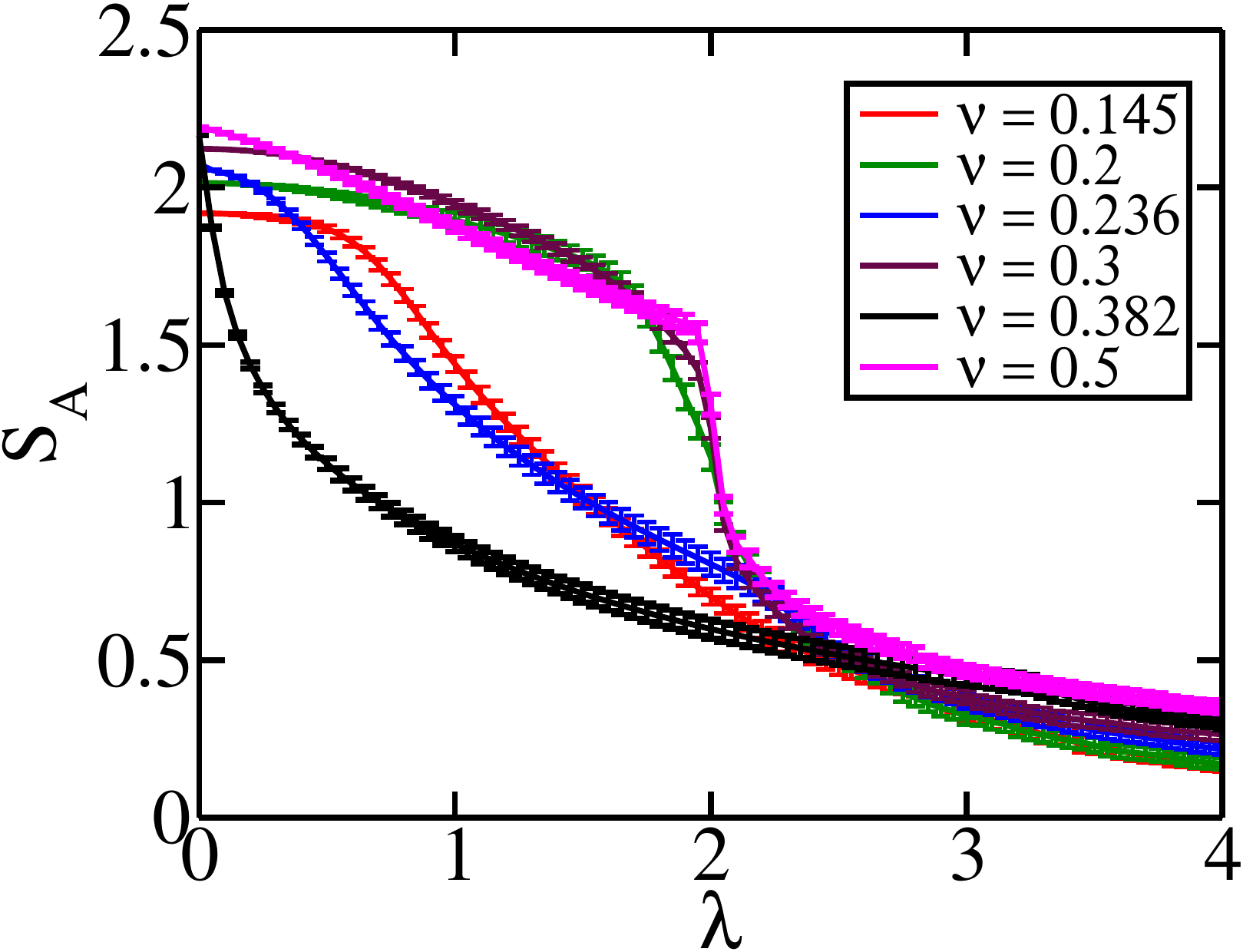}
\caption{Variation of $S_A$ with $\lambda$ for increasing values of filling of fermions $\nu$. For all the plots $L=N/2$ where $N=256$ and an average over $100$ values for $\theta_p$ (running from $0$ to $2\pi$ in uniform steps) is carried out.}
\label{ee_nu}
\end{figure}  
For pure states, von Neumann entropy has established itself as the
standard measure of quantum entanglement, and has been extensively
used to study different many-body phases and to make a distinction amongst
them~\cite{eisert,laflorencie}. Intuitively, one would expect that the
greater the delocalization, the more the entanglement and vice versa,
although exceptions exist~\cite{kannawadi2016persistent}.  We start
with a brief discussion of the calculation of the entanglement entropy
of fermions in the ground state\cite{peschel2003calculation,peschel2009,peschel2012special,prb}. For
the fermionic many-body ground state $\Ket{\Psi_0}$, the density
matrix can be written as $\rho=\Ket{\Psi_0} \Bra{\Psi_0}$. The
entanglement entropy between two subsystems is then given by
$S_A=-Tr(\rho_A \log \rho_A)$, where the reduced density matrix is
$\rho_{A}=Tr_{B}(\rho)$. However, for a single Slater determinant
ground state, Wick's theorem can be exploited to write the reduced
density matrix as $\rho_{A}=\frac{e^{-H_{A}}}{Z}$, where
$H_{A}=\sum\limits_{ij} H_{ij}^A c_{i}^{\dagger}c_{j}$ is called the
entanglement Hamiltonian, and $Z$ is obtained from the condition $Tr
(\rho_{A}) = 1$. The information contained in the reduced density
matrix of size $2^L\times 2^{L}$ can be captured in terms of the
correlation matrix $C$ of size $L\times
L$\cite{peschel2003calculation} within the subsystem $A$, where
$C_{ij}=\left\langle c_{i}^{\dagger}c_{j} \right\rangle $. The
correlation matrix and the entanglement Hamiltonian are related
by\cite{peschel2003calculation,peschel2009,peschel2012special}:
\begin{equation}
  \label{eqn:C_and_H}
  C=\frac{1}{e^{H_A}+1}.
\end{equation} 
Using this relation, the entanglement entropy for free fermions is given by\cite{peschel2009,peschel2012special}
\begin{equation}
S_A=-\sum\limits_{m=1}^{L} [\zeta_m \log \zeta_m + (1-\zeta_m) \log (1-\zeta_m)],
\end{equation}
where $\zeta_m$'s are the eigenvalues of the correlation matrix
$C$. Subsystem scaling of entanglement entropy has been used to
distinguish quantum
phases~\cite{eisert,laflorencie,roy2019quantum}. For free fermions in
$d$ dimension, typically $S_A\propto L^{d-1}\log L$~\cite{swingle} in
metallic phases and $S_A\propto L^{d-1}$ in the localized phase in the
presence of disorder. To produce smoother $S_A$ vs. $L$ plots for the
AAH model, we have done an average of $S_A$ over $100$ (non-random) values of
$\theta_p\in[0,2\pi]$ in uniform steps. Scaling of $S_A$ of the
many-fermionic ground state with $L$ is shown in Fig.~\ref{ee}(a),(b)
and (c) for increasing values of $\nu$ for the delocalized,
multifractal and localized phases respectively. As can be seen from
Fig.~\ref{ee}(a), typically, in the delocalized phase $S_A\propto \log
L$ including the half-filled case~\cite{prb} except when $\nu\approx
\alpha^2, \alpha^3$ and $\alpha^4 (0.382, 0.236$ and $0.145)$ where
$S_A$ becomes a constant abiding by the area-law of entanglement
entropy. In the localized phase, $S_A$ always follows area-law
independent of $\nu$~\ref{ee}(c). In the multifractal phase, generally
$S_A$ scales in some intermediate manner with $L$ whereas the area-law
persists when $\nu=0.145, 0.236, 0.382$. The variation of $S_A$ with
$\nu$ is shown in Fig.~\ref{ee}(d-f) for increasing values of $N$ for
$\lambda=1.0,2.0$ and $3.0$ respectively. In the delocalized phase,
the sudden drops of $S_A$ at $\nu=0.145,0.236,0.382,0.618$ etc. are
clearly visible in Fig.~\ref{ee}(d) coinciding with the appearance of
large gaps in the energy spectrum. In the multifractal and localized
phases, the energy spectrum becomes more fragmented due to appearance
of large gaps. The fluctuations in $S_A$ maximize at the multifractal
point and hence the $S_A$ vs. $\nu$ plots become more complicated
[see Figs~\ref{ee}(e) and \ref{ee}(f)].  However, the value of $S_A$ goes down as
the strength of disorder $\lambda$ increases, which is shown
explicitly in Fig.~\ref{ee_nu}. It is to be noted that for the special
values of $\nu={0.145,0.236,0.382}$ the criticality ($\lambda_c=2$) of
the AAH model is not reflected in the behavior of $S_A$ with $\lambda$
whereas for other values of filling ($\nu={0.2,0.3,0.5}$) there is a
sharp decrease of $S_A$ at $\lambda_c=2$ indicating the inherent
critical nature of the model. It is to be noted that
  the results presented here on the entanglement entropy remains
  unaffected if one chooses to study with the system sizes which are
  Fibonacci numbers $N=34,144,610$ etc., unlike the single particle
  quantities discussed previously. This indicates that the
  many-particle phenomena are not governed by the presence or absence
  of the isolated single particle localized states in the large band
  gaps. Rather many-particle results are the manifestation of the
  large gaps in the single particle energy spectrum.

\subsection{Persistent current}
\begin{figure*}[htpb]
\centering
\includegraphics[width=0.6\columnwidth]{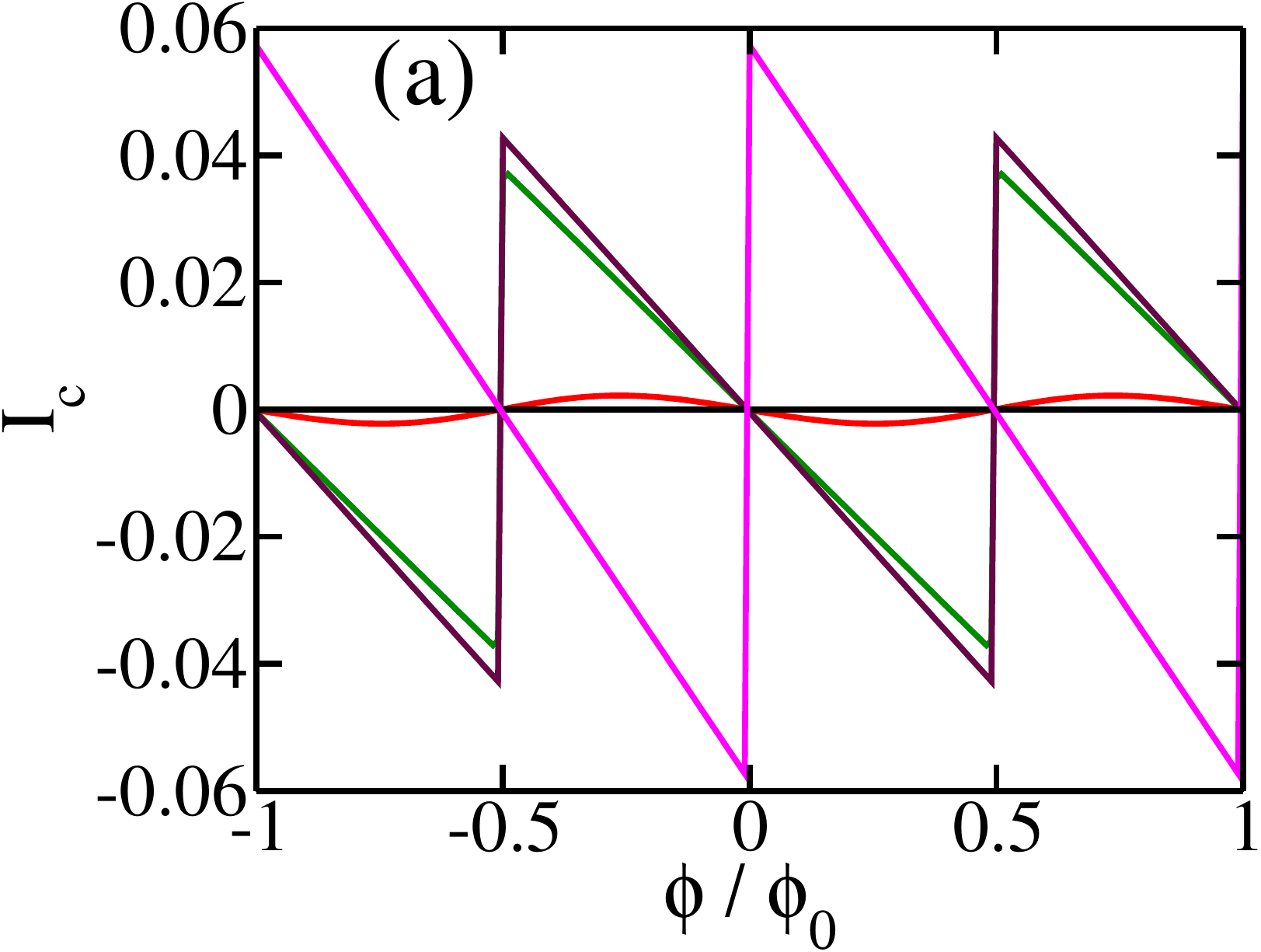}
\includegraphics[width=0.6\columnwidth]{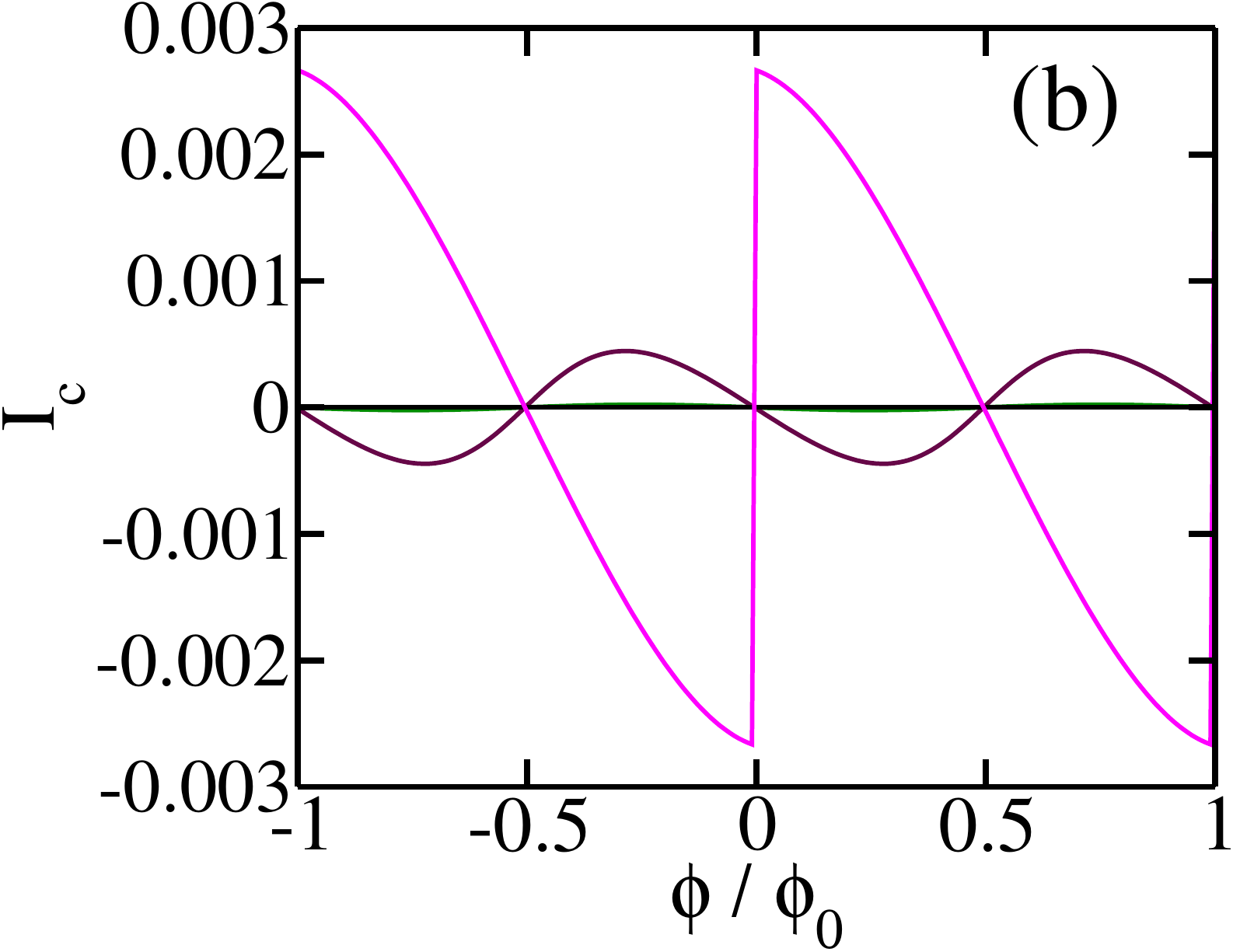}
\includegraphics[width=0.6\columnwidth]{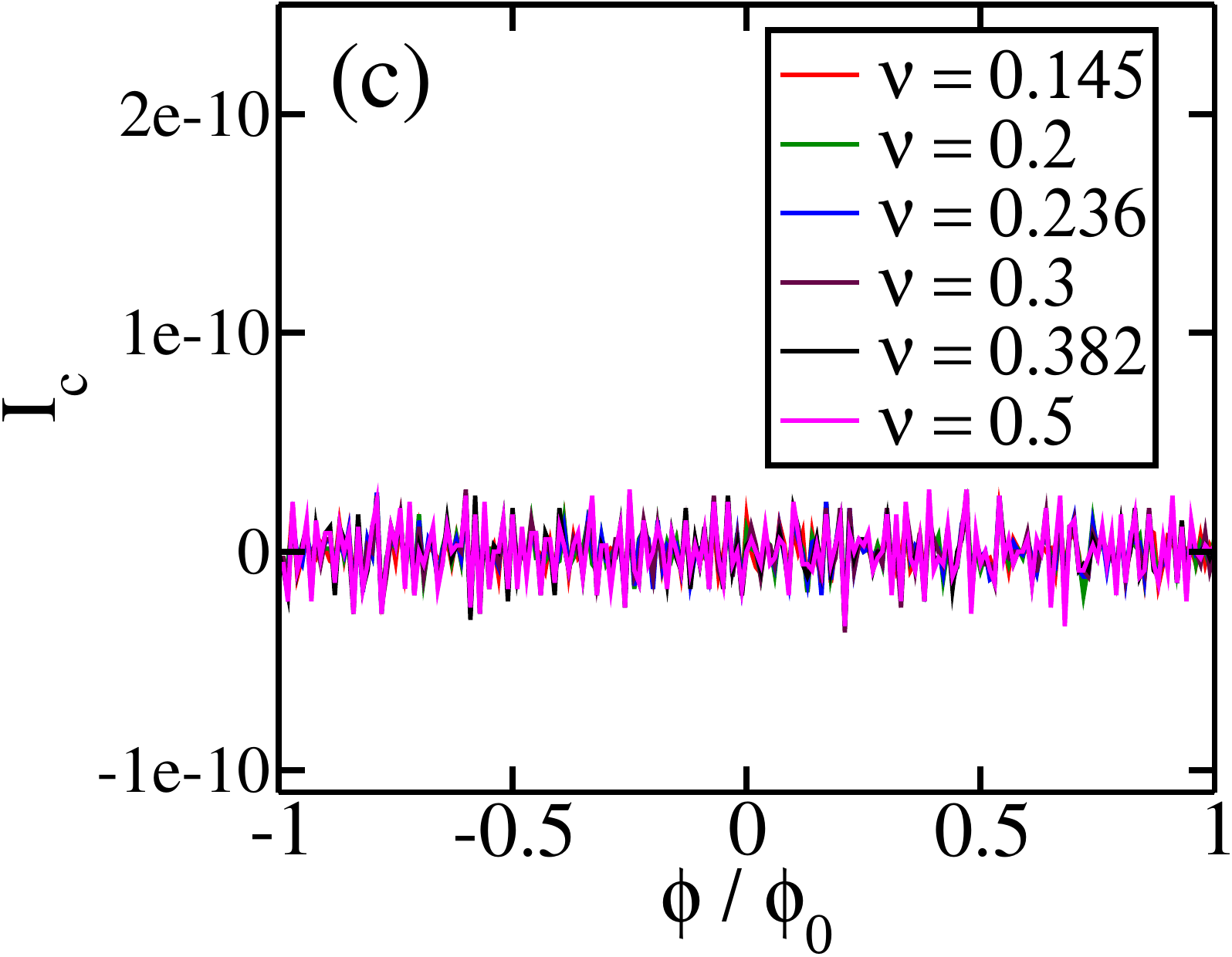}
\includegraphics[width=0.6\columnwidth]{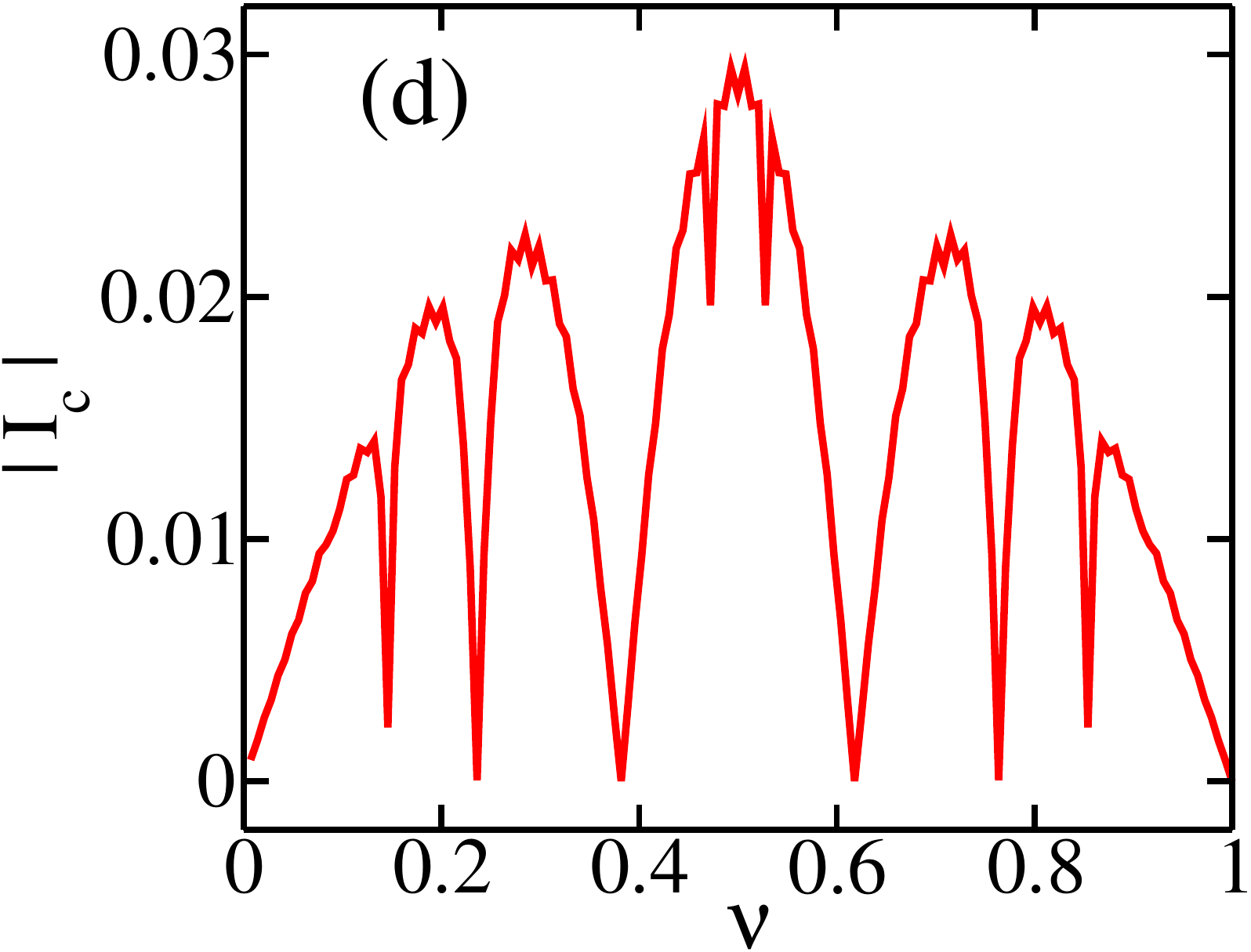}
\includegraphics[width=0.6\columnwidth]{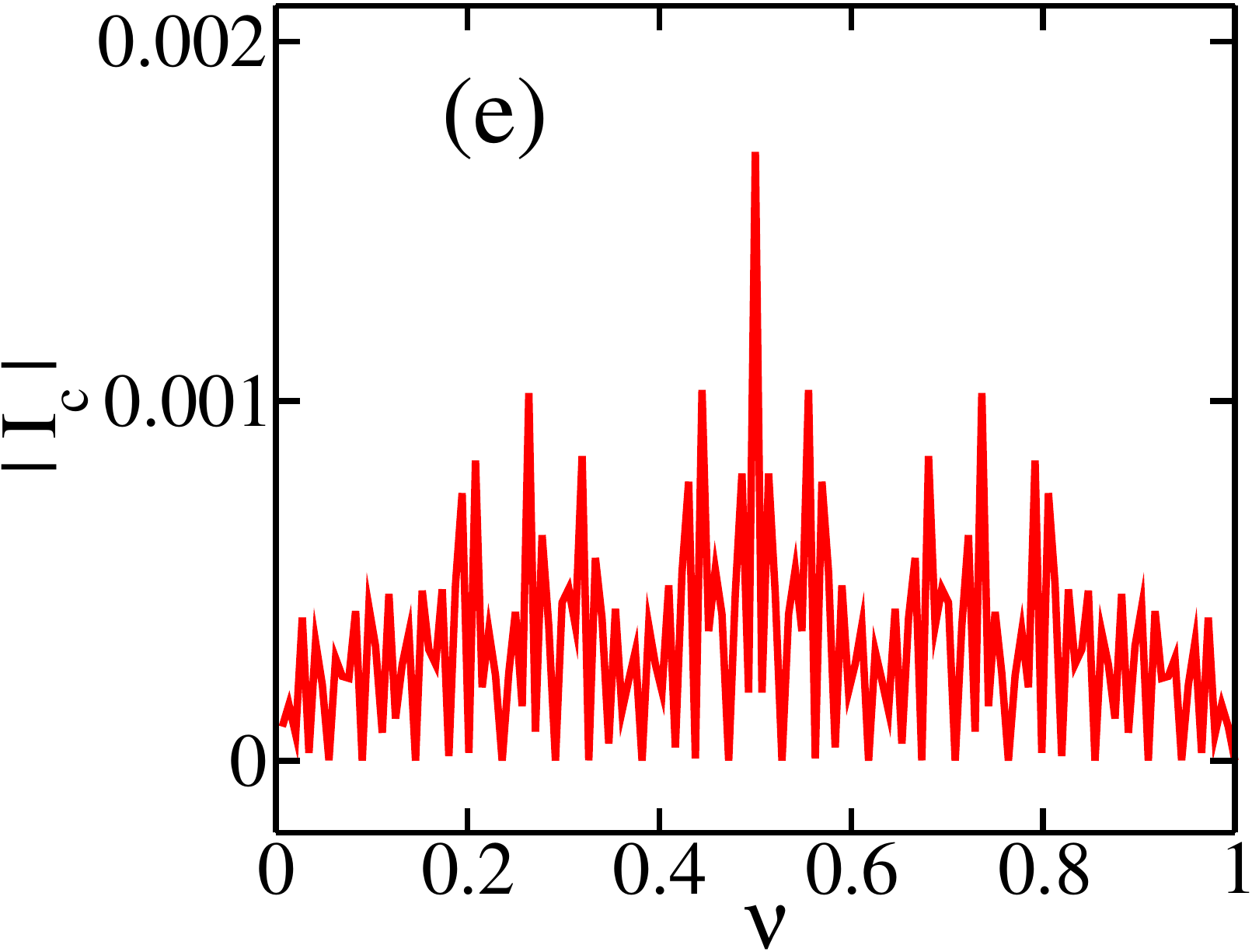}
\includegraphics[width=0.6\columnwidth]{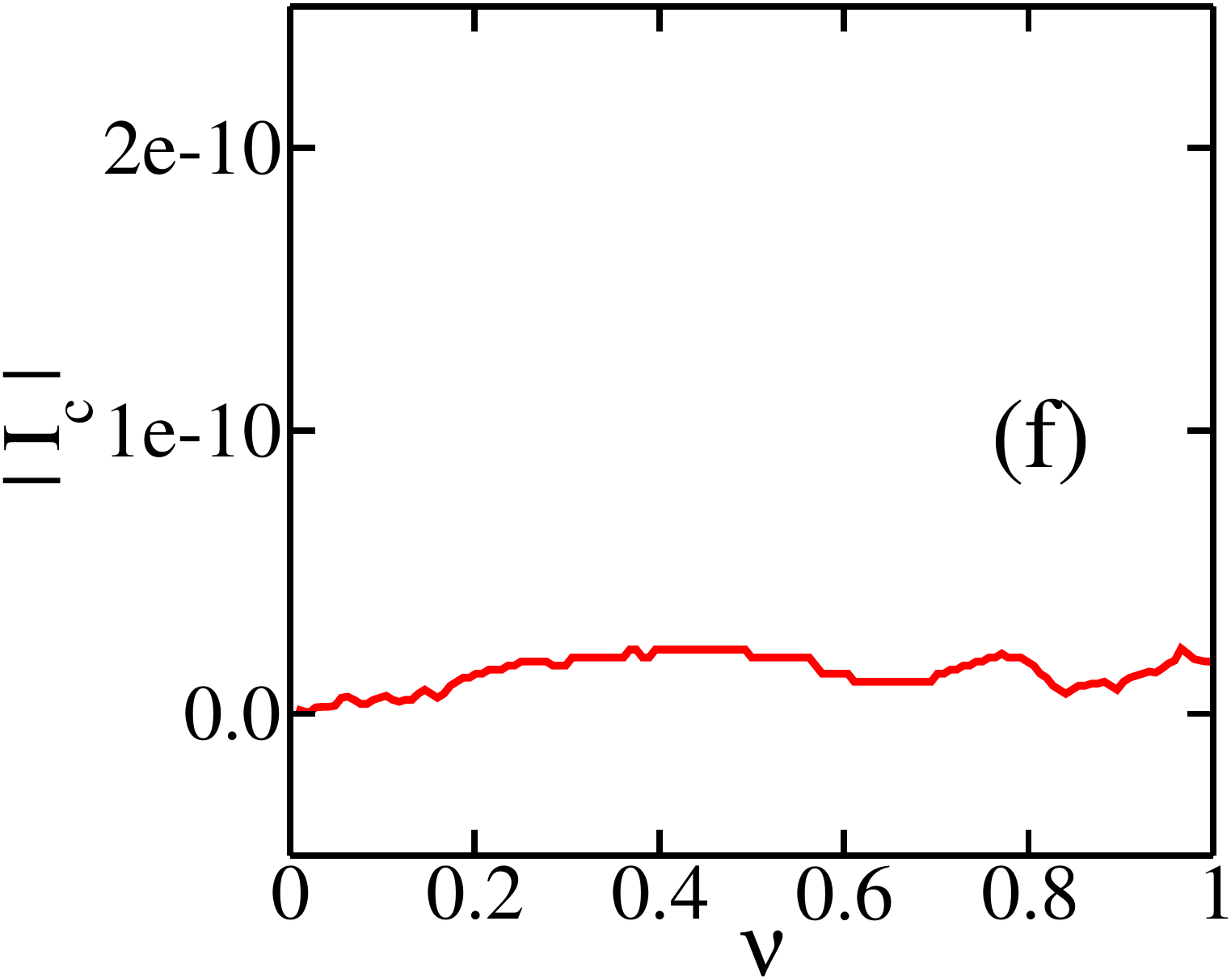}
\caption{(a-c) Persistent current $I_c$ as function of the flux $\phi$
  in unit of $\phi_0$ for increasing values of filling fraction $\nu$
  for $\lambda=1.0,2.0$ and $3.0$ respectively. For all plots (a-c)
  different colors denote different values of $\nu$ as defined in (c).  
  (d-f) $|I_c|$ as function of $\nu$ for
  $\lambda=1.0,2.0$ and $3.0$ respectively at fixed $\phi=0.25
  \phi_0$. For all the plots $\alpha=377/610$ with $N=610$.}
\label{pc2}
\end{figure*}
\begin{figure}
\includegraphics[width=0.85\columnwidth]{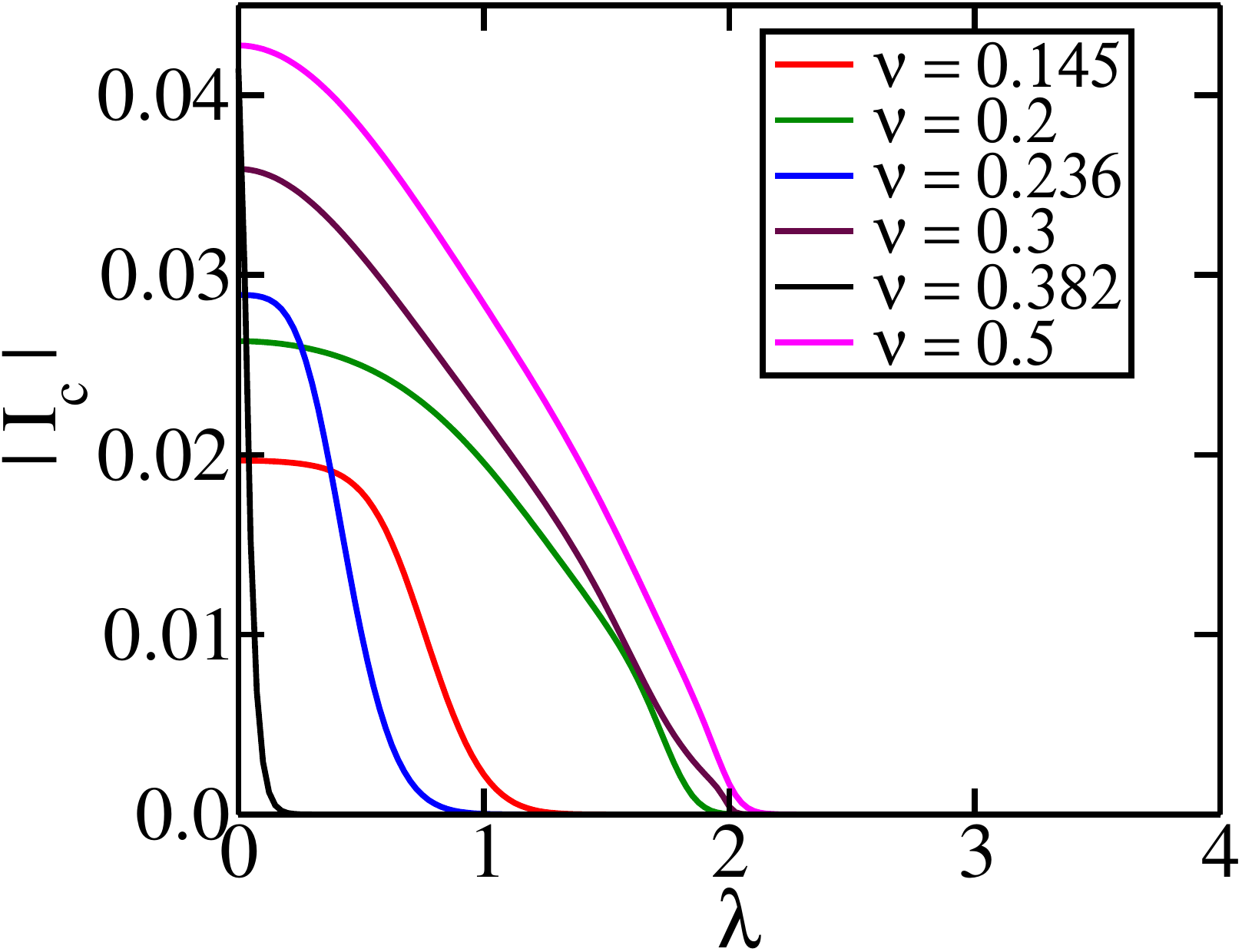}
\caption{The absolute value of the persistent current $|I_c|$ as function of $\lambda$ for increasing values of $\nu$ at fixed $\phi=0.25 \phi_0$. For all the plots $\alpha=377/610$ with $N=610$.}
\label{pc3}
\end{figure}
\begin{figure*}[htpb]
\centering
\includegraphics[width=0.63\columnwidth]{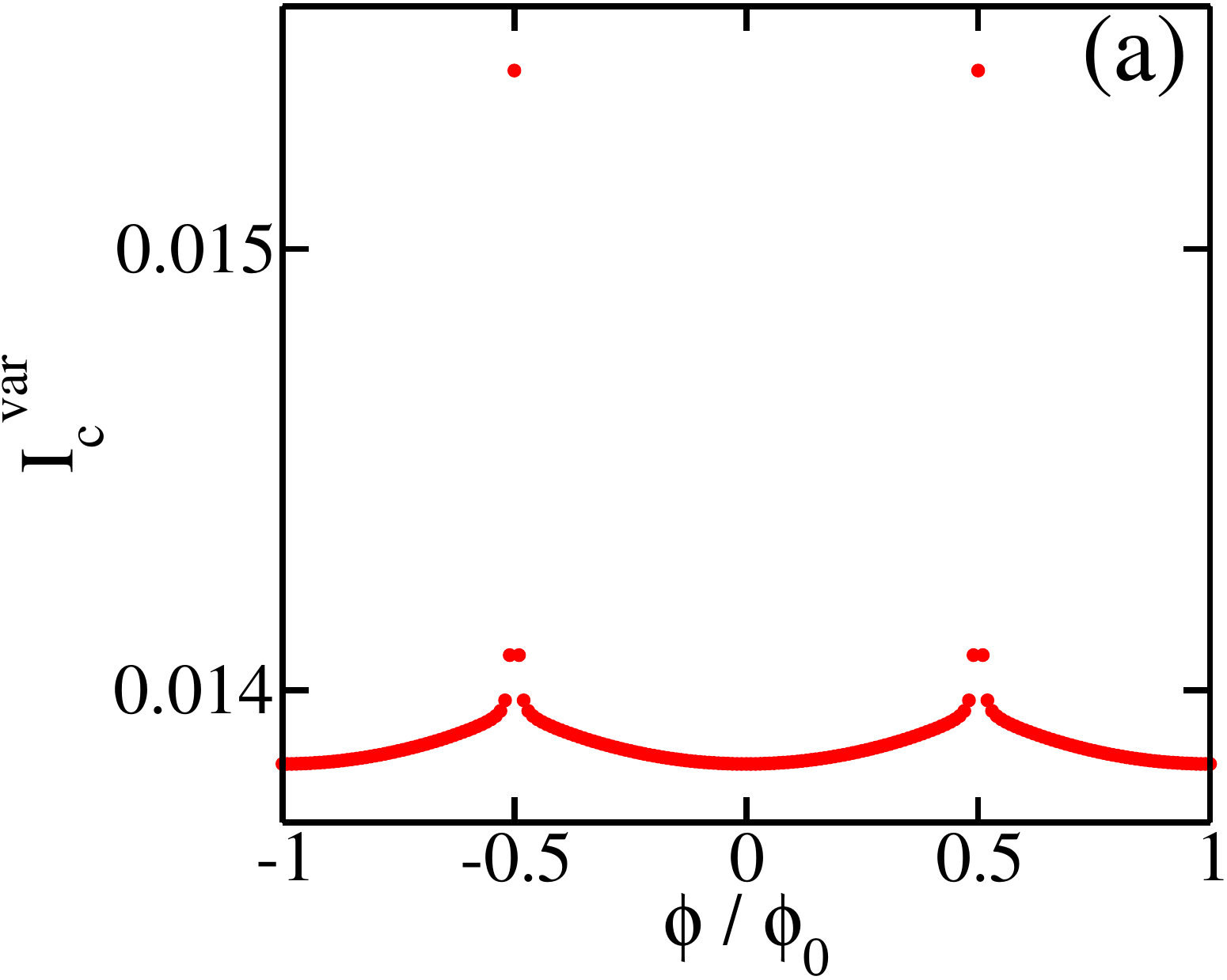}
\includegraphics[width=0.63\columnwidth]{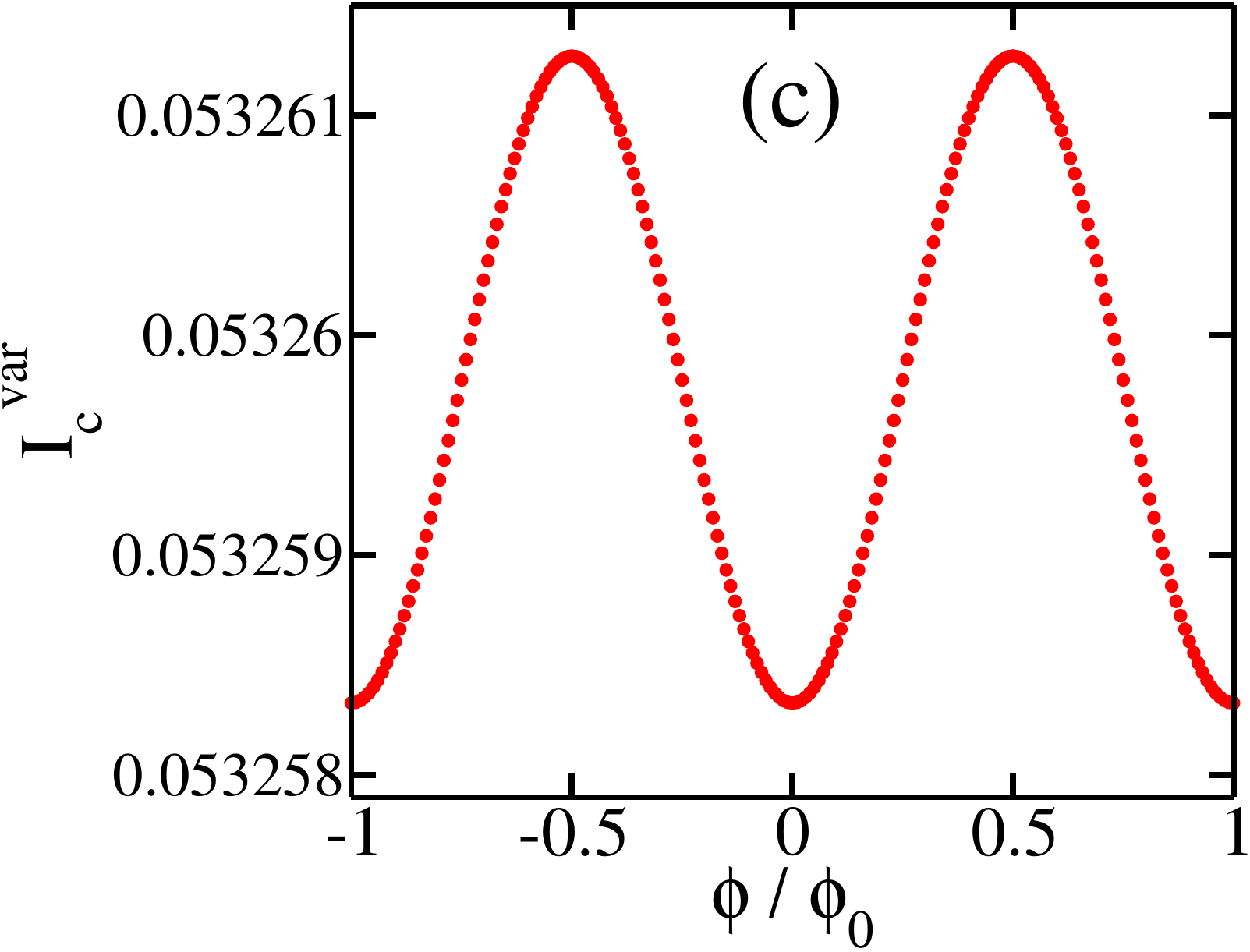}
\includegraphics[width=0.63\columnwidth]{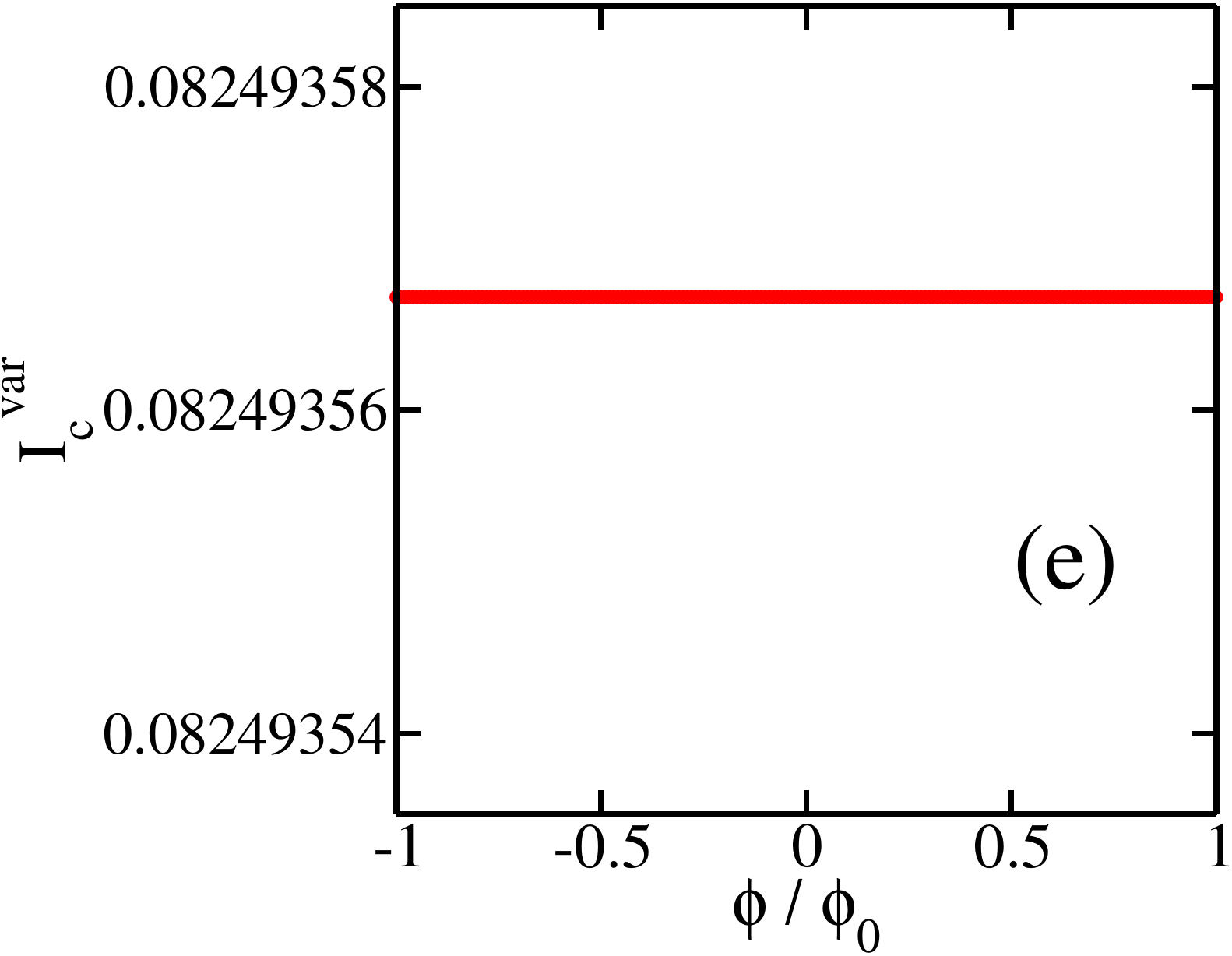}
\includegraphics[width=0.63\columnwidth]{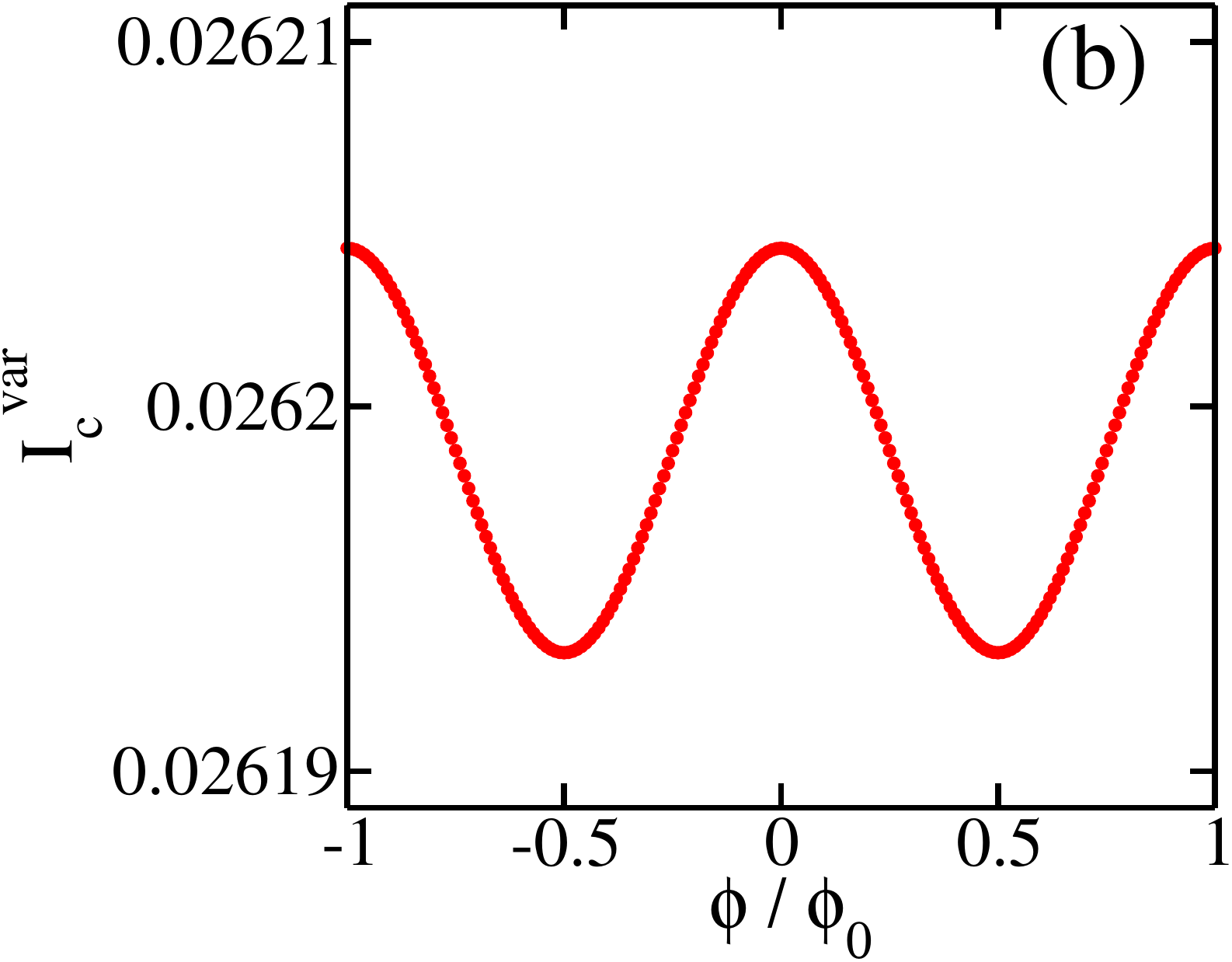}
\includegraphics[width=0.63\columnwidth]{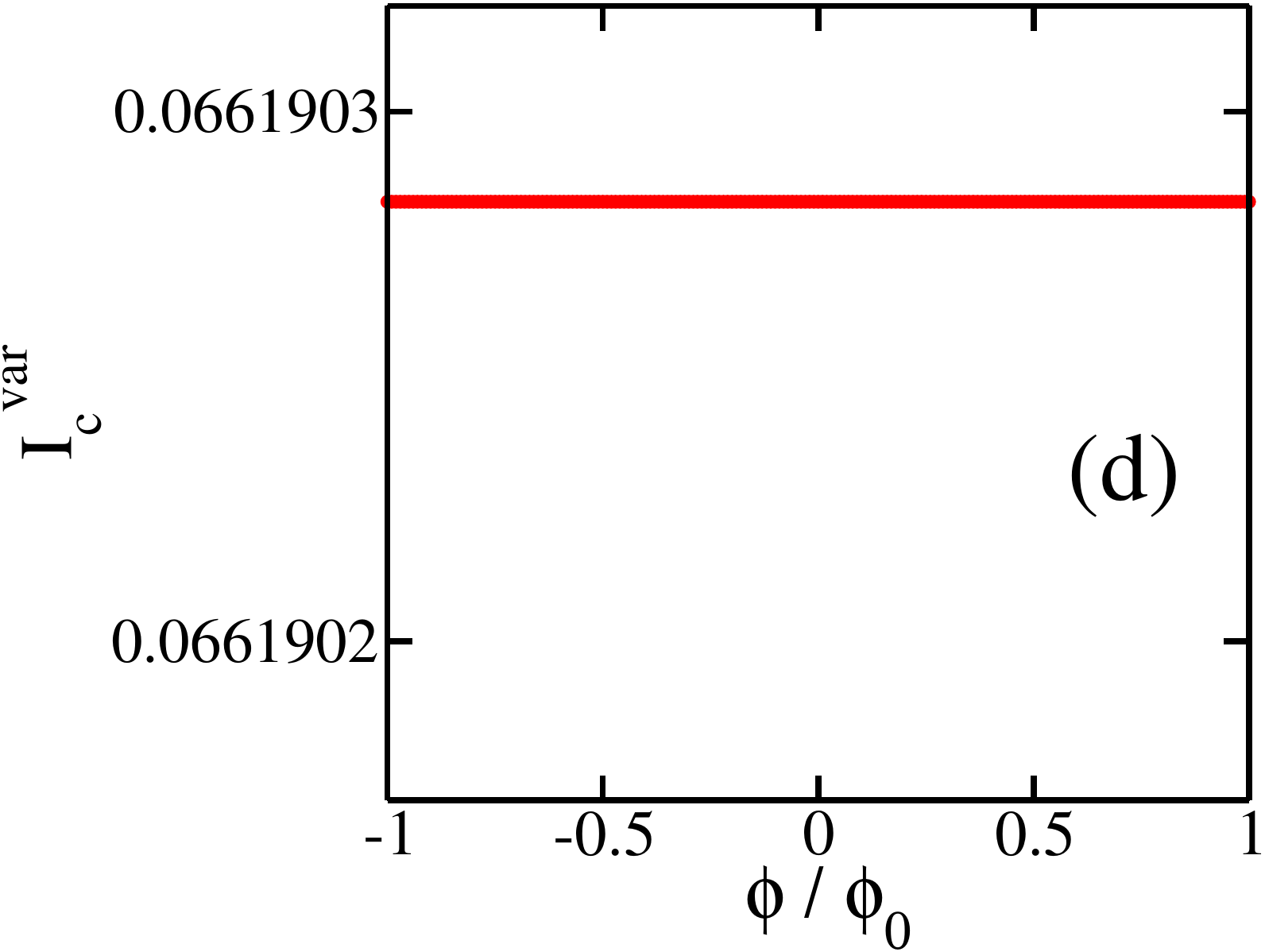}
\includegraphics[width=0.63\columnwidth]{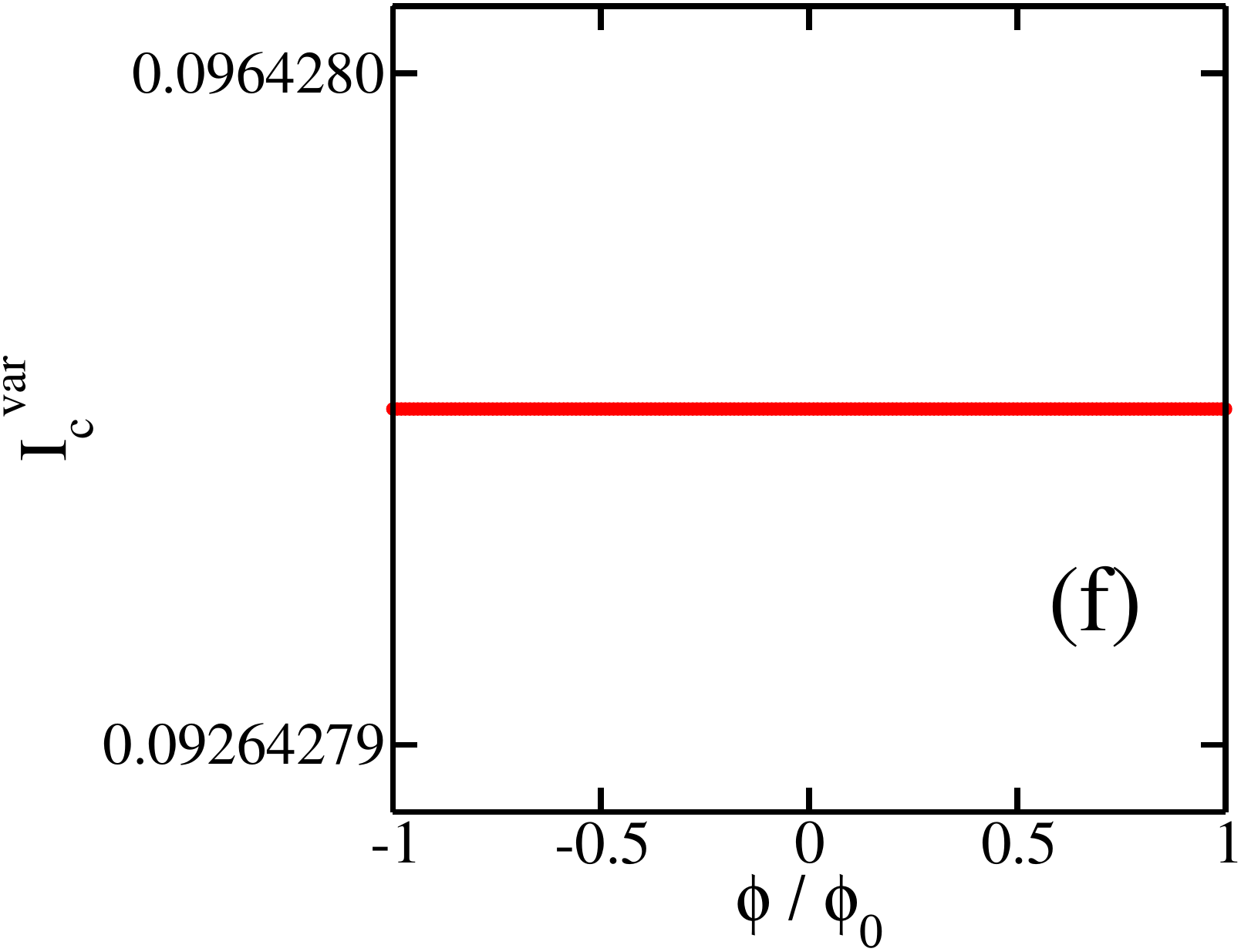}
\caption{Dependence of the variance of persistent current $I_c^{var}$ on the flux $\phi$ (in units of $\phi_0$) for fermions with different filling fractions $\nu$ and for increasing values of the strength $\lambda$ of the quasiperiodic potential. Fig. (a), (c) and (e): $I_c^{var}$ vs $\phi$ plots for $\lambda=1.0,2.0$ and $3.0$ respectively and fixed regular value of $\nu=0.2$. Fig. (b), (d) and (f): $I_c^{var}$ vs $\phi$ plots for $\lambda=1.0,2.0$ and $3.0$ respectively and fixed special value of $\nu=0.236$. Total number of sites $N=144$ and $\alpha=89/144$ for all the plots.}
\label{pc_var}
\end{figure*} 
Next we discuss the persistent current in the fermionic system, which
has also been used to study localization phenomena~\cite{Imry}. Persistent
current can be generated by applying a phase twist at the
boundary. With periodic boundary conditions this is equivalent to
attaching a flux to the fermions moving in a ring. In a mesoscopic
quantum ring a persistent current of electrons can be produced by
applying a magnetic flux $\phi$ inside the ring. In a quantum ring the
current-flux relationship can depend on factors such as band
structure, disorder, interaction, ring geometry, number of particles
etc. In this work, we mainly focus on the current-flux relationship
and its variation with the strength of quasi-periodic disorder
$\lambda$ of the AAH model and the filling fraction $\nu=N_p/N$ of
the $N_p$ fermions in a ring of $N$ sites. We will put $\theta_p=0$
for all the results presented in this section. An irrational number
can be expanded as a continued fraction~\cite{continuedfraction}, which
makes possible a successive rational approximation of it in the form
of $a_0/b_0$, where $a_0$ and $b_0$ are coprime integers.  A
rational approximation of $\alpha$ is given by $\alpha=F_{s-1}/F_{s}$
for two successive members in the series defined as~\cite{fibonacci}:
\begin{eqnarray}
F_s = F_{s-1} + F_{s-2}
\label{fibo}
\end{eqnarray}
with $F_0=0$, $F_1=1$, which converges to the inverse of the
`golden mean' $(\sqrt{5}-1)/2$ in the limit $s\rightarrow\infty$. We
will make a rational approximation over $\alpha$ as described above
along with $N=F_s$ to validate the periodic boundary condition in this
section. In order to calculate the persistent current we consider a
phase-twisted Hamiltonian for fermions, which is given by:
\begin{eqnarray}
H (\theta) = -J\sum_l(e^{-i\theta/N} c_{l}^{\dagger}c_{l+1} + H.c.) +  \lambda\sum\limits_l \cos(2\pi\alpha l) c_{l}^{\dagger}c_l,\nonumber \\
\label{ham_twist}
\end{eqnarray}
where $c_l$ is the fermionic annihilation operator acting on site $l$
and $\theta=2 \pi \frac{\phi}{\phi_0}$, where $\phi_0=h/e$ is the unit
flux quanta. After diagonalization of this Hamiltonian, the single
particle energy levels $\epsilon_{n}(\phi)$ are used to calculate
the persistent current~\cite{Imry,cheung1}:
\begin{eqnarray}
I_c(\phi)=-\frac{\partial E_0}{\partial \phi} 
\label{Ic}
\end{eqnarray}
where $E_0 = \sum_{n}\epsilon_{n}(\phi) \theta(E_{F} - \epsilon_{n})$
is the ground state energy of the system and $E_{F}$ is the Fermi
energy at zero temperature.  In the absence of any potential, the energy
dispersion is given by $\epsilon_n(\phi)=-2J\cos(\frac{2\pi}{N}(n+\frac{\phi}{\phi_0}))$
where $-{N}/2 \leq n < {N}/2$. For $N_p$ fermions in $N$ sites, the
persistent current can be written as\cite{cheung1}:
\begin{equation}
I_c=-I_0 \frac{\sin (\frac{\pi}{N}(2\frac{\phi}{{\phi}_0}  + \eta))}{\sin(\frac{\pi}{N})}, 
\label{pc}
\end{equation}
where $I_0=\frac{4 \pi J}{N {\phi}_0}\sin(N_p \pi/N)$. The persistent
current $I_c$ exhibits periodic variation with flux $\phi/\phi_0$ and
a phase shift $\eta$ is generated due to the parity of the number of
fermions $N$.  For odd $N_p$, $\eta=0$ in region
$-0.5\leq\frac{\phi}{{\phi}_0}<0.5$ and for even $N_p$, $\eta=-1$ in
region $0\leq\frac{\phi}{{\phi}_0}<1$. The persistent current decreases
with system size as $I_0\propto 1/N$ and it is maximum when $\nu=0.5$
since $I_0$ is maximum at the same point.
 
Now as the AAH disorder is turned on, the current-flux plots are shown
in Fig.~\ref{pc2}. In the delocalized phase, $I_c$ is almost vanishing
for all values of $\phi$ when $\nu=0.145, 0.236, 0.382$. Otherwise the
maximum of $I_c$ increases with $\nu$ [Fig.~\ref{pc2}(a))]. The
$I_c-\phi$ plots look like sawtooth curves because when $N$ is very
large, one can use in Eq.~\ref{pc}, the small-angle approximation for
the sine function: $(\sin(x)\approx x)$. In the critical phase $I_c$
is periodic with $\phi$ but its magnitude is very small whereas $I_c$
is still vanishing for special values of $\nu$ as mentioned before
[Fig.~\ref{pc2}(b)]. In the localized phase $I_c$ vanishes for all
values of $\phi$ and $\nu$ [Fig.~\ref{pc2}(c))] as has been reported
for the particular case of half-filling~\cite{jstat}. The absolute value of the persistent current $|I_c|$ as function of $\nu$ for three values of
$\lambda=1.0, 2.0, 3.0$ at fixed value of $\phi=0.25\phi_0$ is plotted
in Figs.~\ref{pc2}(d)-\ref{pc2}(f). The current falls off substantially at the
special filling $\nu=0.145,0.236,382$ even when the system is in the
delocalized phase in Fig.~\ref{pc2}(d) similar to the dependence of
$S_A$ on $\nu$ as shown in Fig.~\ref{ee}(d). In the multifractal
phase, the magnitude of current significantly decreases whereas its
fluctuations increase as function of $\nu$ as shown in
Fig.~\ref{pc2}(e) similar to $S_A$ in multifractal phase
[Fig.~\ref{ee}(e)]. The current vanishes for all values of $\nu$ as
one goes into the localized phase as can be seen from
Fig.~\ref{pc2}(f). The dependence of $|I_c|$ on $\lambda$ for
increasing values of $\nu$ at fixed $\phi$ is shown in
Fig.~\ref{pc3}. The criticality of the AAH model is evident from the
$I_c-\lambda$ plots for normal filling $\nu=0.2, 0.3, 0.5$ as $I_c$
vanishes at $\lambda_c=2$ for those fillings whereas $I_c$ goes to
zero for smaller values of $\lambda<2$ well before $\lambda=2$ for
non-trivial $\nu=0.145, 0.236, 0.382$. This shows that the critical
nature of the model is somehow suppressed at those fillings as is also
clear from Fig.~\ref{ee_nu}.

The persistent current can also be calculated from the expectation value of
the current operator defined as $J^c=-\frac{\partial H}{\partial
  \phi}$. The current operator can thus be written as:
\begin{eqnarray}
J^c=\sum_{l} J^c_{l,l+1}
\label{pc_corr}
\end{eqnarray}
where $J^c_{l,l+1}=-i\frac{2\pi J}{N} (e^{-i\frac{2\pi\phi}{\phi_0N}}
c_{l}^{\dagger}c_{l+1} - e^{i\frac{2\pi\phi}{\phi_0N}}
c_{l+1}^{\dagger}c_{l})$ and $\phi$ is expressed in the units of
$\phi_0$. The two definitions of the persistent current are related
via the Feynman-Hellmann (FH) theorem as $\langle -\frac{\partial
  H}{\partial \phi}\rangle = -\frac{\partial E_0}{\partial \phi}$,
except at points of degeneracies~\cite{FH_zhang,FH_alon,FH_fernandez,FH_vatsya} in the
energy spectrum. At the degenerate points, the system has many
eigenfunctions corresponding to a single energy and any linear
combination of the true eigenfunctions also satisfies the
Schr\"{o}dinger equation. As FH theorem equates two quantities
involving eigenfunctions and eigenvalues respectively, the equality
apparently becomes invalid in the presence of degeneracies in the
eigenvalues (for details, see Appendix~\ref{app}).

\subsection{Variance of persistent current}
\begin{figure}[htpb]
\centering
\includegraphics[width=0.5\columnwidth]{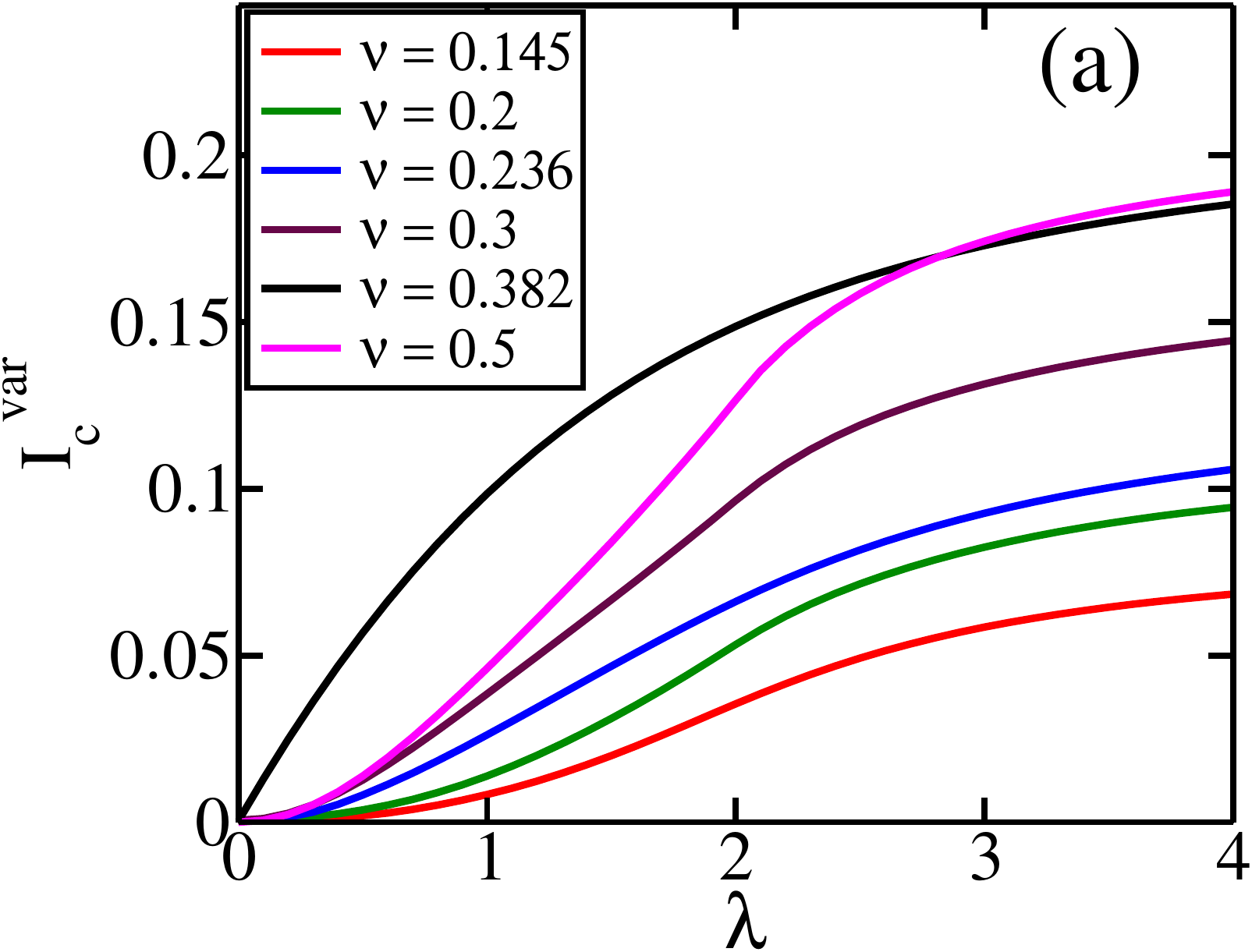}
\includegraphics[width=0.5\columnwidth]{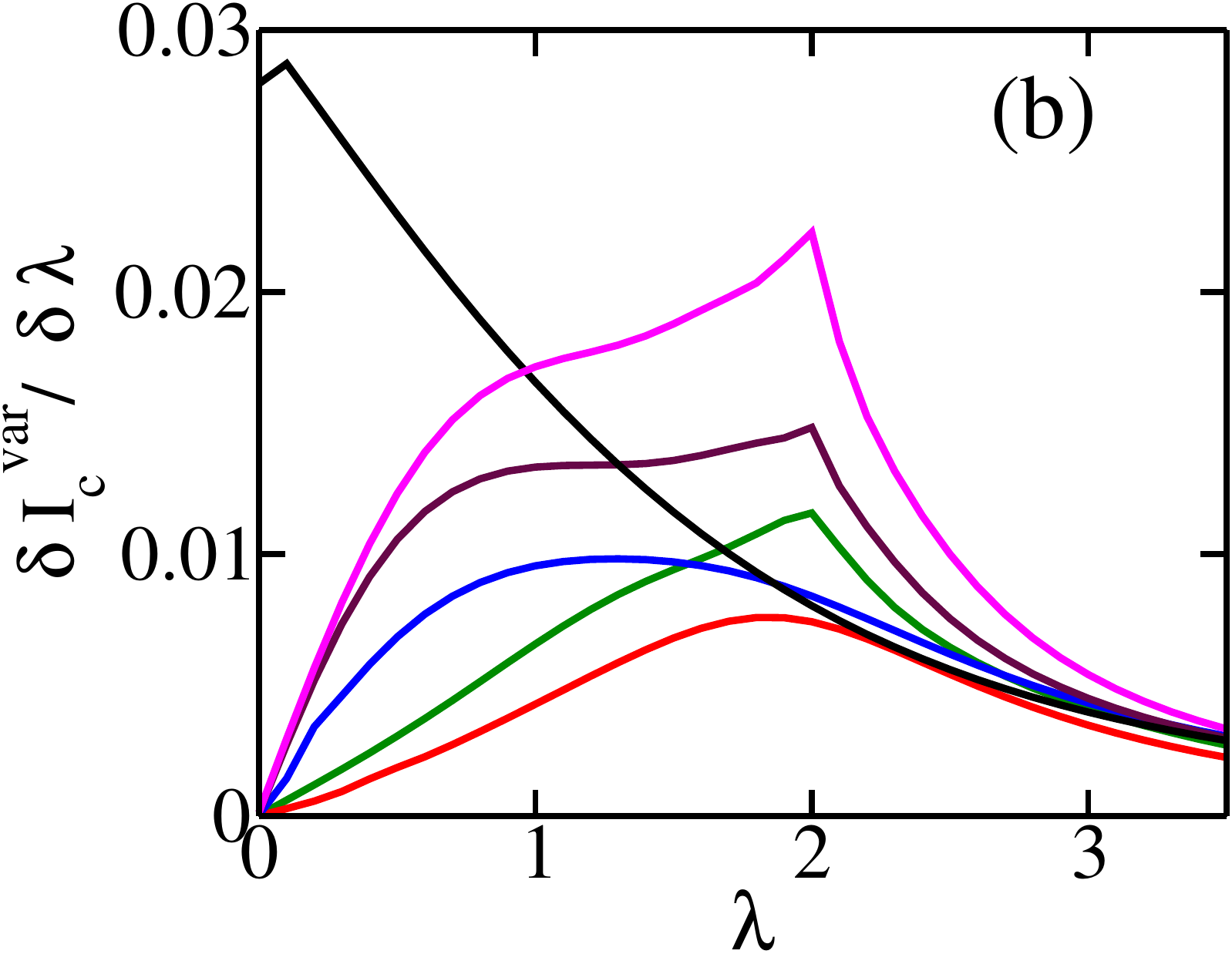}
\caption{(a) The variation of $I_c^{var}$ with $\lambda$ for the fixed $\phi=0.25\phi_0$ for regular and special filling fractions $\nu$ of fermions. (b) The first derivative of $I_c^{var}$ with $\lambda$ for the fixed $\phi=0.25\phi_0$ for increasing filling fractions $\nu$ of fermions.  Total number of sites $N=144$ and $\alpha=89/144$ for all the plots.}
\label{pcvar_l}
\end{figure}
\begin{figure}[htpb]
\centering
\includegraphics[width=0.5\columnwidth]{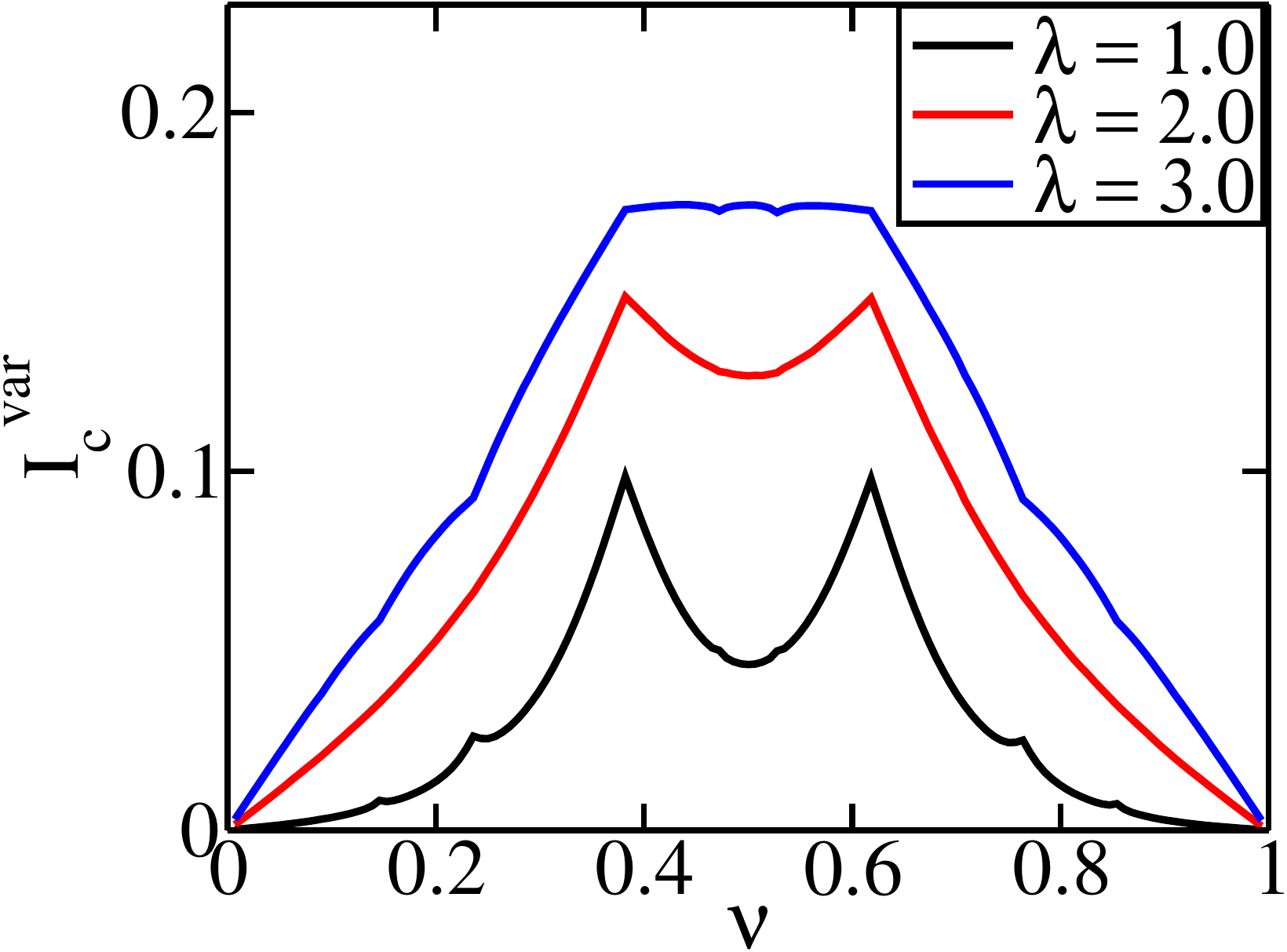}
\caption{Dependence of the variance of persistent current $I_c^{var}$ on the filling fraction $\nu$ for increasing values of the strength $\lambda$ of the quasiperiodic potential and fixed $\phi=0.25\phi_0$. Total number of sites $N=144$ and $\alpha=89/144$ for all the plots.}
\label{pcvar_nu}
\end{figure}
\begin{figure*}[htpb]
\centering
\includegraphics[width=0.66\columnwidth]{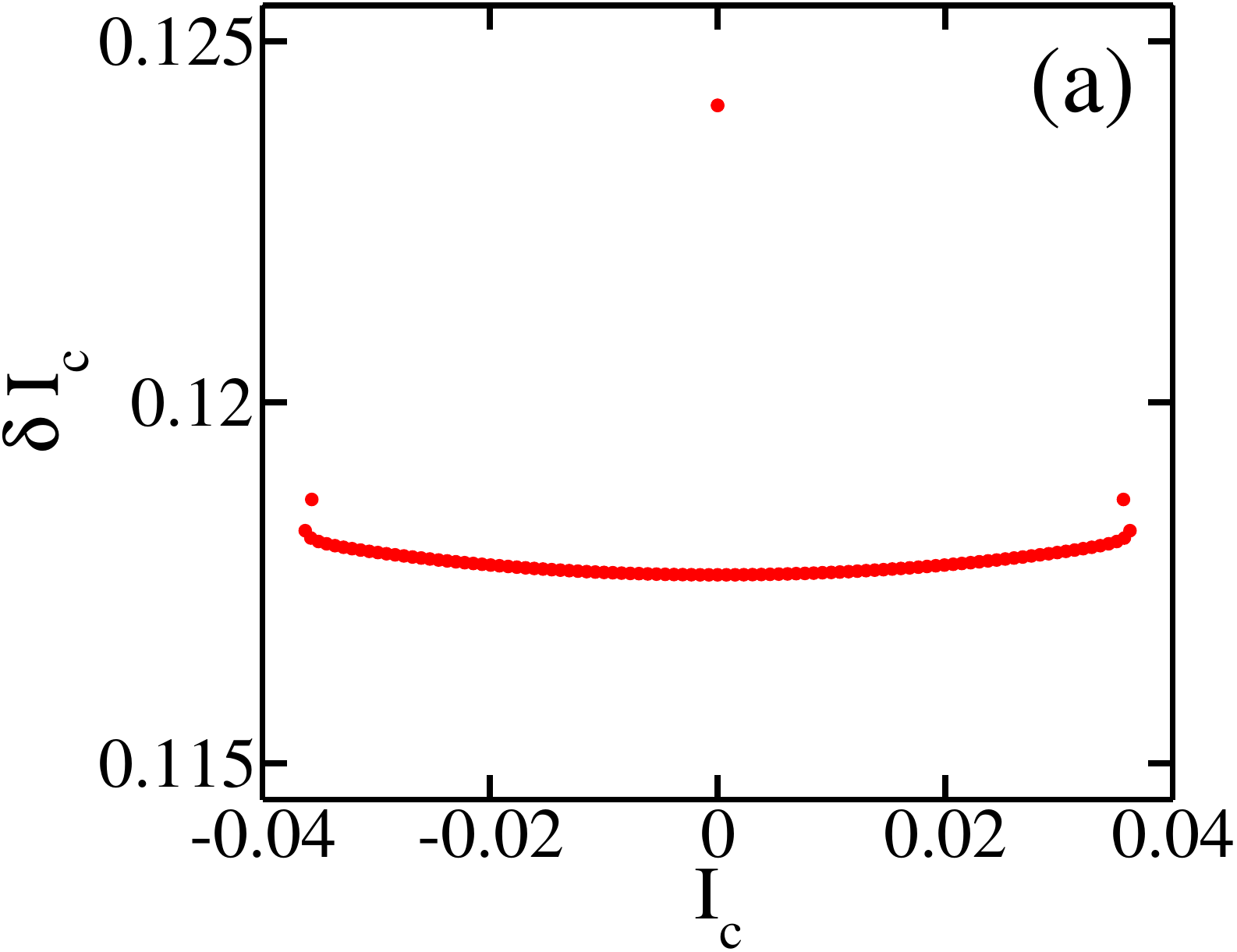}
\includegraphics[width=0.66\columnwidth]{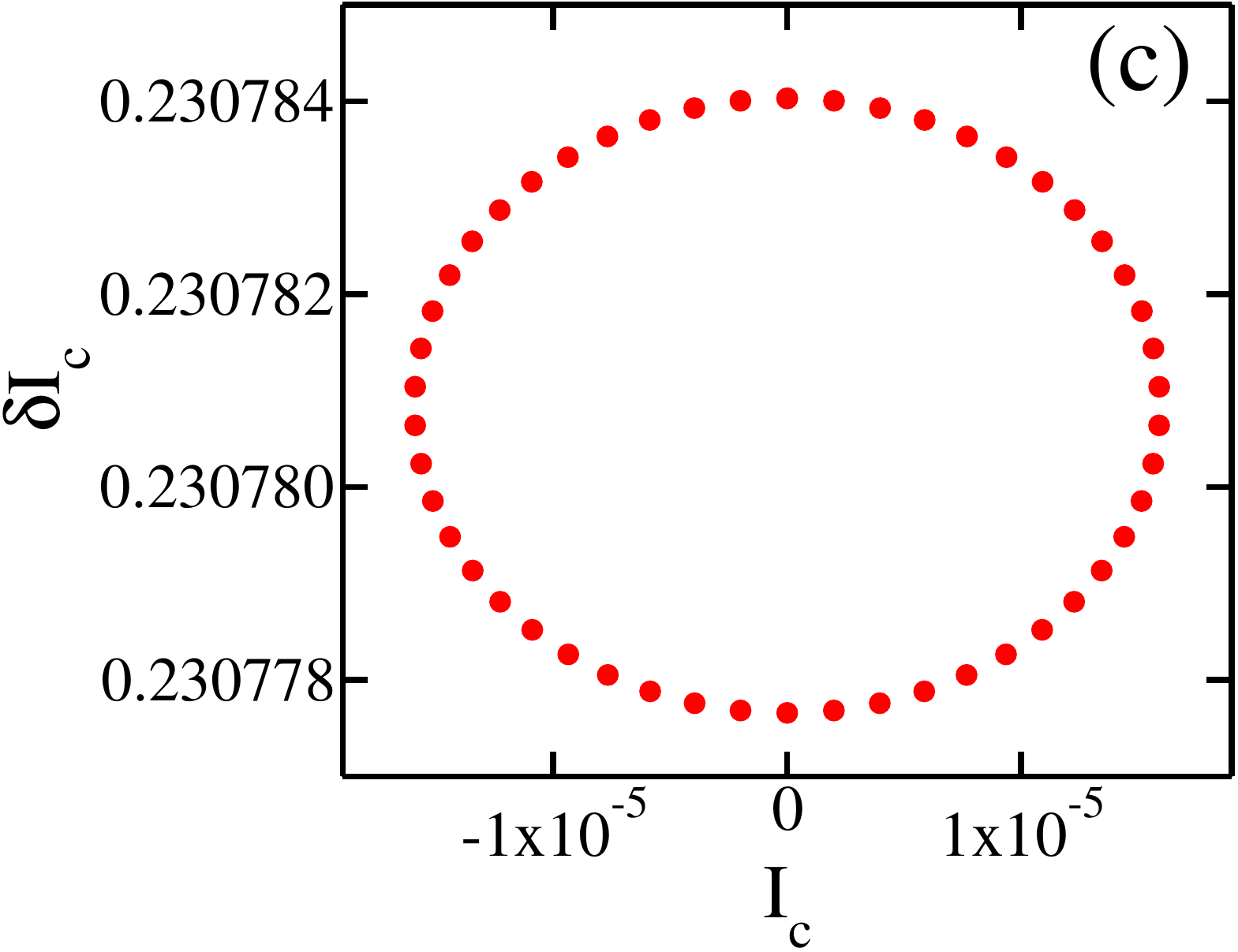}
\includegraphics[width=0.66\columnwidth]{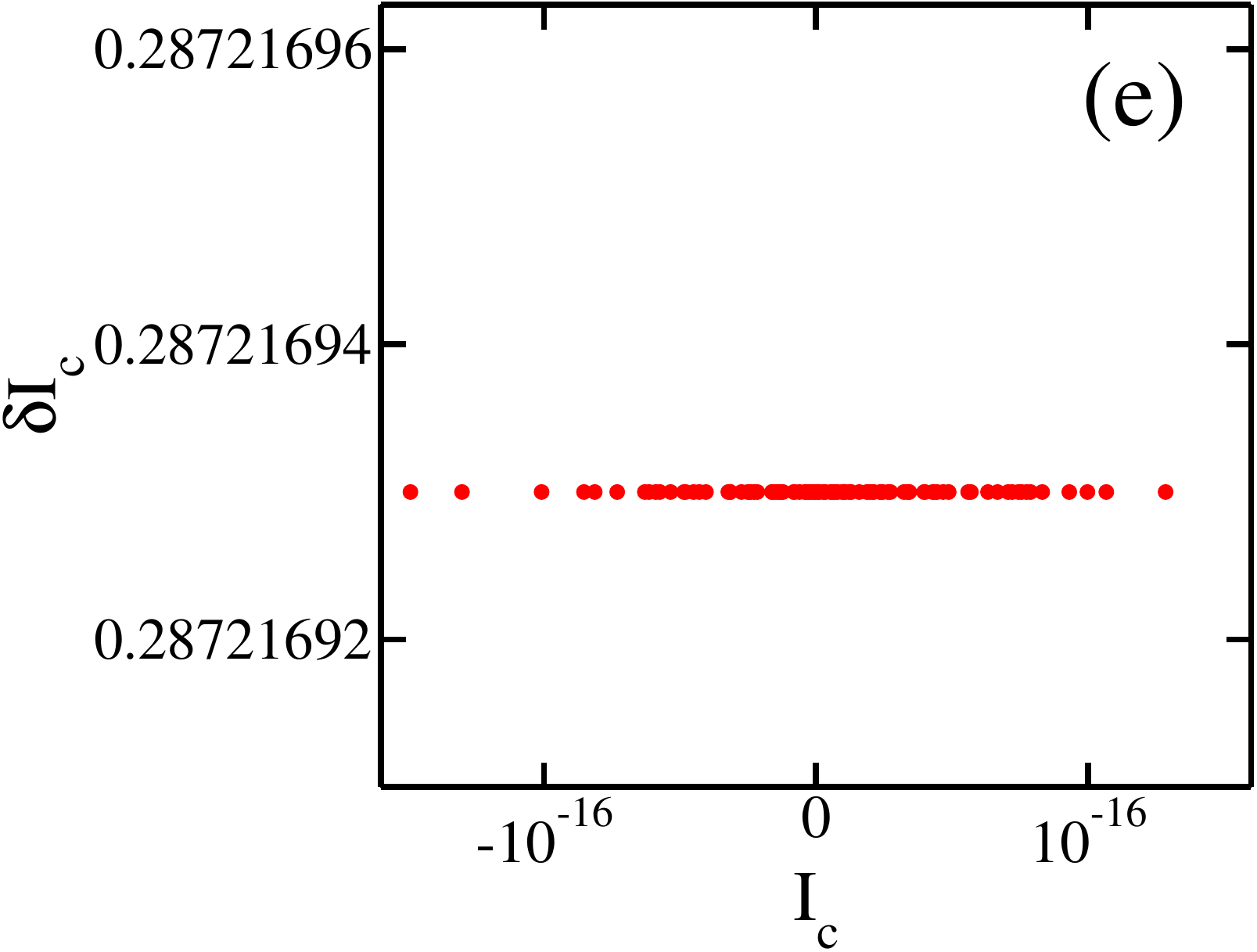}
\includegraphics[width=0.66\columnwidth]{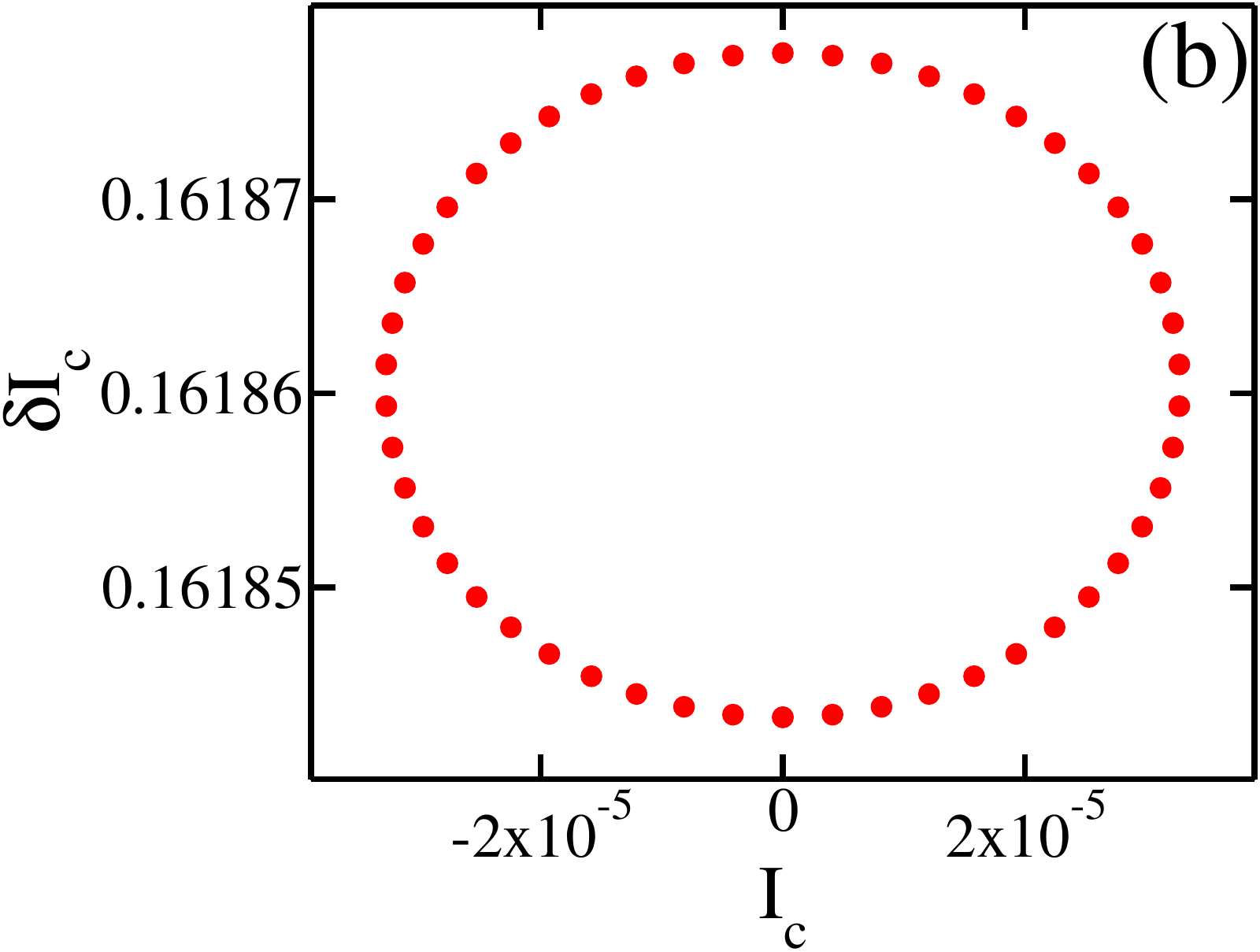}
\includegraphics[width=0.66\columnwidth]{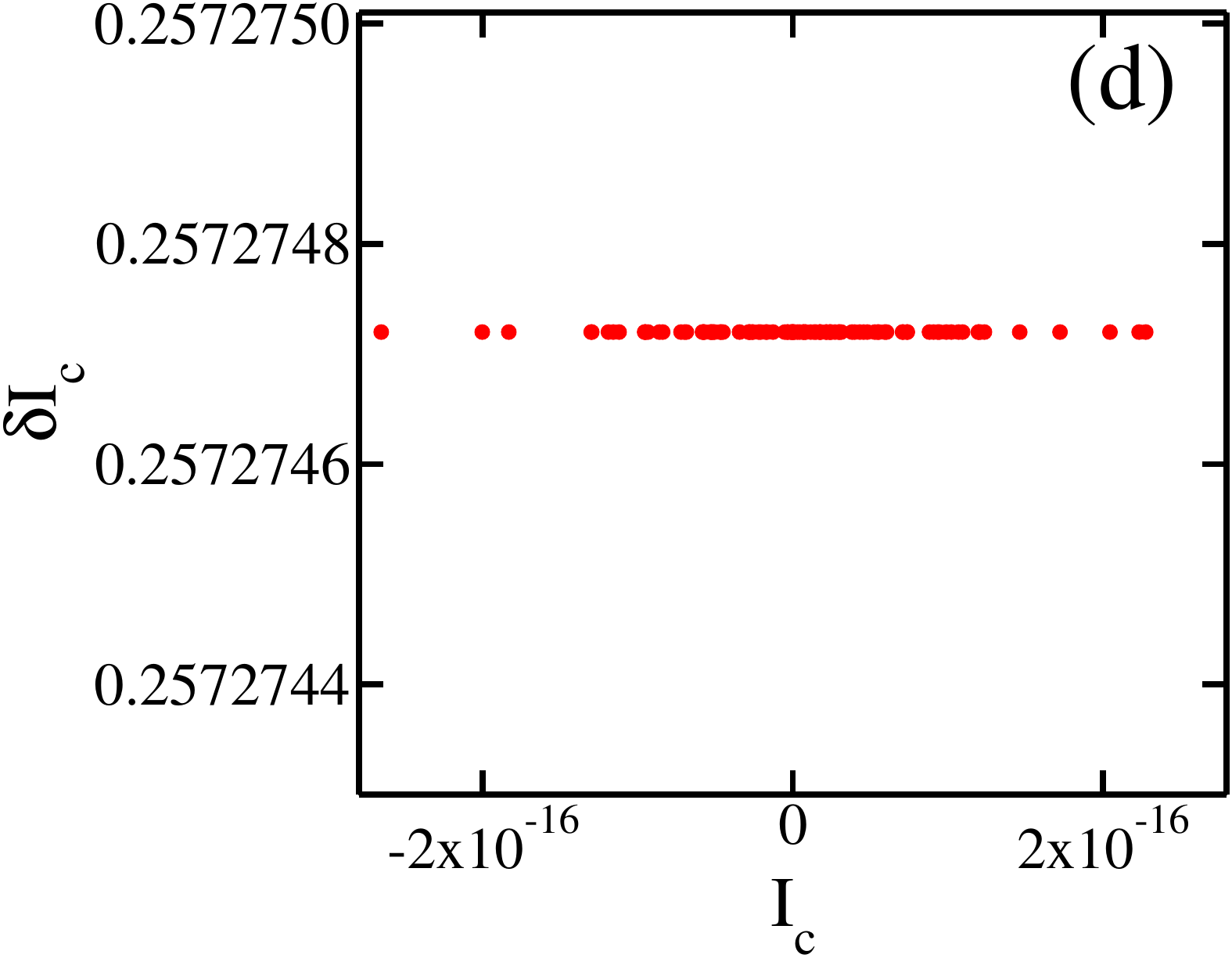}
\includegraphics[width=0.66\columnwidth]{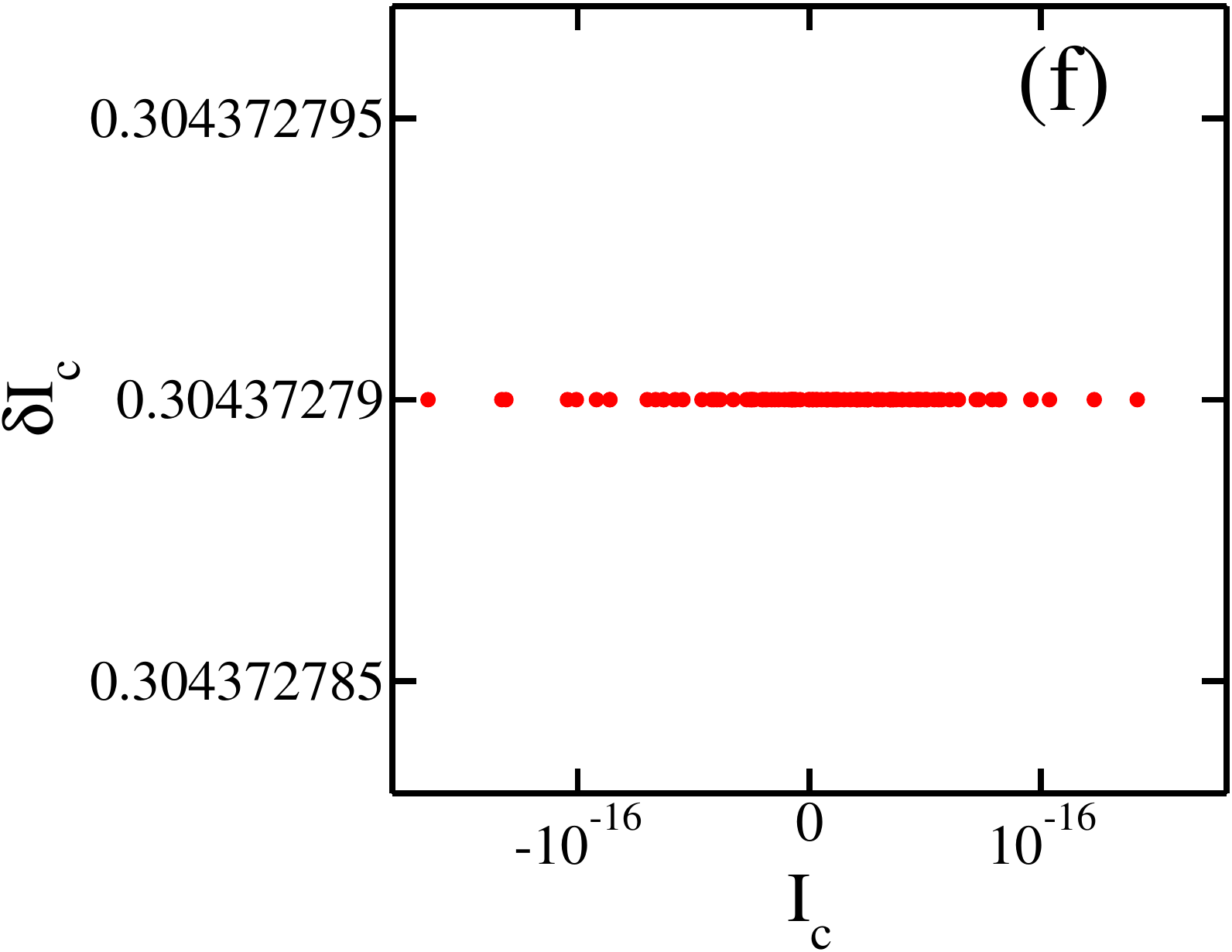}
\caption{The $\delta I_c$-$I_c$ curves for fermions with different filling fractions $\nu$ and for increasing values of the strength $\lambda$ of the quasiperiodic potential. Fig. (a), (c) and (e): $\delta I_c$-$I_c$ curves for $\lambda=1.0,2.0$ and $3.0$ respectively and fixed regular value of $\nu=0.2$ Fig. (b), (d) and (f): $\delta I_c$-$I_c$ curves for $\lambda=1.0,2.0$ and $3.0$ respectively and fixed special value of $\nu=0.236$. Total number of sites $N=144$ and $\alpha=89/144$ for all the plots.}
\label{phase}
\end{figure*}

A study of the fluctuations of observables in quantum systems can be
very profitable, as they carry important information about the
systems, sometimes more than the mean values of the observables. Here
we will study the persistent current, computed using
the definition $I_c=\langle J^c \rangle$, as described in
Eq.~\ref{pc_corr}.  The variance of the persistent current is given
by:
\begin{eqnarray}
I_c^{var}=\langle {J^c}^2\rangle - \langle {J^c}\rangle^2,
\end{eqnarray}
where
\begin{eqnarray}
{J^c}^2=\sum_{l} {J^c_{l,l+1}}^2 + \sum_{l\neq l^\prime}J^c_{l,l+1} J^c_{l^\prime,l^\prime +1} \nonumber\\, 
\end{eqnarray}
and
\begin{eqnarray}
\langle{J^c_{l,l+1}}^2\rangle=&&-{\frac{4\pi^2 J^2}{N^2}}( 2\langle c_l^\dagger c_l \rangle \langle c_{l+1}^\dagger c_{l+1}\rangle \nonumber\\&&- 2\langle c_l^\dagger c_{l+1} \rangle \langle c_{l+1}^\dagger c_{l}\rangle - \langle c_{l+1}^\dagger c_{l+1}\rangle - \langle c_{l}^\dagger c_{l}\rangle ),\nonumber\\
\langle J^c_{l,l+1} J^c_{l^\prime,l^\prime+1} &&+J^c_{l^\prime,l^\prime+1}J^c_{l,l+1}\rangle =-{\frac{4\pi^2 J^2}{N^2}} [(\langle c_l^\dagger c_{l^\prime+1}\rangle\delta_{l^\prime,l+1}\nonumber\\ &&+ 2\langle c_l^\dagger c_{l+1}\rangle \langle c_{l\prime}^\dagger c_{l^\prime+1}\rangle\nonumber
-2\langle c_l^\dagger c_{l^\prime+1}\rangle \langle c_{l^\prime}^\dagger c_{l+1}\rangle)e^{-i2\theta}\nonumber\\ &&- (2\langle c_l^\dagger c_{l+1}\rangle\langle c_{l^\prime+1}^\dagger c_{l^\prime}\rangle
- 2\langle c_l^\dagger c_{l^\prime}\rangle\langle c_{l^\prime+1}^\dagger c_{l+1}\rangle)]\nonumber\\ &&+ h.c.
\label{Ic_var} 
\end{eqnarray}

The variation of $I_c^{var}$ with $\phi$ is shown in
Fig.~\ref{pc_var} for regular filling $\nu=0.2$ and special filling
$\nu=0.236$ in the delocalized $(\lambda=1.0)$, critical
$(\lambda=2.0)$ and localized $(\lambda=3.0)$ phases respectively. The
variance of current varies periodically with flux but it becomes
discontinuous if the Fermi level overlaps with quasi-degenerate levels
$(\phi=\pm 0.5\phi_0)$ for regular fillings ($\nu=0.2$ in the figure)
in the delocalized phase $\lambda<2$ as can be seen from
Fig.~\ref{pc_var}(a). However, for special fillings ($\nu=0.236$ in
Fig.~\ref{pc_var}(b)) $I_c^{var}$ shows a continuous sinusoidal
variation with $\phi$ as the Fermi level does not lie within
degenerate levels for any value of $\phi$. At the critical point
$\lambda=2.0$, $I_c^{var}$ vs. $\phi$ plots become sinusoidal for
regular fillings due to lifting of the degeneracies as shown in
Fig.~\ref{pc_var}(c) for $\nu=0.2$ whereas $I_c^{var}$ becomes
independent of $\phi$ for special fillings (see
Fig.~\ref{pc_var}(d)). In the localized phase $I_c^{var}$ is constant
for both types of fillings as shown in Fig.~\ref{pc_var}(e) and
Fig.~\ref{pc_var}(f) respectively.
For clarity, a fixed value of $\phi=0.25\phi_0$ is chosen for which
the Fermi level is always non-degenerate for any filling fraction $\nu$. For
$\phi=0.25\phi_0$ the variation of $I_c^{var}$ with $\lambda$ is
shown for regular and special values of $\nu$ in
Fig.~\ref{pcvar_l}(a). The variance of current monotonically increases with
$\lambda$ as the fluctuations of quantum observables
increase with localization. The first derivative of $I_c^{var}$ with
respect to $\lambda$ is calculated as a function of $\lambda$ to show
how the slope of the plots in Fig.~\ref{pcvar_l}(a) change across the
delocalization-localization transition. It turns out that for the
regular fillings, the change of slope shows a maximum at the
transition point $\lambda=2$ whereas no such peaks appear at the same
point in the plots for special fillings [see
Fig.~\ref{pcvar_l}(b)]. The variation of $I_c^{var}$ with filling
fraction $\nu$ is shown in Fig.~\ref{pcvar_nu} for $\lambda=1.0,2.0$
and $3.0$ respectively. In the delocalized phase ($\lambda=1.0$), the
plot shows sudden increase of $I_c^{var}$ at the values of special
fillings whereas those maxima disappear as one gets into the
localized phase. Such increase of quantum fluctuations is a property
of localized systems which indicates that the fermionic system behaves like
a localized one for special fillings even when $\lambda<2$.

\subsection{Relation between persistent current and its fluctuations}
\begin{figure}[htpb]
 \centering
 \includegraphics[width=0.68\columnwidth]{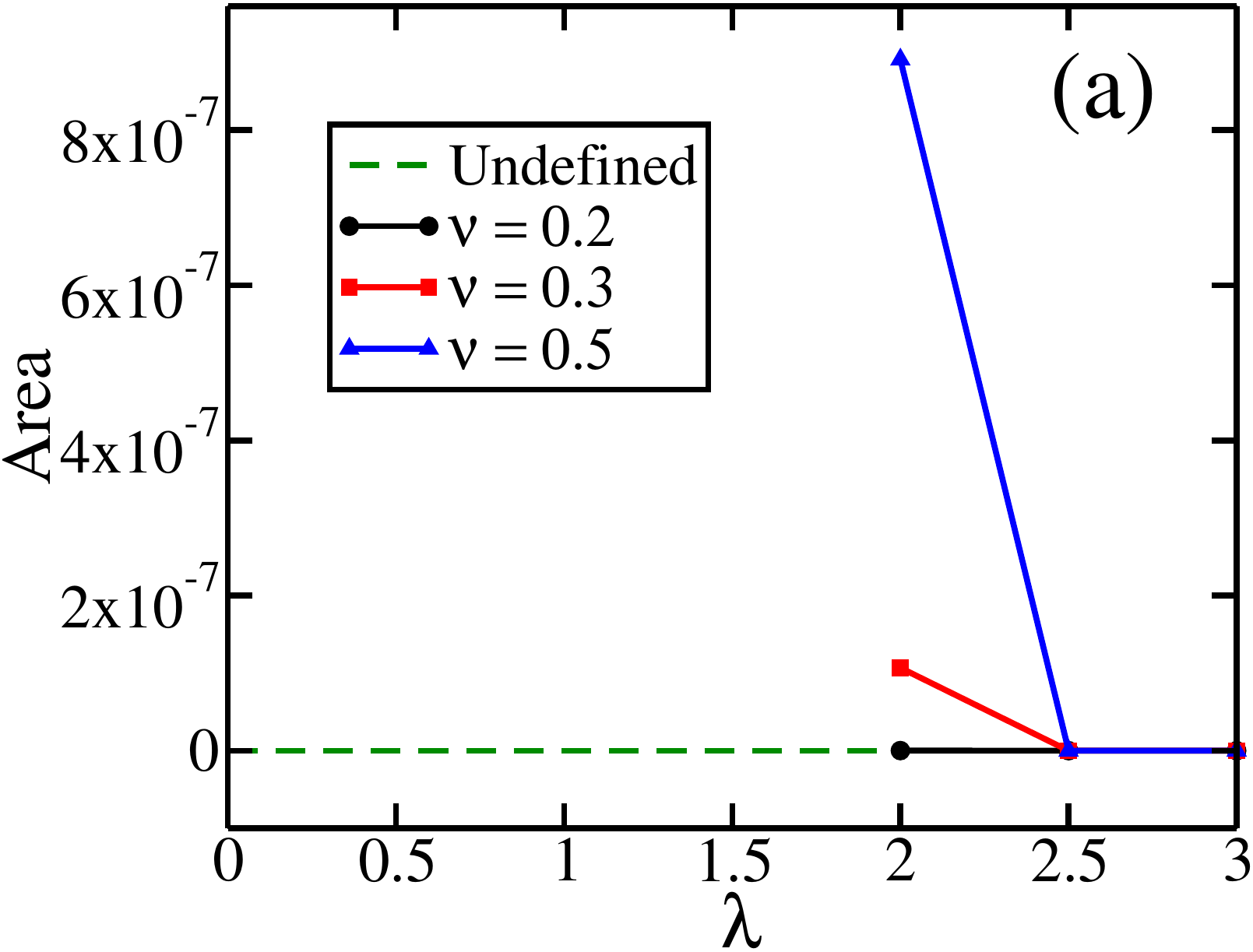}
 \includegraphics[width=0.68\columnwidth]{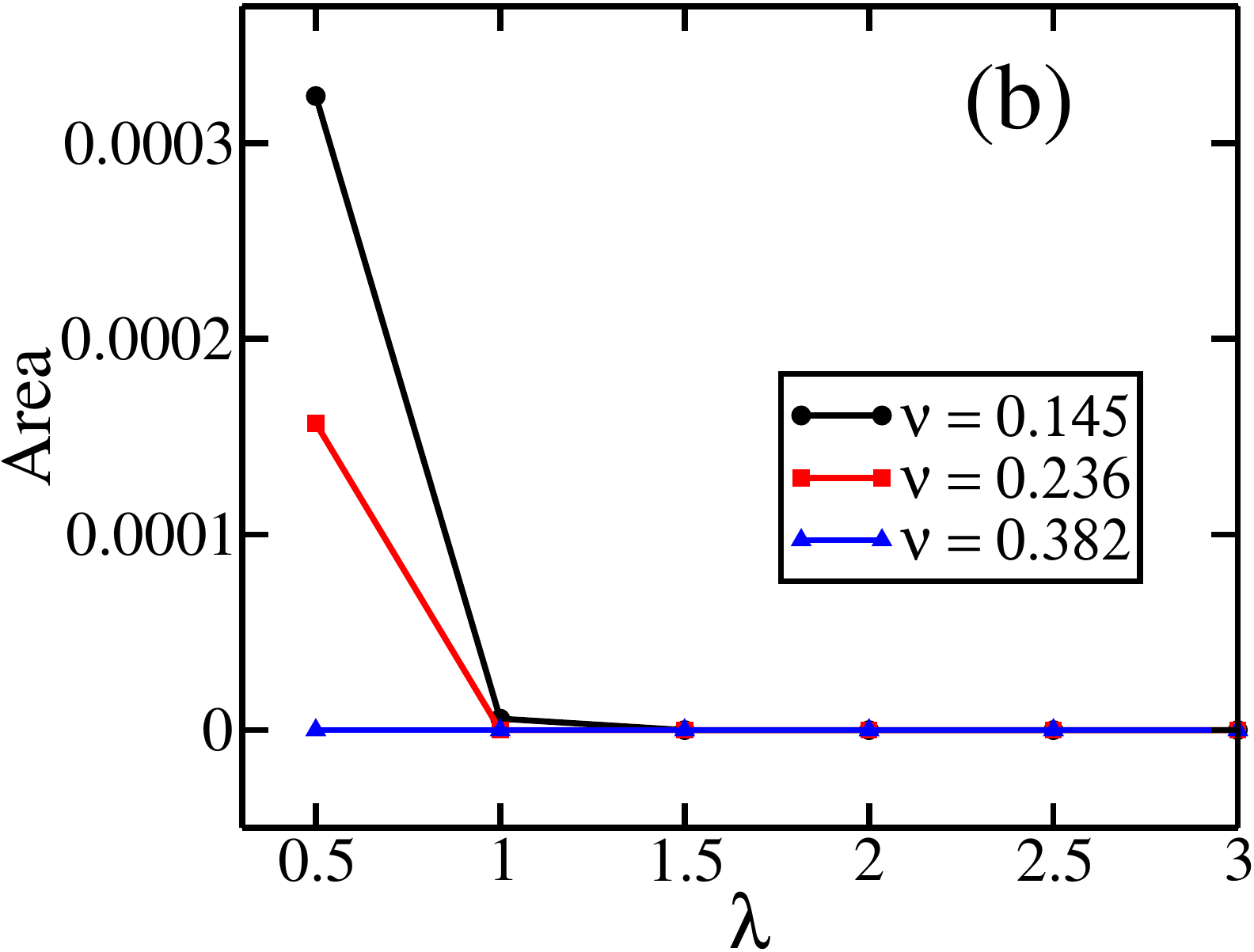}
 \caption{(a) The area enclosed by the $\delta I_c$-$I_c$ curves as function of $\lambda$ for non-special filling fractions. The green dotted line represents region where area connot be defined as a closed curve is not obtained due to quasi-degeneracies. (b) The area enclosed by the $\delta I_c$-$I_c$ curves as function of $\lambda$ for special filling fractions. For all the plots $N=144$ and $\alpha=89/144$. The $\delta I_c$-$I_c$ curves are plotted by varying $\phi \in [-\pi,\pi]\phi_0$.}
 \label{area}
 \end{figure}

In this subsection, we will explore the relationship between the
persistent current $I_c$ and its standard deviation $\delta I_c$, which are related by,
\textcolor{red}{
\begin{eqnarray}
\delta I_c=\sqrt{I_c^{var}}
\end{eqnarray}
}
The $\delta I_c$ vs. $I_c$ plots are shown in Fig.~\ref{phase}. These
plots are obtained by varying the flux uniformly in the entire range $\phi\in[-\pi\phi_0,\pi\phi_0]$.
In the delocalized phase, the $\delta I_c$-$I_c$ curves are open and
discontinuous for the non-special values of $\nu$ due to the presence
of quasi-degeneracies in the energy spectrum, as shown in
Fig.~\ref{phase}(a) whereas these are closed continuous curves for the
special fillings as can be seen from Fig.~\ref{phase}(b) due to no
degeneracies. Such discontinuous $\delta I_c$-$I_c$ curves at the
degenerate points have been recently reported for a one-dimensional
model without any disorder~\cite{pc_metcalf}. At the critical phase,
the $\delta I_c$-$I_c$ curves are closed ones for the non-special
$\nu$ due to the lifting of degeneracies [Fig.~\ref{phase}(c)] and the
curves are straight lines for special values of $\nu$
[Fig.~\ref{phase}(d)]. In the localized phase, the $\delta I_c$-$I_c$
curves are horizontal lines for any value of $\nu$ [see
Figs.~\ref{phase}(e) and \ref{phase}(f)]. The area enclosed by the
$\delta I_c$-$I_c$ curves can be calculated to distinguish between the
non-special and special fillings across the
delocalization-localization transition in the AAH model. For the
non-special values of $\nu$, the area is undefined in the delocalized
phase for $\lambda<2$ due to quasi-degeneracies at the specific values
of $\phi$. At $\lambda=2$, for the non-special values of $\nu$, the
$\delta I_c$-$I_c$ curves are closed ones and the area enclosed by
these are finite whereas the area goes to zero in the localized phase
$\lambda>2$ as shown in Fig.\ref{area}(a). For the special fillings,
for small values of $\lambda$ the area is finite but it goes to zero
as $\lambda$ increases well before it reaches the critical point
$\lambda=2$ as can be seen in Fig.~\ref{area}(b). This
once again reinforces the finding that the criticality of the AAH model is reflected by the
non-special fillings but not by the special fillings.

It is noteworthy that although the isolated single
  particle localized states are absent in single-particle spetra when
  the system size is chosen to be a Fibonacci number ($N=144,610,...$), the many-particle
  results on the persistent current depend on whether the values of
  filling fraction $\nu$ are special ones or not. Hence, this
  establishes that the contrasting behavior between the two kinds of filling
  fractions is a manifestation of the large gaps in the single-particle spectra rather than the presence or absence of the isolated
  localized single particle states.  We would like to mention that
  there are large gaps above the Fermi level when
  $\nu=\alpha,\alpha^2,\alpha^3, \alpha^4$, which approximately
  evaluates to $\nu=0.618,0.382,0.236,0.145$. One has to choose the
  numerical values of $\alpha$ carefully so that there is a large gap
  above the Fermi level. The results remain qualitatively unchanged
  even for smaller system sizes. However, the
  delocalization-localization transition at $\lambda=2$ for
  non-special filling fractions becomes sharper as the system size $N$
  increases. For special filling fractions the system gets localized
  at $\lambda<2$. As system size increases, these localization
  transition points shift to lower values of $\lambda$. Hence
  quantitatively, the distinctions between special fillings and
  non-special fillings as shown in Fig.5, Fig.7, Fig.9, Fig.12
  etc. become sharper as one goes from smaller to larger system sizes.

\section{Conclusion}
We have studied the effect of the single particle energy gaps on the
transport properties (single-particle and many-particle fermionic) of
the AAH potential in one dimension. Even in the well
  known delocalized phase ($\lambda<2$), the IPR and von Neumann
  entropy of the single-particle eigenstates drastically rise and fall
  off respectively for non-Fibonacci system sizes when the eigenstate
  index $=\alpha,\alpha^2,\alpha^3, \alpha^4 (\approx
  0.618,0.382,0.236,0.145)$. The single particle spectrum contains
  large gaps above the energy levels with these special indices.  The
  eigenstates with the special indices are isolated states with a
  different kind of localization properties as compared to the other
  eigenstates. However, these isolated localized states disappear when
  the system sizes are Fibonacci numbers such as $N=144,610$ etc. For
  Fibonacci system sizes, all the single particle eigenstates have the
  same kind of localization properties and the peaks (falls) in the
  IPR (von Neumann entropy) vanish.

  Next we have considered the many-particle properties of the free
  fermionic ground states. The entanglement entropy falls off
  substantially when the filling fraction is $\nu=\alpha^2,\alpha^3,
  \alpha^4(\approx 0.382,0.236,0.145)$. Also the subsystem scaling of
  $S_A$ follows the area-law at the same special values of $\nu$
  whereas in the delocalized phase $S_A$ shows logarithmic violation
  of the area-law for normal fillings $\nu=0.2, 0.3, 0.5$. The
  entanglement entropy of the fermionic ground states with special
  values of $\nu$ do not seem to show any signatures of criticality at
  $\lambda_c=2$ unlike the non-special fillings. All these
  filling-dependent features on many-particle entanglement entropy are
  completely independent of whether or not the system size is a
  Fibonacci number. This proves that the contrasting behavior
  of the entanglement entropy for the special and non-special fillings
  is a manifestation of the large gaps rather than the isolated
  localized states in the single particle spectra. The persistent
current of fermions exhibits similar striking filling-fraction
dependent behavior. It is vanishingly small even in the delocalized
phase for those special fillings and the criticality of the AAH model
is not reflected in the behavior of the current with $\lambda$. We
have also calculated the variance of the persistent current and
explored its relations with the mean value of persistent current in
our model. In the delocalized phase, the standard deviation vs. mean
persistent current curves are discontinuous for the non-special values
of filling fractions and continuous (closed) for the special values of
filling fractions whereas these curves become straight lines in the
localized phase for both types of filling fractions. The area enclosed
by these curves can be used to distinguish between the non-special
filling fractions and special filling fractions across the
delocalization-localization transition. In the delocalized phase, the
area is undefined for non-special fillings and finite for special
fillings whereas the area is zero in the localized phase for all
fillings. 

The persistent current depends on whether
  the values of filling fraction are special ones or not even when system
  sizes are Fibonacci numbers when the isolated single particle
  localized states are absent in the single particle spectra. Hence, this again establishes that the
  interesting many particle phenomena depend on the locations of large
  gaps in the single particle energy spectra rather than on the
  presence or absence of the isolated localized states. The effect of
energy gap structures on transport properties as described in this
work has been hitherto unexplored. The connection between the
persistent current and its quantum fluctuations is very interesting
and may be extended to other contexts like the many-body
localization. The filling fraction
  dependent persistent current is one of the striking aspects of our work. This may
  potentially be tested in cold atom based
  experiments~\cite{pc0,pc1,pc2,pc4} by creating a synthetic flux in
  bichromatic optical lattices, given that the results are
  independent of system sizes and even valid for small system
  sizes. The non-equilibrium dynamics of a wave-packet with specific
  energy may be used to probe the single-particle transport properties
  as it has been done in Ref.~\onlinecite{ASinha2018}. A similar
  non-equilibrium study of a cleverly defined many-particle
  wave-packet may be useful in this regard, although this work
  is based on the equilibrium transport properties. Our results open
up the possibility of further exploration of non-trivial
filling-fraction dependent localization or delocalization phenomena in
the AAH model. Hopefully our work will encourage the community to
revisit the energy spectra of quantum systems with quasi-periodic
potential and thus help fine-tune our understanding of
filling-fraction dependence.

\section*{Acknowledgements}
N.R is grateful to the University Grants Commission (UGC), India for his Ph.D fellowship. 
A.S is grateful to SERB for the startup grant (File Number: YSS/2015/001696) and to DST-INSPIRE Faculty Award [DST/INSPIRE/04/2014/002461]. We would like to thank Archak Purkayastha for useful comments.

\bibliography{refs}
\appendix

\section{Degeneracy and persistent current: no disorder case}\label{app}
\begin{figure}[h!]
\includegraphics[width=0.75\columnwidth]{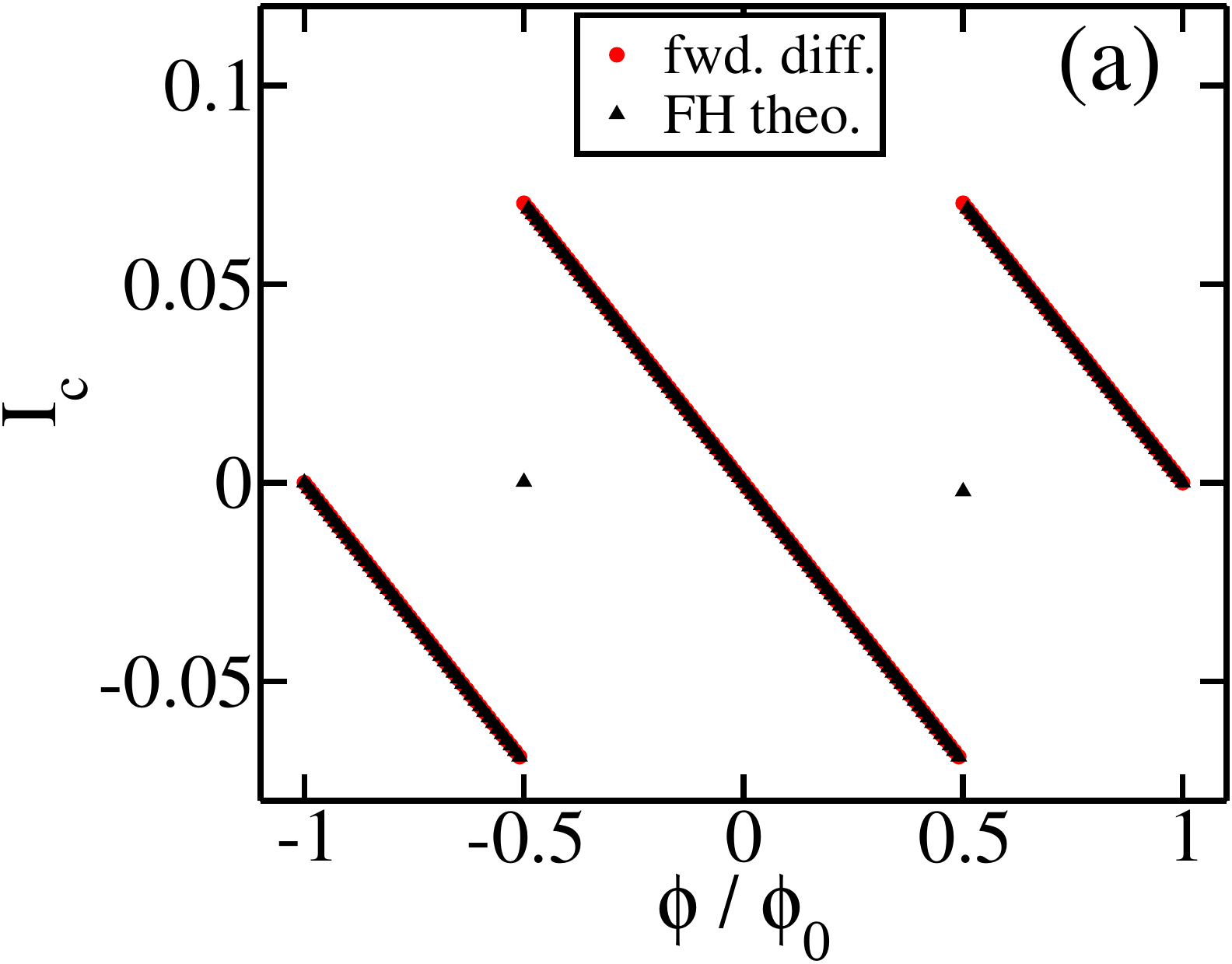}
\includegraphics[width=0.75\columnwidth]{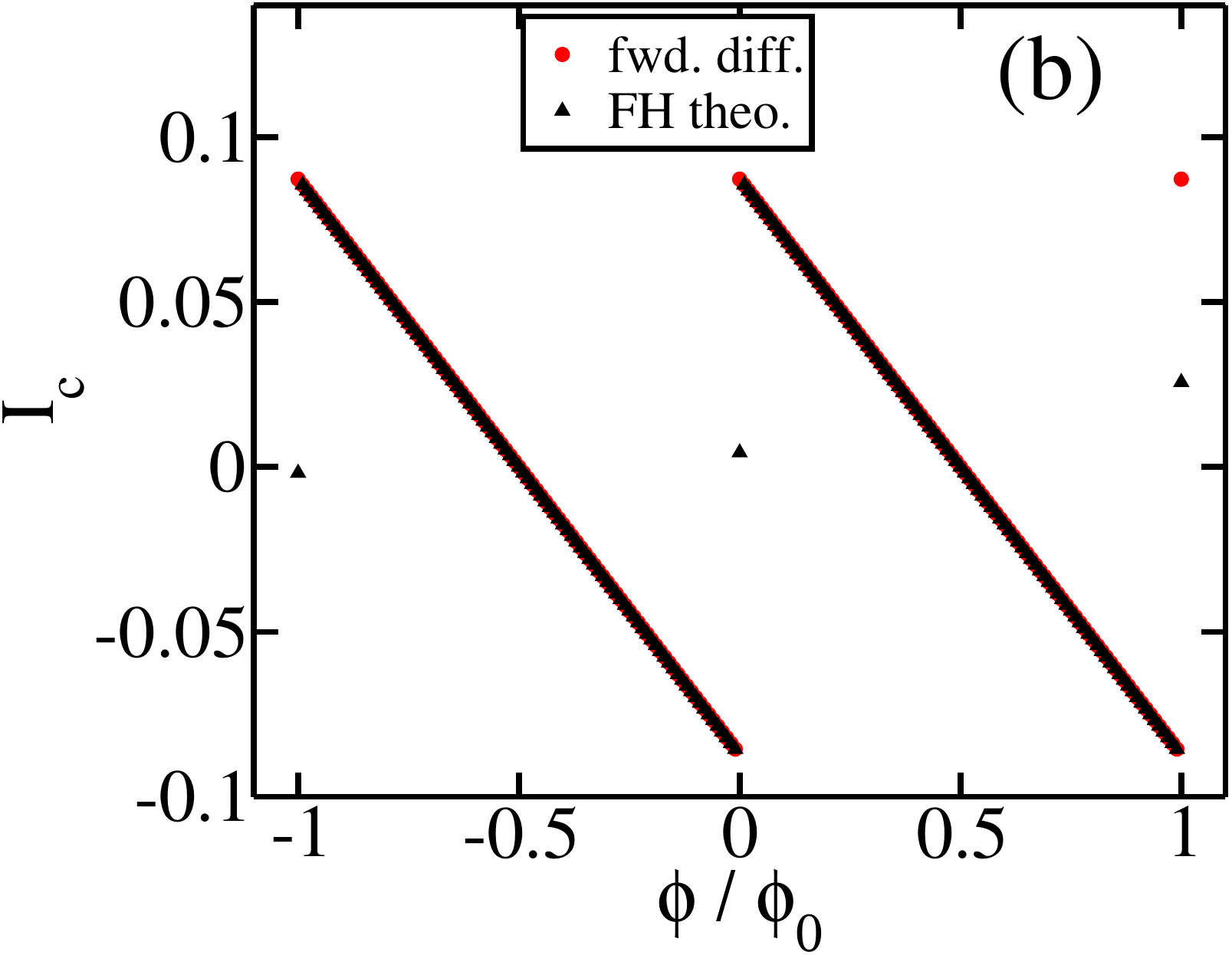}
\caption{Comparison of two approaches to compute the persistent current. (a) The variation of persistent current $I_c$ with the flux $\phi$ (in units of $\phi_0$) for fermions with filling fraction $\nu=0.3$ calculated by taking derivative of the ground state energy using ``forward difference" method and by taking the expectation of the current operator, which is connected to the first method via `FH' theorem. (b) The same plot for $\nu=0.5$. Total number of sites $N=144$ and disorder strength $\lambda=0$ for all the plots.}
\label{pc_app}
\end{figure}
\begin{figure}
\includegraphics[width=0.75\columnwidth]{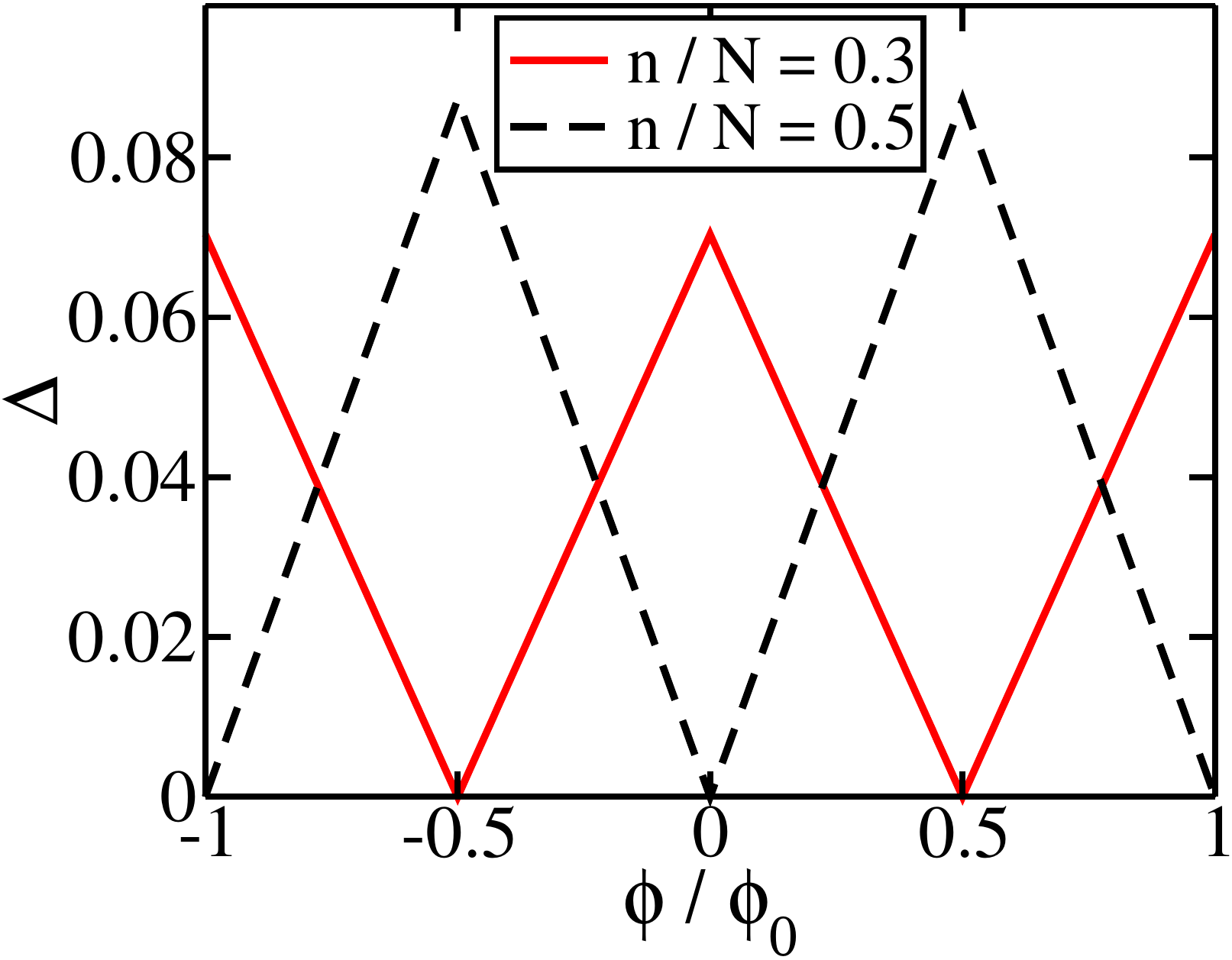}
\caption{Dependence of the single particle gaps $\Delta$ on the flux $\phi$ (in units of $\phi_0$) at the $n\textsuperscript{th}$ energy level where $n/N=0.3$ and $0.5$ respectively. Total number of sites $N=144$ and disorder strength $\lambda=0$ for all the plots.}
\label{gap_app}
\end{figure}   
In this appendix we will discuss breakdown of the Feynman-Hellmann (FH) theorem in presence of degeneracies and how that shows up in the  calculation of the persistent current and its variance for a one-dimensional tight binding chain without any disorder. According to FH theorem
\begin{eqnarray}
\langle -\frac{\partial H}{\partial \phi}\rangle = -\frac{\partial E_0}{\partial \phi},
\end{eqnarray} 
 which simplifies to
 \begin{eqnarray}
 \langle J^c\rangle=-\frac{\partial E_0}{\partial \phi},
 \label{FHT}
 \end{eqnarray}
 where $E_0$ is energy of the many-fermionic ground state, flux $\phi$ (in units of $\phi_0$) and $J^c$ is defined in the main text Eq.~\ref{pc_corr}, which is calculated using the elements of the correlation matrix. So the left-hand side of Eq.~\ref{FHT} involves the fermionic ground state wavefunction whereas the right-hand side of the same involves the ground state energy. The single-particle energy dispersion for a one-dimensional periodic lattice without any disorder is given by,
 \begin{eqnarray}
 \epsilon_n = -2J\cos\bigg(\frac{2\pi}{N}\bigg(n + \frac{\phi}{\phi_0}\bigg)\bigg),
 \end{eqnarray}
 where $n=0,\pm1,\pm2,\pm3,...$. For a non-degenerate spectra, using $E_0 = \sum_{n}\epsilon_{n}(\phi) \theta(E_{F} - \epsilon_{n})$ one obtains the expression of the persistent current written in Eq.~\ref{pc} in the main text, using both the approaches depicted in Eq.~\ref{FHT}.    
Figs.~\ref{pc_app}(a) and \ref{pc_app}(b) show the variation of
the persistent current $I_c$ with flux $\phi$ for filling fraction
$\nu=0.3$ and $0.5$ respectively, computed using both the approaches
for comparison. It is to be noted that there is an exact overlap
between two methods except at $\phi=\pm0.5 \phi_0$ and $\phi=0,\pm1
\phi_0$ in Figs.~\ref{pc_app}(a) and \ref{pc_app}(b)
respectively. The $I_c-\phi$ curves look different for different
fillings as number of fermions is odd in one case and even in the
other case and as $I_c$ depends on whether the number of fermions is
odd or even.
When the Fermi level lies in the degenerate levels, the current by taking the derivative of the ground state energy calculated using the forward difference method happens to give the known exact results whereas the calculation of expectation of the current operator involving the ground state wavefunction gives a different result. The single-partcle energy gap $\Delta$ at the $n\textsuperscript{th}$ level is shown as function of flux $\phi$ for $n/N=0.3$ and $0.5$ respectively in Fig.~\ref{gap_app}. The gap vanishes at $\phi=\pm0.5 \phi_0$ and $\phi=0,\pm \phi_0$ respectively.

\begin{figure}[htpb]
\centering
\includegraphics[width=0.75\columnwidth]{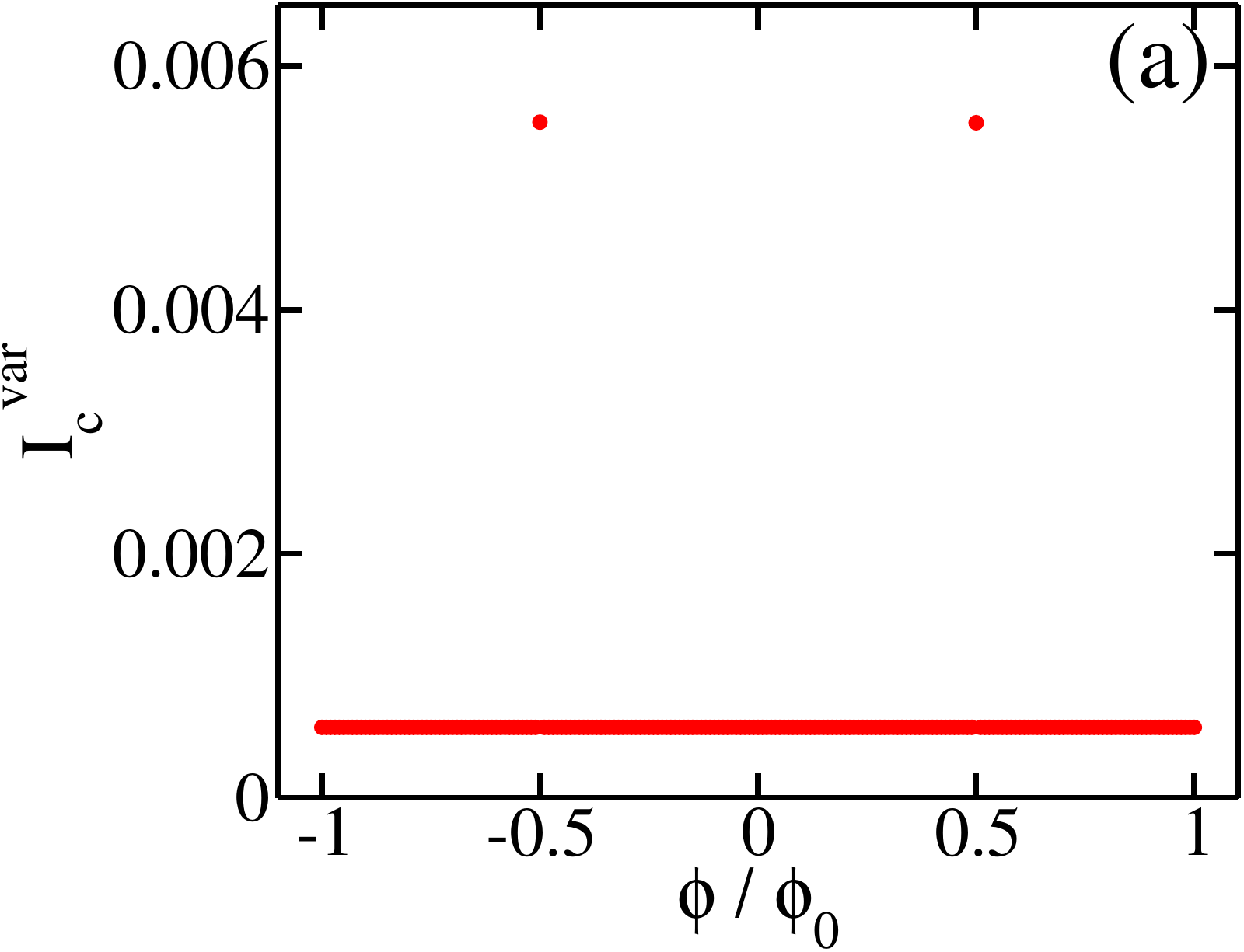}
\includegraphics[width=0.75\columnwidth]{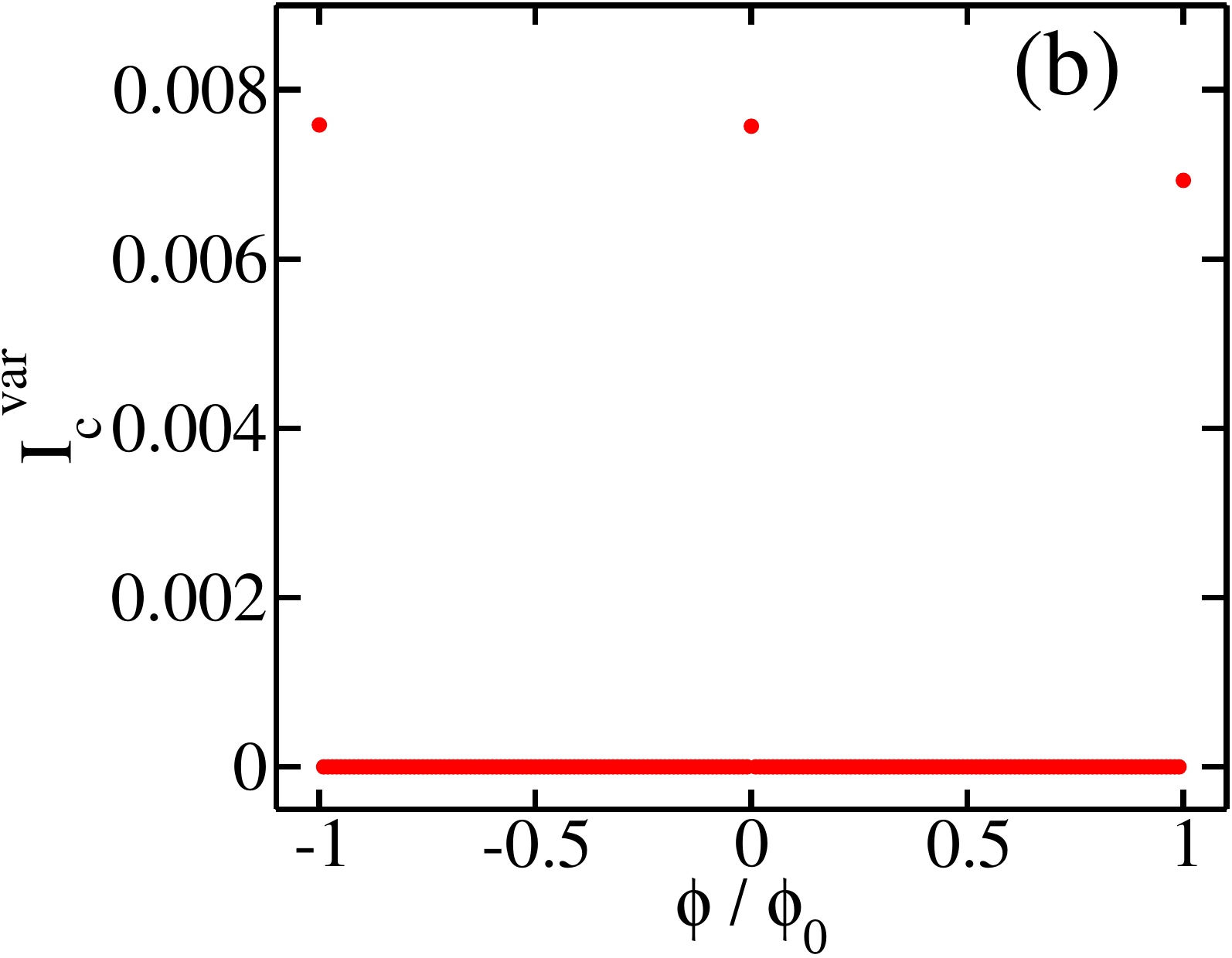}
\caption{(a) The variation of $I_c^{var}$ with the flux $\phi$ (in units of $\phi_0$) for fermions with filling fraction $\nu=0.3$ calculated using Eq.~\ref{Ic_var}. (b) The same plot for $\nu=0.5$. Total number of sites $N=144$ and disorder strength $\lambda=0$ for all the plots.}
\label{pcvar_app}
\end{figure} 
The variance of the persistent current $I_c^{var}$ as function of flux $\phi$ is shown in Fig.~\ref{pcvar_app}(a) and Fig.~\ref{pcvar_app}(b) for filling fractions $\nu=0.3$ and $0.5$ respectively. The curves show discontinuities exactly at the point of degeneracies as mentioned earlier.  
At the point of degeneracies, any linear superposition $\ket{\Psi_0}$ of the true ground states $\ket{\Phi_0}$ is also a valid solution of the Schr$\ddot{o}$dinger equation. On numerical computation, one does not obtain the true degenerate ground state and the FH theorem as mentioned in Eq.~\ref{FHT} breaks down in presence of degeneracies~\cite{FH_zhang,FH_alon,FH_fernandez,FH_vatsya}. The corrected FH theorem for degenerate case can be described by adding one additional condition in the degenerate subspace, written as~\cite{FH_fernandez,FH_alon,FH_vatsya}
\begin{eqnarray}
&&\bra{\Phi_{0,i}} -\frac{\partial H}{\partial \phi} \ket{\Phi_{0,j}} = 0 \hspace{0.25cm} \text{for}\hspace{0.25cm} i\neq j\nonumber \\
&& \bra{\Phi_{0,i}} -\frac{\partial H}{\partial \phi} \ket{\Phi_{0,i}} = -\frac{\partial E_0}{\partial \phi}
\label{FHT_deg}
\end{eqnarray}
where $i,j=1,...,q$ for the q-degenerate subspace. After numerically computing the ground states $\ket{\Psi_0}$'s at the degenerate point, one essentialy constructs matrix A given by
\begin{eqnarray}
A_{ij}=\bra{\Psi_{0,i}} -\frac{\partial H}{\partial \phi} \ket{\Psi_{0,j}},
\end{eqnarray}   
which contains off-diagonal elements. Then one diagonalizes the matrix A to obtain the correct linear superposition
\begin{eqnarray}
\ket{\Phi_{0,i}}=\sum\limits_{j=1}^{q}C_{ij} \ket{\Psi_{0,j}}
\end{eqnarray}
that satisfies the conditions shown in Eq.~\ref{FHT_deg}. The degenerate points should be treated specially under the treatment charted out here. These details are included in the interest of completion, although we have not used such treatments in our work.  
\end{document}